\documentclass[10pt,aps,prd,twocolumn,superscriptaddress,floatfix,amsmath,amssymb,amsfonts,longbibliography,nofootinbib]{revtex4-2}

\usepackage[english]{babel}
\usepackage{graphicx}
\usepackage{dcolumn}
\usepackage{bm}

\usepackage{physics}
\usepackage{amsmath,amssymb,mathtools}
\usepackage{booktabs}
\usepackage{tikzsymbols}
\usepackage[utf8]{inputenc}
\usepackage[T1]{fontenc}
\usepackage{array}
\usepackage{booktabs}

\usepackage{tikz}
\usetikzlibrary{arrows.meta, positioning}
\usepackage{lipsum}
\usepackage{hyperref}

\usepackage{pgfplots}
\pgfplotsset{compat=1.18}

\hypersetup{
    colorlinks=true,
   linkcolor=blue,
    filecolor=red,      
    urlcolor=cyan,
}

\begin{document}

\title{State-Selective Signatures of Quantum and Classical Gravitational Environments}

\author{Partha Nandi}
\email{pnandi@sun.ac.za}
\affiliation{Department of Physics, University of Stellenbosch, Stellenbosch, 7600, South Africa}
\affiliation{National Institute of Theoretical and Computational Sciences (NITheCS), Stellenbosch, 7604, South Africa}

\author{Sankarshan Sahu}
\email{sankarshan.sahu@sorbonne-universite.fr}
\affiliation{Sorbonne Universit\'e, CNRS, Laboratoire de Physique
Th\'eorique de la Mati\`ere Condens\'ee, LPTMC, 75005 Paris, France}

\author{Bibhas Ranjan Majhi}
\email{bibhas.majhi@iitg.ac.in}
\affiliation{Department of Physics, Indian Institute of Technology Guwahati, Guwahati 781039, Assam, India}

\author{Francesco Petruccione}
\email{petruccione@sun.ac.za}
\affiliation{National Institute of Theoretical and Computational Sciences (NITheCS), Stellenbosch, 7604, South Africa}
\affiliation{School of Data Science and Computational Thinking, University of Stellenbosch, Stellenbosch 7600, South Africa}


\date{\today}

\begin{abstract}
A unified framework is developed for determining whether a gravitational-wave (GW) background behaves as a classical field or as a genuinely quantum environment. \emph{Unified} here means that both descriptions originate from the same tidal coupling derived from geodesic deviation, which yields an identical quadratic interaction Hamiltonian for the detector; the only distinction lies in whether the GW degrees of freedom are modeled as classical phase-randomized coherent states or as quantized graviton modes. Within this common framework, the reduced dynamics of a quantum harmonic oscillator exhibit a sharp structural contrast: a quantized graviton bath preserves coherence within the lowest phonon-number manifold, forming a protected sector at leading order, whereas a classical stochastic GW field inevitably induces decoherence even inside this subspace. This difference provides an operational criterion for diagnosing the classical or quantum nature of gravitational waves using mesoscopic optomechanical systems. Our results establish decoherence structure—not merely its magnitude—as a sensitive probe of gravitational quantumness and delineate the experimental regimes under which such tests may become feasible. 

\end{abstract}

\maketitle


\section{Introduction}

The first direct detection of gravitational waves by LIGO in 2015~\cite{abbott,KAGRA:2021vkt} confirmed the predictions of Einstein’s general relativity with remarkable precision and established gravitational radiation as a tangible physical reality. Yet a deeper question persists: do gravitational waves themselves admit a fundamentally quantum description, and if so, can this be established operationally? 
More generally, the interpretation of quantum states of the gravitational field can be subtle: configurations that appear as superpositions of spacetime geometries may in some cases admit equivalent descriptions on a single classical background with appropriately transformed quantum degrees of freedom \cite{Foo:2025}. This issue lies at the heart of the interface between quantum mechanics and gravitation.

In practice, the statement that ``gravitational waves are classical''
is typically understood to mean that the radiation field can be
described by coherent states with large occupation numbers, which
represent the quantum states that most closely resemble classical wave
configurations~\cite{Manikandan:2025ykr}. In quantum optics, coherent
states exhibit minimal quantum fluctuations and the field operators act
effectively as $c$-numbers in expectation values, reproducing the
behavior of classical radiation fields. From this perspective,
deviations from coherent-state statistics—such as thermal, squeezed, or
other nonclassical states—would signal intrinsically quantum features
of gravitational radiation.


This viewpoint has recently been sharpened by Manikandan and Wilczek, who proposed operational tests of the coherent-state hypothesis using resonant bar detectors and counting statistics~\cite{Manikandan:2025hlz, Manikandan:2025qgv}. Their analysis highlights that departures from coherent-state behavior could provide direct evidence for a quantum structure underlying gravitational waves.

In the present work, we adopt a complementary dynamical perspective.
Rather than analyzing the counting statistics of the gravitational field
itself, we investigate how a mesoscopic quantum
system~\cite{Marti:2023abu} evolves when coupled to a gravitational-wave
background. Within the framework of open quantum
systems~\cite{BreuerPetruccione2006}, the gravitational field plays the
role of an environment whose microscopic nature—whether a classical
stochastic field or a quantized graviton bath—can imprint distinct
structural signatures on the reduced dynamics of a quantum probe.

More generally, connections between coherent-state phase-space
representations and effective hydrodynamic descriptions of quantum
dynamics have long been recognized. As an illustrative example,
Ghosh \textit{et al.} showed that the Schr\"odinger equation can be
mapped to Euler and Navier--Stokes–type equations in a coherent-state
representation when environmental modes are included
\cite{Ghosh:1977SchrodingerFluid}. This perspective highlights how
coupling to unobserved degrees of freedom can induce effective
dissipative dynamics even in fundamentally quantum systems.

Related open-system formulations have been employed to study
decoherence induced by stochastic fluctuations of quantum spacetime,
where Planck-scale geometric noise acts as an effective environment
for quantum systems~\cite{Nandi:2025qyj}. Open-system ideas have also
been explored in cosmological contexts, where tracing over
inhomogeneous fluctuations can render the homogeneous sector
effectively open and lead to the emergence of dark-sector components
from quantum openness~\cite{Nandi:OpenUniverse2026}. In parallel,
stochastic formulations of quantum gravity have been investigated in
which geometric degrees of freedom may arise from probabilistic
processes on underlying microscopic structures~\cite{Nandi:2025hqe}.
Gravity-induced decoherence and entanglement redistribution in
many-body gravitational systems have likewise been studied in various
contexts~\cite{Zhang:2025ozb,PhysRevD.103.026017}
(see also \cite{Guerreiro:2025sge,Kaku:2024lgs,Jones:2024npd,Lin:2025ipk}).

These developments are conceptually related to the broader framework
of semiclassical gravity, where quantum fields propagate on a classical
spacetime and the geometry evolves according to the expectation value
of the stress–energy tensor~\cite{SinghPadmanabhan1987}. Within this
picture, quantum fluctuations of matter fields can influence the
effective spacetime dynamics even when the gravitational sector itself
is treated classically.

Our aim is to formulate and analyze this scenario in a minimal yet
experimentally relevant setting, focusing on qualitative differences in
the structure of decoherence rather than on the absolute magnitude of
decoherence rates.

Direct probes of gravitational-wave–induced motion face fundamental
limitations. Even in advanced interferometers such as LIGO, a strain
$h\sim10^{-21}$ displaces a $40$\,kg mirror by only
$\sim10^{-18}\,\mathrm{m}$, comparable to quantum and technical noise
levels~\cite{LIGOSuspension2012,Sigg1998GravitationalWaves,
Nandi:2024jyf,Nandi:2024zxp}. For realistic astrophysical backgrounds
($h\sim10^{-23}$–$10^{-26}$), the induced displacement becomes many
orders of magnitude smaller than the mirror’s zero-point fluctuations,
making displacement-based access to quantum features of gravity
essentially impossible.

Mesoscopic mechanical systems offer a conceptually different
opportunity. Levitated or optomechanical resonators can act as
high-sensitivity probes of weak forces and have been proposed
as detectors of high-frequency gravitational waves
\cite{Arvanitaki2013}. Although the GW-induced displacement of a micron-scale
resonator is likewise extremely small—far below its zero-point
amplitude—the key advantage of mesoscopic platforms lies in their
ability to prepare and measure well-defined quantum states of motion.
State-of-the-art optomechanical membranes, phononic-crystal resonators,
and circuit-QED hybrids can access the lowest phonon-number
manifolds~\cite{OConnell2010,Teufel2011,Chan2011,Miki:2023nce,Wan:2017thesis},
enabling coherence-based probes of gravitational physics rather than
displacement sensing.

To illustrate the scale, consider a GHz resonator such as a
high-overtone bulk acoustic resonator (HBAR)~\cite{Chu:2017koi} or an
optomechanical crystal (OMC)~\cite{OConnell2010,Teufel2011,Chan2011}
with a characteristic mode length $L\lesssim1\,\mu\mathrm{m}$. 
In a representative HBAR device, the mechanical mode has
$\omega/2\pi\simeq5.96~\text{GHz}$ and effective mass
$m_{\rm eff}\simeq16.2~\mu\text{g}$, yielding a zero-point amplitude
$x_{\rm zpf}\simeq2.9\times10^{-19}\,\mathrm{m}$ for this
particular GHz-scale device \cite{PhysRevD.111.026009}. In other
optomechanical platforms with smaller masses and lower frequencies,
the zero-point displacement can be significantly larger
(typically $x_{\rm zpf}\sim10^{-17}\,\mathrm{m}$), bringing it closer
to gravitational-wave strain scales. For representative gravitational-wave strain amplitudes (e.g.\ $h\sim10^{-23}$), the direct GW-induced displacement
$\Delta x_{\rm GW}=hL$ is therefore only
$\sim10^{-29}\,\mathrm{m}$ for micron-scale devices, so that
$\Delta x_{\rm GW}/x_{\rm zpf}\sim10^{-11}$. Such signals are
far too small to classicalize the oscillator or populate phonons;
the device therefore remains fully quantum. In this regime, the
relevant signature of the gravitational environment is not the
displacement amplitude but its \emph{statistical structure}.

Motivated by these developments, we model the detector as a generic
quantum mechanical resonator—formally, a single quantum harmonic
oscillator mode coupled to a gravitational-wave background
\cite{Nandi:2022sjy,Nandi:2024zxp,Nandi:2024jyf,Dutta:2025bge,Dutta:2025ouy}.
Within a unified open-quantum-system framework, we compare two physically
distinct descriptions of the gravitational field:

\begin{enumerate}
    \item \textbf{Quantum gravitational waves}: the gravitational field is
    treated as a Gaussian quantum environment, taken either in the vacuum
    state or in an effective thermal (Gibbs) state. Here the term
    ``thermal'' refers to the form of the quantum density matrix and does
    not imply ongoing graviton thermalization through interactions \cite{Cho:2021gvg}. Such
    a state may arise, for example, as a relic of early-universe initial
    conditions (e.g.\ a radiation-dominated era preceding inflation
    \cite{Bhattacharya:2006prl, PhysRevLett.111.021302}),
    or may be employed as a convenient parametrization of Gaussian
    quantum fluctuations.

    \item \textbf{Classical gravitational waves}: the gravitational field
    is modeled as a classical stochastic background, represented by an
    ensemble of phase-randomized coherent states. This description
    captures incoherent superpositions of classical gravitational waves
    emitted by many independent sources and provides a natural classical
    noise limit of the theory.
\end{enumerate}

The choice of gravitational field states is motivated by the distinct physical
regimes they represent. A pure coherent state of the graviton field possesses
a nonvanishing expectation value $\langle h_{ij} \rangle$ and thus corresponds
to a deterministic classical gravitational wave~\cite{Manikandan:2025ykr}. Thermal and phase-randomized
coherent states, by contrast, have vanishing one-point functions but nonzero
two-point correlators, capturing stochastic gravitational-wave backgrounds with
either quantum or classical origin.
Within the same microscopic detector--gravity coupling, these different field
states give rise to qualitatively distinct reduced dynamics for the harmonic
oscillator. In particular, the structure of the induced master equation---and the existence or absence of protected subspaces---allows one to distinguish between quantum-vacuum fluctuations and classical stochastic gravitational noise, without relying solely on the magnitude of decoherence rates.

In both cases, the interaction arises from the same geodesic-deviation
tidal coupling, leading to an identical quadratic detector–gravity
Hamiltonian; the difference lies solely in the statistical properties of
the gravitational sector. Building on this unified starting point, we
derive the associated master equations and analyze their
dynamics using perturbative and Liouvillian techniques.
Our central result is that, for a quantized graviton field
in the vacuum state, the reduced dynamics exhibits a
Fock-space \emph{protected subspace}, within which coherence
in the lowest phonon-number manifold is preserved at leading
order. This protection is absent for classical
phase-randomized backgrounds and is also lifted in the
presence of finite graviton occupation (e.g.\ thermal
states), where decoherence becomes unavoidable.
This sharp structural distinction provides a clear
operational method for discriminating between classical and
genuinely quantum-vacuum gravitational fluctuations using
mesoscopic mechanical probes. In addition, by estimating the
magnitude of graviton-induced decoherence for realistic GHz
mechanical platforms, we derive model-independent bounds on
possible non-vacuum graviton occupation numbers at the
detector frequency.


\noindent\textbf{Organization of the paper.}  
Section~II introduces the oscillator–gravitational-wave interaction and
derives the effective Hamiltonian.  
Section~III formulates the open-system dynamics for a quantized graviton
bath.  
Section~IV analyzes the decoherence produced by a classical,
phase-randomized background and contrasts it with the quantum case.
Sections~V and~VI discuss experimental implications, coherence-time
estimates, and prospects for detection.  
Appendices provide supporting derivations and technical details.


\section{Quantum Mechanics on a Gravitational Background}

We consider weak gravitational waves propagating on a flat Minkowski background. The spacetime metric is written as
\begin{equation}
g_{\mu\nu} = \eta_{\mu\nu} + h_{\mu\nu}, 
\qquad |h_{\mu\nu}| \ll 1,
\end{equation}
where \( h_{\mu\nu} \) is a small perturbation around the flat metric \( \eta_{\mu\nu} \).  
In the \emph{transverse–traceless} (TT) gauge, the physical degrees of freedom are isolated by imposing
\begin{equation}
h_{0\mu} = 0, \qquad 
\partial^i h_{ij} = 0, \qquad 
h^{i}{}_{i} = 0.
\end{equation}
Under these conditions, the quadratic (free) Einstein–Hilbert action becomes
\begin{equation}
S_{\mathrm{grav}} = 
\frac{1}{64\pi G}
\int d^4x \,
\big(
\dot{h}_{ij}\dot{h}^{ij}
- \partial_k h_{ij}\,\partial_k h^{ij}
\big),
\end{equation}
describing two tensorial modes satisfying the vacuum wave equation
\begin{equation}
\square h^{TT}_{ij}(t, \vec{x}) = 0, 
\qquad i,j = 1,2,3,
\end{equation}
corresponding to freely propagating gravitational waves with ``$+$’’ and ``$\times$’’ polarizations.

To analyze the interaction of gravitational waves with matter, we consider a congruence of timelike geodesics \( x^{\mu}(\tau) \) centered around a freely falling detector. The analysis is performed in the \emph{proper detector frame}, a locally inertial coordinate system constructed along the detector’s worldline~\cite{Antoniou2016PropagationGW}. In this frame, the metric is locally Minkowskian at the origin, while curvature effects manifest as tidal forces proportional to spatial separations. The four-velocity of the detector is \( u^\mu = (1, 0, 0, 0) \), and the spatial separation vector \( \xi^\mu \), defined by \( \xi^\mu u_\mu = 0 \), represents the physical displacement between nearby geodesics~\cite{maggiore2008}.  

The relative acceleration between neighboring geodesics is given by the geodesic deviation equation~\cite{MTW:1973}:
\begin{equation}
\frac{D^2 \xi^\mu}{D\tau^2}
= - R^\mu_{\ \nu\rho\sigma} \, \xi^\rho \, u^\nu u^\sigma,
\label{eq:geodesic-deviation}
\end{equation}
where \( D/D\tau \) denotes the covariant derivative along the trajectory.  
In the weak-field limit, the Riemann tensor is linear in \( h_{\mu\nu} \):
\begin{equation}
R^\mu_{\ \rho\sigma\nu} = 
\frac{1}{2} \eta^{\mu\lambda}
\big(
\partial_\sigma \partial_\rho h_{\nu\lambda}
- \partial_\sigma \partial_\lambda h_{\nu\rho}
- \partial_\nu \partial_\rho h_{\sigma\lambda}
+ \partial_\nu \partial_\lambda h_{\sigma\rho}
\big).
\end{equation}

In the \emph{long-wavelength approximation}, valid when the gravitational wavelength is much larger than the detector size, spatial variations can be neglected:
\begin{equation}
h^{TT}_{ij}(t, \vec{x}) \simeq h^{TT}_{ij}(t).
\end{equation}
Substituting this into Eq.~\eqref{eq:geodesic-deviation} and working in the detector frame yields
\begin{equation}
\frac{d^2 \xi^i}{dt^2}
= \frac{1}{2} \ddot{h}^{TT}_{ij}(t)\, \xi^j,
\end{equation}
where we have used
\begin{equation}
R^j_{\ 0k0} = -\frac{1}{2} \ddot{h}^{TT}_{jk}(t).
\end{equation}
For a gravitational wave propagating along the \( z \)-axis, the perturbation can be expressed as
\begin{equation}
h^{TT}_{ij}(t)
= \chi(t)
\big(
\epsilon_\times \, \sigma^1_{ij}
+ \epsilon_+ \, \sigma^3_{ij}
\big),
\end{equation}
where \( \chi(t) \) represents the time-dependent amplitude and \( \epsilon_\times, \epsilon_+ \) are the polarization components associated with the standard basis tensors \( \sigma^1_{ij} \) and \( \sigma^3_{ij} \), the components of Pauli matrices.

To describe the response of a realistic detector, we consider a mesoscopic optomechanical mirror~\cite{PhysRevA.87.043832} whose center-of-mass (COM) displacement \( \xi_1(t) \) along the \( x \)-axis experiences both optical trapping and gravitationally induced tidal accelerations. We project the dynamics onto the mechanically relevant center-of-mass mode
along the $\xi_{1}(t)$-axis; for a gravitational wave propagating along $z$, the
$\times$ polarization does not excite this mode at linear order in the absence of transverse displacement, and quantum fluctuations in orthogonal directions are neglected as they do not couple to the readout channel \cite{Parikh:2020fhy}.

In Fermi normal coordinates centered at the optical trap, the mirror obeys
\begin{equation}
m_0 \ddot{\xi}_1(t)
+ m_0 \omega_c^2 \xi_1(t)
= \frac{m_0}{2} \ddot{h}^{TT}_{11}(t) \xi_1(t),
\end{equation}
where \( m_0 \) is the mirror mass and \( \omega_c \) the trapping frequency.  
The corresponding effective Hamiltonian is
\begin{equation}
\begin{split}
H_D &= 
\frac{1}{2m_0}
\left[
p_1 + m_0 \dot{h}^{TT}_{11}(t)\, \xi_1
\right]^2
+ \frac{1}{2}m_{0}\omega^{2}_{c}\xi^{2}_{1} \\
&\simeq
\frac{p_1^2}{2m_0}
+  \dot{h}^{TT}_{11}(t)\, \xi_1 p_1
+ \frac{1}{2}m_{0}\omega^{2}_{c}\xi^{2}_{1},
\end{split}
\end{equation}
where \( p_1 \) denotes the canonical momentum.  
This Hamiltonian, first introduced in~\cite{Speliotopoulos:1995} and later used in quantum-gravity-inspired detector models~\cite{PhysRevLett.127.081602, Saha:2018}, forms the basis for our subsequent quantum analysis of the mirror’s motion in a gravitational-wave background.

In large-scale detectors such as LIGO, the end mirrors are suspended pendulums with masses around $m_0\sim$ \( 40\,\mathrm{kg} \) and fundamental suspension frequencies near $\omega_0\sim$ \( 0.74\,\mathrm{Hz} \)~\cite{LIGOSuspension2012}.  
The gravitational-wave-induced displacements (\( \sim 10^{-18}\,\mathrm{m} \)) are about an order of magnitude larger than the quantum zero-point fluctuations, yet the mirror’s high thermal occupation suppresses any observable quantum signatures. Consequently, mirror motion in LIGO is accurately modeled classically, and intrinsically quantum effects—such as gravitationally induced Aharonov–Bohm–type phase shifts—remain undetectable.
In contrast, mesoscopic optomechanical systems employ oscillators with effective masses \( m_0 \sim 10^{-12}\text{–}10^{-9}\,\mathrm{kg} \) and frequencies \( \omega_c \sim 10^5\text{–}10^6\,\mathrm{rad/s} \)~\cite{vonLupke2022, Aspelmeyer2014, Vitali2007}.  
Their quantum zero-point displacement,
\begin{equation}
\Delta x_{\mathrm{zpf}}
= \sqrt{\frac{\hbar}{2 m_0 \omega_c}},
\end{equation}
typically reaches \( \Delta x_{\mathrm{zpf}} \sim 10^{-17}\,\mathrm{m} \), comparable to expected gravitational-wave strains.  
State-of-the-art experiments~\cite{Teufel2011, Chan2011, vonLupke2022, Delic2020} achieve near-ground-state cooling (\( \langle n \rangle < 1 \)), providing an experimentally accessible regime where quantum mechanics and weak spacetime curvature can be jointly explored. 

Thus, in the weak-field approximation, the dynamics of a mesoscopic quantum system  under a GW traveling along the \(z\)-axis is governed by the time-dependent Hamiltonian \cite{Nandi:2024jyf}:
\begin{eqnarray}
\hat{H}(t) &=&  \left( \alpha \hat{p}_1^2 + \beta {\hat{\xi}_{1}^2} \right)+
\gamma(t)\left( \hat{\xi}_1 \hat{p}_1 + \hat{p}_1 \hat{\xi}_1 \right),
\label{eq:GW_Hamiltonian}
\end{eqnarray}

where 
$\alpha = 1/(2m_{0}), \quad \beta = (1/2) m_{0} \omega_{c}^2, \quad \gamma(t) = \frac{\dot{\chi}(t)}{2} \epsilon_+$.

\section{Open Quantum Graviton Environment: Master Equation}

The interaction between a mesoscopic quantum system—modeled here as a single mechanical mode described by a quantum harmonic oscillator—and a quantized gravitational environment induces effective open-system dynamics. In this section, we derive the corresponding master equation governing the decoherence of the detector due to its coupling with an environment of quantized, plus-polarized gravitational wave modes (gravitons).


We begin with the total Hamiltonian,
\begin{align}\label{H}
\hat{H} &= 
\underbrace{\sum_{k} \hbar \omega^{g}_{k} :\hat{b}^{\dagger}_{k}\hat{b}_{k}:}_{\hat{H}_g} \otimes \mathbf{I}_{d}  
+ \mathbf{I}_{g} \otimes 
\underbrace{\hbar \omega_{c} \Big(\hat{a}^{\dagger}_{1}\hat{a}_{1} + \frac{1}{2}\Big)}_{\hat{H}_d} \nonumber\\
&\quad + \underbrace{i\hbar \sum_{k} \gamma^{k}(t) \otimes (\hat{a}^{\dagger 2}_{1}-\hat{a}^{2}_{1})}_{\hat{H}_{\rm int}(t)}.
\end{align}
Here, $\hat{H}_g$ and $\hat{H}_d$ denote the free Hamiltonians of the
gravitational field and the detector, respectively, while
$\hat{H}_{\rm int}(t)$ describes their interaction.
The quantity $\omega^{g}_{k}$ is the angular frequency of the $k^{\text{th}}$ mode of
the linearized gravitational field (graviton), and $\omega_c$ is the natural
frequency of the detector, modeled as a single quantum harmonic oscillator
mode. The operators $\hat{b}_k$ and $\hat{b}_k^{\dagger}$ are the annihilation
and creation operators of the graviton modes, satisfying bosonic commutation
relations, while $\hat{a}_1$ and $\hat{a}_1^{\dagger}$ are the corresponding
annihilation and creation operators for the detector mode, obeying the standard bosonic Fock algebra.
Normal ordering is employed in $\hat{H}_g$ to remove the irrelevant vacuum
energy contribution of the graviton field, thereby avoiding unphysical
divergences (see Appendix~\ref{AppA} for details). The interaction coefficient
$\gamma^{k}(t)$ is defined as

\begin{equation}
\gamma^{k}(t)
=
i\,\mathcal{C}_{\omega^{g}_{k}}
\left(
\hat{b}^{\dagger}_{k} e^{i \omega^{g}_{k} t}
-
\hat{b}_{k} e^{-i \omega^{g}_{k} t}
\right),
\qquad
\mathcal{C}_{\omega^{g}_{k}}
=
\frac{C_{\omega^{g}_{k}}}{\sqrt{L^{3}}},
\label{f}
\end{equation}
where $C_{\omega^{g}_{k}}
=
\sqrt{\frac{\omega^{g}_{k}\, c\, l_{p}^{2}}{2}}$.

Passing to the interaction picture with respect to the free Hamiltonian
$\hat{H}_0 = \hat{H}_g + \hat{H}_d$, the interaction Hamiltonian becomes
\begin{align}
\hat{H}^{I}_{\rm int}(t) &=
\hbar \sum_k \mathcal{C}_{\omega^g_k}
\Big[
\hat{b}_k \hat{a}_1^{\dagger 2}
e^{i(2\omega_c - 2\omega_k^g)t}
- \hat{b}_k \hat{a}_1^{2}
e^{-i(2\omega_c + 2\omega_k^g)t}
\nonumber\\
&\quad
+ \hat{b}_k^{\dagger} \hat{a}_1^{2}
e^{i(2\omega_k^g - 2\omega_c)t}
- \hat{b}_k^{\dagger} \hat{a}_1^{\dagger 2}
e^{i(2\omega_c + 2\omega_k^g)t}
\Big].
\label{ha}
\end{align}
This expression follows from the standard interaction-picture
transformation
\begin{equation}
\hat{H}^{I}_{\rm int}(t)
=
e^{\frac{i}{\hbar}(\hat{H}_g + \hat{H}_d)t}
\hat{H}_{\rm int}(t)
e^{-\frac{i}{\hbar}(\hat{H}_g + \hat{H}_d)t}.
\end{equation}
The time evolution of the total density matrix is governed by the von-Neumann equation,
\begin{equation}\label{VN}
\frac{d}{dt} \hat{\rho}^{\rm tot}(t) = -\frac{i}{\hbar} [\hat{H}^{I}_{\rm int}(t), \hat{\rho}^{\rm tot}(t)].
\end{equation}
Formally integrating Eq.~\eqref{VN} gives
\begin{equation}
\hat{\rho}^{\rm tot}(t) = \hat{\rho}^{\rm tot}_0 - \frac{i}{\hbar} \int_0^t ds \, [\hat{H}^{I}_{\rm int}(s), \hat{\rho}^{\rm tot}(s)],
\end{equation}
which, upon substitution, yields
\begin{align}\label{VN1}
\frac{d}{dt} \hat{\rho}^{\rm tot}(t) 
    &= - \frac{i}{\hbar} [\hat{H}^I_{\rm int}(t), \hat{\rho}^{\rm tot}_0] \nonumber\\
    &\quad - \frac{1}{\hbar^2} \int_0^t ds \, [\hat{H}^I_{\rm int}(t), [\hat{H}^I_{\rm int}(s), \hat{\rho}^{\rm tot}(s)]].
\end{align}
In the above $\hat{\rho}^{\text{tot}}_0$ is the initial composite density matrix.

Applying the Born approximation (i.e. the interaction between the graviton modes and the detector is very weak, so that 
\(\hat{\rho}^{\rm tot}(s) \approx \hat{\rho}_d(s) \otimes \rho_g\)),  
and then tracing over the gravitational field degrees of freedom, we obtain
\begin{equation}\label{VN2}
\frac{d}{dt} \hat{\rho}_d(t) = - \frac{1}{\hbar^2} \int_0^t ds \, 
\operatorname{Tr}_g \Big[ \hat{H}^I_{\rm int}(t), [\hat{H}^I_{\rm int}(s), \hat{\rho}_d(s) \otimes \rho_g] \Big].
\end{equation}

The first-order contribution 
$\operatorname{Tr}_g\!\left[\hat{H}^I_{\rm int}(t),\hat{\rho}^{\rm tot}_0\right]$
vanishes under the standard assumption that the initial total state is
factorized, $\hat{\rho}^{\rm tot}_0=\hat{\rho}_d(0)\otimes\rho_g$, and that
the gravitational environment has vanishing first moments,
$\mathrm{Tr}_g(\hat{b}_k\,\rho_g)=\mathrm{Tr}_g(\hat{b}_k^\dagger\,\rho_g)=0$.
Since the interaction Hamiltonian~\eqref{ha} is linear in the graviton
operators $\hat b_k$ and $\hat b_k^\dagger$, the trace over the gravitational
sector produces expectation values of these operators, all of which vanish for
vacuum and thermal states. Consequently,
\begin{equation}
\operatorname{Tr}_g\!\left[\hat{H}^I_{\rm int}(t),\hat{\rho}^{\rm tot}_0\right]=0,
\end{equation}
and the leading contribution to the reduced dynamics arises at second order in
the interaction.

Extending the upper integration limit to \(t \to \infty\), the master equation in the Born--Markov approximation becomes
\begin{equation}\label{FE}
\frac{d}{dt} \hat{\rho}_d(t) = - \frac{1}{\hbar^2} \int_0^\infty ds \, 
\Tr_g \Big[ \hat{H}^{I}_{\rm int}(t), [\hat{H}^{I}_{\rm int}(t-s), \hat{\rho}_d(t) \otimes \rho_g] \Big],
\end{equation}
where the graviton thermal state is
\begin{equation}
\rho_{g}\rightarrow \rho^{th}_g = \frac{e^{-\beta_{g} \hat{H}_{G} }}{\tr (e^{-\beta_{g} \hat{H}_{G}})}.
\end{equation}
Furthermore, this allows us to compute the explicit form of the thermal correlation functions for the following bilinears:
\begin{align}
    &\langle \hat{b}_l^{\dagger} \hat{b}_m^{\dagger} \rangle_g = 0, & \quad \langle \hat{b}_l \hat{b}_m \rangle_g = 0,\nonumber\\
    &\langle \hat{b}_l^{\dagger} \hat{b}_m \rangle_g = \delta_{lm} \bar{n}_l, & \quad \langle \hat{b}_l \hat{b}_m^{\dagger} \rangle_g = \delta_{lm}(1 + \bar{n}_l),\nonumber\\
     &\langle \hat{b}_l^{\dagger} \hat{b}_m^{\dagger} \rangle^{*}_g = 0, & \quad \langle \hat{b}_l \hat{b}_m \rangle^{*}_g = 0,\nonumber\\
    &\langle \hat{b}_l^{\dagger} \hat{b}_m \rangle^{*}_g = \delta_{lm}(1 + \bar{n}_l), & \quad \langle \hat{b}_l \hat{b}_m^{\dagger} \rangle_g = \delta_{lm}\bar{n}_l.
    \label{cv}
\end{align}
where the expectation value is taken over the graviton thermal state, where $\langle \hat{A} \hat{B} \rangle_g \equiv \mathrm{Tr}_g \big( \hat{A} \hat{B} \, \hat{\rho}_g \big)$, $\langle\hat{A}\hat{B}\rangle^{*}_{g}\equiv\rm Tr_{g}(\hat{A}~\hat{\rho}_{g}~\hat{B})$, and $\bar{n}_l = \frac{1}{e^{\hbar \omega_l^g / k_B T_{g}} - 1}$ 
is the thermal occupation number of mode $l$ with $\beta_{g}= 1/(k_B T_{g})$.

Following a standard coarse-graining procedure, we assume that the graviton 
reservoir modes are closely spaced in frequency and pass to the continuum limit. 
In a cubic quantization volume $V=L^3$, the discrete momentum sum is replaced by
\begin{equation}
\sum_{\vec k}
\;\longrightarrow\;
\frac{V}{(2\pi)^3}\int d^3k
=
\frac{V}{2\pi^2}\int_0^\infty d\omega_g\, f(\omega_g) .
\end{equation}
where $f(\omega_g)$ denotes the spectral profile of the gravitational reservoir, 
reducing to $f(\omega_g)=\omega_g^2$ for free gravitons in $3+1$ dimensions.

Also we use Sokhotski-Plemelj theorem (see Eq. (4.15) of \cite{Barman:2021bbw})
\begin{eqnarray}
    \int_0^\infty \mathrm{d}s\, e^{\pm i(2\omega^g_l - 2\omega_c)s} &=& \pi \delta(\omega^g_l - \omega_c) 
    \nonumber
    \\
    &-& i\mathcal{P}\Big(\frac{1}{2(\omega_l^g - \omega_c)}\Big),
\end{eqnarray}
where $\mathcal{P}$ represents the principal value. Now in the Markovian regime, the Dirac-delta contribution dominates because it corresponds to real, energy-conserving transitions, while the principal-value term represents virtual off-resonant processes whose net effect is only a small frequency shift. This yields, for the term containing the nested commutator in Eq.~(\ref{FE}),
\begin{multline}
\int_0^\infty \mathrm{d}s\, 
\langle \hat{H}_{\rm int}(t) \hat{H}_{\rm int}(t{-}s) \hat{\rho_{d}}(t) \rangle_g 
= \frac{\pi}{2} \hbar^2 f(\omega_c) C_{\omega_c}^2 \\
\times\bigg[ (1 + \bar{n}_c) \big( 
\hat{a}_1^{\dagger 2} \hat{a}_1^2 \hat{\rho_{d}}(t) 
- \hat{a}_1^2 \hat{a}_1^2 \hat{\rho_{d}}(t) e^{-i 4 \omega_c t} \big) 
+ \\
\bar{n}_c \big( 
\hat{a}_1^2 \hat{a}_1^{\dagger 2} \hat{\rho_{d}}(t) 
- \hat{a}_1^{\dagger 2} \hat{a}_1^{\dagger 2} \hat{\rho_{d}}(t) e^{i 4 \omega_c t} \big) 
\bigg]~,
\label{eq:term1}
\end{multline}
where $f(\omega_c)$ denotes the spectral weight of the graviton bath
evaluated at the detector frequency $\omega_c$ (see Appendix~\ref{AppB}
for details).

Carrying out the analogous steps for the remaining contributions in
Eq.~(\ref{FE}) and collecting all terms, the reduced density matrix in
the Schr\"odinger picture satisfies
\begin{align}
\frac{d}{dt}\hat{\rho}_{d}(t)
= \mathcal{L}_{u}\,\hat{\rho}_{d}(t)
+ \mathcal{L}_{g}\,\hat{\rho}_{d}(t),
\label{eom_total_PRD}
\end{align}
which is equivalent to a Lindblad-type master equation, with
$\mathcal{L}_{u}$ and $\mathcal{L}_{g}$ denoting the unitary and
graviton-induced dissipative contributions, respectively.
The unitary part is given by
\begin{align}
\mathcal{L}_{u}\hat{\rho}_{d}(t)
= -\frac{i}{\hbar}\big[\hat{H}_{0},\,\hat{\rho}_{d}(t)\big],
\label{Lu_PRD}
\end{align}
while the graviton-induced dissipative contribution reads
\begin{multline}
\mathcal{L}_{g}\hat{\rho}_{d}(t)
= -\frac{\pi}{2} f(\omega_{c}) C_{\omega_{c}}^{2}
\Big\{
(1+\bar{n}_{c})\Big[
\acomm{\hat{a}_{1}^{\dagger 2}\hat{a}_{1}^{2}}{\hat{\rho}_{d}}
- 2\hat{a}^{2}_{1}\hat{\rho}_{d}(t)\hat{a}^{\dagger 2}_{1} \\
+ \hat{a}_{1}^{2}\hat{\rho}_{d}\hat{a}_{1}^{2} e^{-i4\omega_{c} t}
+ \hat{a}_{1}^{\dagger 2}\hat{\rho}_{d}\hat{a}_{1}^{\dagger 2} e^{i4\omega_{c} t}
- \hat{a}_{1}^{2}\hat{a}_{1}^{2}\hat{\rho}_{d} e^{-i4\omega_{c} t}
- \hat{\rho}_{d}\hat{a}_{1}^{\dagger 2}\hat{a}_{1}^{\dagger 2} e^{i4\omega_{c} t}
\Big] \\
+\, \bar{n}_{c}\Big[
\acomm{\hat{a}_{1}^{2}\hat{a}_{1}^{\dagger 2}}{\hat{\rho}_{d}}
- 2\hat{a}_{1}^{\dagger 2}\hat{\rho}_{d}\hat{a}_{1}^{2}
+ \hat{a}_{1}^{2}\hat{\rho}_{d}\hat{a}_{1}^{2} e^{-i4\omega_{c} t}
+ \hat{a}_{1}^{\dagger 2}\hat{\rho}_{d}\hat{a}_{1}^{\dagger 2} e^{i4\omega_{c} t} \\
- \hat{\rho}_{d}\hat{a}_{1}^{2}\hat{a}_{1}^{2} e^{-i4\omega_{c} t}
- \hat{a}_{1}^{\dagger 2}\hat{a}_{1}^{\dagger 2} \hat{\rho}_{d} e^{i4\omega_{c} t}
\Big]
\Big\}.
\label{Lg_PRD}
\end{multline}

To study the decoherence induced by the graviton environment, we now
solve the master equation~\eqref{eom_total_PRD} perturbatively to
leading order in the gravitational interaction.
Formally, the solution of Eq.~\eqref{eom_total_PRD} can be written as
\begin{align}
\hat{\rho}_d(t) = \hat{\Phi}(t)\,\hat{\rho}_d(0),
\label{df}
\end{align}
where the dynamical map $\hat{\Phi}(t)$ satisfies
\begin{align}
\hat{\Phi}(t)
= e^{\mathcal{L}_u t}
\left[
1 + \int_0^t d\tau\,
e^{-\mathcal{L}_u \tau}\,
\mathcal{L}_g\,
\hat{\Phi}(\tau)
\right].
\end{align}
To first order in $\mathcal{L}_g$, this expression reduces to
\begin{align}
\hat{\Phi}(t)
\simeq
e^{\mathcal{L}_u t}
\left[
1 + \int_0^t d\tau\,
e^{-\mathcal{L}_u \tau}\,
\mathcal{L}_g\,
e^{\mathcal{L}_u \tau}
\right].
\end{align}
This approximation corresponds to treating the graviton-induced
dissipative contribution perturbatively, while the unitary evolution
generated by $\mathcal{L}_u$ is taken into account exactly. In other
words, we work in the interaction picture with respect to
$\mathcal{L}_u$.

Using the explicit form
$\mathcal{L}_u\hat{\rho}
= -\frac{i}{\hbar}[\hat{H}_0,\hat{\rho}]$,
the action of the unitary propagator on an arbitrary operator $A$ is
given by
\begin{align}
e^{\mathcal{L}_u t} A
= e^{-\frac{i}{\hbar}\hat{H}_0 t}\,
A\,
e^{\frac{i}{\hbar}\hat{H}_0 t},
\end{align}
which follows directly from the Baker--Campbell--Hausdorff lemma.
Consequently, the first-order perturbative solution for the reduced
density matrix takes the form
\begin{align}
\hat{\rho}_d(t)
&= e^{\mathcal{L}_u t}\,\hat{\rho}_d(0)
\nonumber\\
&\quad
+ e^{\mathcal{L}_u t}
\int_0^t d\tau\,
e^{\frac{i}{\hbar}\hat{H}_0 \tau}\,
\mathcal{L}_g
\!\left[
e^{-\frac{i}{\hbar}\hat{H}_0 \tau}\,
\hat{\rho}_d(0)\,
e^{\frac{i}{\hbar}\hat{H}_0 \tau}
\right]
e^{-\frac{i}{\hbar}\hat{H}_0 \tau}.
\label{perturbative_solution}
\end{align}
Substituting the explicit form of $\mathcal{L}_g$, we obtain
\begin{align}
\hat{\rho}_d(t) &= e^{\mathcal{L}_u t}\,\hat{\rho}_d(0)
\nonumber\\
&\quad - \frac{\pi}{2}\,f(\omega_c)\,C_{\omega_c}^2\,
e^{\mathcal{L}_u t}\int_0^t d\tau \, \Biggl[ 
(1+\bar{n}_c)\Big\{ \nonumber\\
&\qquad \acomm{\hat{a}_1^{\dagger 2}\hat{a}_1^2}{\hat{\rho}_d(0)}
- 2\,\hat{a}_1^{ 2}\hat{\rho}_d(0)\hat{a}_1^{\dagger 2}
- e^{-i4\omega_c\tau}\,\hat{a}_1^2\hat{a}_1^2\,\hat{\rho}_d(0) \nonumber\\
&\qquad - e^{\,i4\omega_c\tau}\,\hat{\rho}_d(0)\,\hat{a}_1^{\dagger 2}\hat{a}_1^{\dagger 2}
+ e^{-i4\omega_c\tau}\,\hat{a}_1^2\hat{\rho}_d(0)\hat{a}_1^2 \nonumber\\
&\qquad 
+ e^{\,i4\omega_c\tau}\,\hat{a}_1^{\dagger 2}\hat{\rho}_d(0)\hat{a}_1^{\dagger 2}
\Big\} + \bar{n}_c\Big\{ \acomm{\hat{a}_1^2\hat{a}_1^{\dagger 2}}{\hat{\rho}_d(0)}
\nonumber
\\
&\qquad - 2\,\hat{a}_{1}^{\dagger 2}\hat{\rho}_d(0)\hat{a}_1^{2}
- e^{\,i4\omega_c\tau}\,\hat{a}_1^{\dagger 2}\hat{a}_1^{\dagger 2}\,\hat{\rho}_d(0) \nonumber\\
&\qquad  - e^{-i4\omega_c\tau}\,\hat{\rho}_d(0)\,\hat{a}_1^2\hat{a}_1^2
+ e^{-i4\omega_c\tau}\,\hat{a}_1^2\hat{\rho}_d(0)\hat{a}_{1}^2
+ \nonumber\\
&\qquad  e^{\,i4\omega_c\tau}\,\hat{a}_1^{\dagger 2}\hat{\rho}_d(0)\hat{a}_1^{\dagger 2}
\Big\}
\Biggr].
\label{eq:perturbative_solution_prd}
\end{align}
It is worth emphasizing that, throughout this analysis, we deliberately neglect all conventional environmental dissipation channels—such as thermal relaxation, cavity losses, and mechanical damping \cite{PhysRevA.56.2249}—in order to isolate decoherence arising solely from the interaction with the gravitational field.

As a first example, consider the initial state 
$\hat{\rho}_{d}(0)=\ket{0}\bra{0}$. 
Using Eq.~\eqref{eq:perturbative_solution_prd}, 
the corresponding diagonal matrix element evolves, 
to leading order in the perturbative expansion, as
\begin{equation}
\label{decoher}
\bra{0}\hat{\rho}_{d}(t)\ket{0}
= 1-\frac{t}{\tau_{G}},
\qquad
\frac{t}{\tau_G}\ll 1,
\end{equation}
where the characteristic graviton-induced transition time is defined by
\begin{equation}
\label{tauG}
\tau_{G}^{-1}
=
2\pi\,\bar n_{c}\, f(\omega_{c})\, C_{\omega_{c}}^{2}.
\end{equation}
This expression describes graviton-induced excitation out of the
ground state and is valid in the short-time regime where the
Born--Markov expansion applies.

For comparison, starting from the first excited state
$\hat{\rho}_{d}(0)=\ket{1}\bra{1}$,
Eq.~\eqref{eq:perturbative_solution_prd} yields
\begin{equation}
\label{decoher1}
\bra{1}\hat{\rho}_{d}(t)\ket{1}
= 1-3\frac{t}{\tau_{G}},
\qquad
\frac{t}{\tau_G}\ll 1,
\end{equation}
reflecting the quadratic structure of the detector–gravity coupling. For a quantum-vacuum gravitational environment
($\bar n_c=0$), the rate vanishes identically.

We next consider a coherent superposition of the ground and first excited states,
\begin{equation}
\hat{\rho}_{d}(0)
=
\frac{1}{2}
\big(
\ket{0}\bra{0}
+
\ket{0}\bra{1}
+
\ket{1}\bra{0}
+
\ket{1}\bra{1}
\big),
\label{s}
\end{equation}
in order to examine how graviton-induced decoherence affects coherence between energy eigenstates.
Applying Eq.~\eqref{eq:perturbative_solution_prd}, the relevant matrix elements evolve as
\begin{align}
\bra{0}\hat{\rho}_{d}(t)\ket{0}
&= \frac{1}{2}\left(1-\frac{t}{\tau_{G}}\right),\\
\bra{1}\hat{\rho}_{d}(t)\ket{1}
&= \frac{1}{2}\left(1-3\frac{t}{\tau_{G}}\right),\\
\bra{0}\hat{\rho}_{d}(t)\ket{1}
&= \frac{1}{2}\left(1-2\frac{t}{\tau_{G}}\right)e^{i\omega_{c}t} = 
\Big(\bra{1}\hat{\rho}_{d}(t)\ket{0}\Big)^*.
\label{vc}
\end{align}

Remarkably, for any initial density matrix supported entirely within the subspace spanned by
$\{\ket{0},\ket{1}\}$—that is, for any pure state of the form
$\ket{\psi(0)}=\alpha_{1}\ket{0}+\beta_{1}\ket{1}$ with $\alpha_{1},\beta_{1}\in\mathbb{R}$—
we find that graviton-induced decoherence vanishes in the zero-temperature limit.
In this sense, vacuum fluctuations of the linearized gravitational field are effectively
\emph{blind} to coherent superpositions within the $\{\ket{0},\ket{1}\}$ sector:
such states remain fully coherent at $T=0K$, provided all other environmental couplings are suppressed.
This behavior changes qualitatively once higher excited states are involved.
For example, taking the initial state $\hat{\rho}_{d}(0)=\ket{2}\bra{2}$ and applying
Eq.~\eqref{eq:perturbative_solution_prd}, we obtain
\begin{equation}
\label{decoherence}
\bra{2}\hat{\rho}_{d}(t)\ket{2}
=
1-2\pi f(\omega_{c})\, C^{2}_{\omega_{c}}\, t\, \bigl(1+7\bar n_{c}\bigr).
\end{equation}
Equation~\eqref{decoherence} demonstrates that, unlike the protected
$\{\ket{0},\ket{1}\}$ sector, decoherence persists even in the zero-temperature
limit ($\bar n_{c}=0$).

The introduction of a thermal state for the gravitational field at this stage serves a dual purpose.
First, it provides a physically well-defined mixed quantum state that interpolates smoothly
between vacuum fluctuations and classical stochastic gravitational-wave noise.
In particular, in the high-occupation limit, the thermal graviton state reproduces the
field correlators of a classical Gaussian stochastic background, while at low temperatures
it retains intrinsically quantum features.
Second, unlike the vacuum state, a thermal environment can induce decoherence even when
spontaneous emission channels are kinematically suppressed, thereby lifting the protection
observed in the $\{\ket{0},\ket{1}\}$ subspace.
In this sense, the thermal graviton bath furnishes a controlled and physically motivated
framework for comparing genuinely quantum gravitational fluctuations with their classical
stochastic counterpart within a unified open-system description \cite{Caldeira:1982iu}.
Such thermal or mixed graviton states may arise, for instance, if inflation is preceded by a
radiation-dominated phase, or through stimulated emission during inflation
\cite{Bhattacharya:2006prl,Ghayour:2012Thesis}, and may also be understood more generally
as effective descriptions of stochastic gravitational-wave backgrounds with finite
occupation numbers \cite{BreuerPetruccione2006,BirrellDavies1982}.

One can even compute the time evolution of a coherent superposition of the second-excited state with the ground state, i.e. we consider the initial state:

\begin{equation}
\hat{\rho}_{d}(0)
=
\frac{1}{2}
\big(
\ket{0}\bra{0}
+
\ket{0}\bra{2}
+
\ket{2}\bra{0}
+
\ket{2}\bra{2}
\big),
\label{ds2}
\end{equation}
the short–time evolution of the density–matrix elements obtained from
Eq.~\eqref{eq:perturbative_solution_prd} reads

\begin{align}
\bra{0}\hat{\rho}_{d}(t)\ket{0}
&= \frac{1}{2}\!\left(1+2\pi f(\omega_c) C_{\omega_c}^{2} t\right),\\
\bra{2}\hat{\rho}_{d}(t)\ket{2}
&= \frac{1}{2}\!\left(1-2\pi f(\omega_c) C_{\omega_c}^{2} t\,[1+6\bar n_c]\right),\\
\bra{2}\hat{\rho}_{d}(t)\ket{0}
&= \frac{1}{2}\!\left(1-\pi f(\omega_c) C_{\omega_c}^{2} t\,[1+8\bar n_c]\right)e^{-2i\omega_c t}
\nonumber\\
&\quad
-\pi f(\omega_c) C_{\omega_c}^{2}(2\bar n_c+1)\frac{\sin(2\omega_c t)}{2\omega_c}
\nonumber\\
&=\Big(\bra{0}\hat{\rho}_{d}(t)\ket{2}\Big)^* .
\label{fd}
\end{align}
These expressions reveal a qualitative difference compared with the case of a
superposition confined to the lowest manifold $\{\ket{0},\ket{1}\}$.
In particular, even in the vacuum limit $\bar n_c=0$, the matrix elements
$\bra{2}\hat{\rho}_{d}(t)\ket{2}$ and
$\bra{0}\hat{\rho}_{d}(t)\ket{2}$ decay irreversibly, indicating that this
sector is not protected against gravitationally induced dissipation.
The origin of this behavior lies in the quadratic detector–gravity coupling,
which enforces a selection rule $\Delta n=\pm2$ and therefore allows
vacuum–fluctuation–induced transitions between $\ket{2}$ and $\ket{0}$ even
when the gravitational field is in its ground state. By contrast,
superpositions supported entirely within the $\{\ket{0},\ket{1}\}$ manifold do
not couple to the gravitational environment at leading order in the vacuum,
so that no decoherence occurs at zero temperature.

At the same time, the ground–state population
$\bra{0}\hat{\rho}_{d}(t)\ket{0}$ increases linearly at short times. This
increase does not signal a restoration of coherence or any violation of
probability conservation. Rather, it reflects the redistribution of population
within the truncated Hilbert space: transitions from $\ket{2}$ into
$\ket{0}$ are permitted by the quadratic coupling, while the reverse
process is absent in the vacuum because it would require absorption from an
unoccupied gravitational mode. Consequently, population flows
unidirectionally toward the ground state at zero temperature, producing the
observed growth of $\bra{0}\hat{\rho}_{d}(t)\ket{0}$.
The trace of the density matrix remains conserved, and the reduced dynamics
remains completely positive and trace preserving.

Such behavior is generic in open quantum systems: decay of coherence in one
sector of Hilbert space is typically accompanied by population accumulation in
another. What is specific to the present model is that this redistribution
occurs only once the detector state involves levels separated by two quanta,
consistent with the $\Delta n=\pm2$ selection rule imposed by the quadratic
detector–gravity interaction. In this sense, the $\{\ket{0},\ket{1}\}$ manifold
forms a protected subspace of the vacuum dynamics, whereas higher manifolds do
not.

We observe that a thermal graviton state induces decoherence in the harmonic oscillator through genuine quantum fluctuations. To clarify the distinction between classical and quantum gravitational environments, we next model the gravitational field by a phase-randomized coherent-state ensemble, which represents a classical stochastic background with identical average intensity but different fluctuation structure.


\section{Classical Limit of Gravitational Decoherence}
To isolate the effect of classical gravitational waves on the quantum detector,
we model the gravitational field as a classical stochastic background represented
by an ensemble of phase--randomized coherent states. As already emphasized at the
beginning of the previous section, the gravitational field is quantized directly
in the physical transverse--traceless sector. Consequently, the graviton creation
and annihilation operators obey standard bosonic commutation relations,
\begin{equation}
[\hat b_k , \hat b_{k'}^\dagger] = \delta_{kk'},
\end{equation}
without the introduction of indefinite--metric structures or pseudo--commutators.

A multimode coherent state of the gravitational field is defined as a displaced
vacuum~\cite{Botke1974Coherent, kv1t-j27m},
\begin{equation}
\ket{\{\alpha_k\}}
=
\hat{D}(\{\alpha_k\})\ket{0}_{g}; \,\
\hat{D}(\{\alpha_k\})
=
\exp\!\left(
\sum_k \alpha_k \hat{b}_k^\dagger
-
\alpha_k^* \hat{b}_k
\right),
\end{equation}
where $\hat{b}_k$ and $\hat{b}_k^\dagger$ are the annihilation and creation
operators of the physical plus polarized graviton mode $k$, $\ket{0}_{g}$ denotes the graviton
vacuum, and $\alpha_k = |\alpha_k| e^{i\phi_k}$ specifies the amplitude and phase
of each mode. Each mode satisfies
$\hat{b}_k\ket{\{\alpha_k\}}=\alpha_k\ket{\{\alpha_k\}}$, so that the state
factorizes as $\ket{\{\alpha_k\}}=\bigotimes_k \ket{\alpha_k}$.

The state of the gravitational sector is then taken to be
\begin{equation}
\label{Gch}
\rho^{\rm coh}_{g}
=
\bigotimes_{k}
\left(
\int_{0}^{2\pi} \frac{d\phi_k}{2\pi}
\, |\alpha_k\rangle\langle \alpha_k|
\right),
\end{equation}
where $\rho^{\mathrm{coh}}_g$ denotes the density matrix corresponding to coherent
graviton states with randomized phases. The phases $\phi_k$ associated with
different modes are assumed to be uncorrelated and uniformly distributed over
the interval $[0,2\pi)$.

The use of phase-randomized coherent states to represent a classical
stochastic field has a long history in quantum field theory. In
particular, it was shown in early studies of multiparticle production
that a wide class of physically relevant density operators can be written
in diagonal form in the coherent--state representation, with classical or
chaotic sources corresponding to ensembles that are insensitive to the
phases of the coherent amplitudes \cite{Botke1974Coherent}. In such cases,
one--point functions vanish while phase--invariant two--point correlators
remain finite, capturing the statistical character of the underlying
field. The phase averaging in Eq.~(\ref{Gch}) is therefore not a dynamical
assumption, but a minimal representation of an incoherent classical
background.

A closely related phase--averaging mechanism appears in coherent states
of charged bosons, where superselection rules enforce projection onto
fixed--charge sectors and eliminate phase--sensitive observables
\cite{Bhaumik1976ChargedCoherent}. In the present gravitational setting,
however, the averaging reflects classical stochasticity rather than an
exact symmetry constraint.

We emphasize that this construction also differs from early covariant coherent--state
formulations of electromagnetism and linearized gravity based on
indefinite--metric Hilbert spaces, where pseudo--commutation relations appear
prior to the imposition of subsidiary conditions
\cite{Gomatam1971Coherent}. Here, the analysis is performed entirely within the
physical graviton sector, and no pseudo--operator structures arise.

Physically, a pure coherent state corresponds to a deterministic classical
gravitational wave (\ref{pc}) with a well-defined amplitude and phase. The uniform phase
averaging in Eq.~\eqref{Gch} removes any global phase reference while preserving
the classical energy content of each mode. As a result, the gravitational field
is described as an incoherent classical background characterized by statistical
properties rather than phase coherence.
Such a phase--randomized coherent ensemble provides a minimal quantum
representation of classical stochastic gravitational-wave backgrounds, as
expected for radiation produced by many independent classical sources, such as
unresolved astrophysical binaries or other incoherent emitters. In this case,
the randomness originates from classical uncertainty in the phases rather than
from intrinsic quantum fluctuations. Consequently, the gravitational field acts
as a classical stochastic process rather than a quantum environment, even though
it is conveniently represented using coherent states.


As a bookkeeping device for the classical limit, we distinguish the Planck constants in the gravitational and detector sectors by the substitutions $\hbar \rightarrow \hbar_G$ for gravitons, while $\hbar \rightarrow \hbar_D$ for the detector. This separation is introduced purely as a bookkeeping
device to track the classical limit of the gravitational
sector while keeping the detector fully quantum.
Then in the interaction picture, and under the Born approximation, the reduced density matrix satisfies the second–order equation
\begin{equation}
\frac{d}{dt}\hat{\rho}_{d}(t)
=
-\frac{1}{\hbar_{D  }^2}
\int_{0}^{\infty} ds\;
\mathrm{Tr}_g 
\Big[
\hat{H}_{\mathrm{int}}(t),\,
\big[
\hat{H}_{\mathrm{int}}(t-s),\,
\hat{\rho}_{d}(t)\otimes \rho_g
\big]
\Big].
\end{equation}
Averaging over the coherent–state phases yields
\begin{align}\label{bilina}
\langle \hat{b}_l^{\dagger}\hat{b}_m^{\dagger} \rangle_g^{\alpha}=0, \qquad
\langle \hat{b}_l^{\dagger}\hat{b}_m \rangle_g^{\alpha}=\delta_{lm}|\alpha_l|^2,\nonumber \\
\langle \hat{b}_l \hat{b}_m^{\dagger} \rangle_g^{\alpha}=\delta_{lm}(1+|\alpha_l|^2),
\qquad
\langle \hat{b}_l \hat{b}_m \rangle_g^{\alpha}=0,\nonumber\\
\langle \hat{b}_l^{\dagger}\hat{b}_m^{\dagger} \rangle_g^{*\alpha}=0, \qquad
\langle \hat{b}_l^{\dagger}\hat{b}_m \rangle_g^{*\alpha}=\delta_{lm}(1+|\alpha_l|^2),\nonumber \\
\langle \hat{b}_l \hat{b}_m^{\dagger} \rangle_g^{*\alpha}=\delta_{lm}|\alpha_l|^2,
\qquad
\langle \hat{b}_l \hat{b}_m \rangle_g^{*\alpha}=0,
\end{align}
so that only the phase–neutral contributions survive.
Proceeding exactly as in the fully quantum calculation, the detector density matrix
(\ref{df}) evolves as
\begin{align}\label{coherME}
\hat{\rho}_{d}(t)
&= 
e^{\hat{\mathcal{L}}_{1} t}\, 
\hat{\rho}_{d}(0)\,
e^{\hat{\mathcal{L}}_{1} t}
\nonumber\\
&\quad
-\frac{\pi}{2}\,
f(\omega_{c})\, C_{\omega_{c}}^{2}
\int_{0}^{t} d\tau\,
\bigg[
\big(1 + |\alpha|^{2}\big)\,\nonumber\\
&\quad
\mathcal{D}_{+}\!\left[\hat{\rho}_{d}(\tau)\right]
+
|\alpha|^{2}\,
\mathcal{D}_{-}\!\left[\hat{\rho}_{d}(\tau)\right]
\bigg].
\end{align}
Here $|\alpha|^{2}=|\alpha_{k}|^{2}\vert_{k=\omega_{c}}$ denotes the occupation
number of the resonant gravitational-wave mode, and the dissipators
$\mathcal{D}_{\pm}$ are written explicitly following the same procedure as in
Eq.~(\ref{de}) of Appendix~\ref{AppC}, including the long-time averaging (although in the main text we presented the result without explicitly displaying this averaging step). The overall strength of the
$\alpha$-dependent dissipative contribution in Eq.~\eqref{coherME} is governed
by the effective rate
\begin{equation}
\Gamma_{\alpha}
\equiv
\pi\, f(\omega_{c})\, C_{\omega_{c}}^{2}\,|\alpha|^{2},
\end{equation}
which plays the role of a gravitationally induced damping (or decoherence) rate
for the detector.



For the initial state given in Eq.~(\ref{s}),
the density matrix elements evolve as
\begin{align}
\bra{0}\hat{\rho}_{d}(t)\ket{0}
&=
\frac{1}{2}\left(1-\frac{t}{\tau_\alpha}\right),\\
\bra{1}\hat{\rho}_{d}(t)\ket{1}
&=
\frac{1}{2}\left(1-3\frac{t}{\tau_\alpha}\right),\\
\bra{0}\hat{\rho}_{d}(t)\ket{1}
&=
\frac{1}{2}\left(1-2\frac{t}{\tau_\alpha}\right)e^{i\omega_c t},\\
\bra{1}\hat{\rho}_{d}(t)\ket{0}
&=
\frac{1}{2}\left(1-2\frac{t}{\tau_\alpha}\right)e^{-i\omega_c t},
\label{e}
\end{align}
with the decoherence rate
\begin{equation}
\frac{1}{\tau_\alpha} = 2 |\alpha|^2 \pi f(\omega_c) C_{\omega_c}^2.
\label{x}
\end{equation}

The classical limit is implemented at the end of the calculation by scaling
\begin{equation}
C_{\omega_c} \sim \sqrt{\hbar_G}.
\end{equation}
In the classical regime of the graviton sector, where the occupation number is large, $|\alpha_k|^2 \to \infty$, one has
\begin{equation}
\lim_{\substack{\hbar_G \to 0 \\ |\alpha|^2 \to \infty}} C_{\omega_c}^2 |\alpha|^2 \sim \text{constant}~.
\end{equation}
In this limit, the decoherence rate remains finite as $\hbar_G \to 0$, leading to a stochastic decoherence that is independent of temperature and vacuum fluctuations.  
Physically, this decoherence originates from the classical randomness of the gravitational field: each mode is represented by a coherent state with a fixed amplitude but a randomly distributed phase. It is worth emphasizing that this behavior is qualitatively distinct from that
of a pure coherent gravitational-wave state with a fixed phase. In that case,
the $\alpha$-dependent contributions to the reduced dynamics are purely
oscillatory and, upon long-time (secular) averaging in the appropriate
classical limit, the dissipative terms vanish identically. Consequently, the
reduced evolution of the detector remains unitary and no decoherence is
induced for any detector state. We refer the reader to Appendix~\ref{AppBB} for a detailed
demonstration.


Importantly, decoherence induced by a phase--randomized coherent background
occurs irrespective of the initial state of the detector; even a detector
prepared in a coherent superposition is not protected in the same way as the
$\{\ket{0},\ket{1}\}$ subspace. Consequently, decoherence generated by a
classical, phase--randomized coherent state does not constitute direct
evidence for the intrinsic quantum nature of gravitational waves. This
behavior stands in sharp contrast to the fully quantum scenario, where a
zero--temperature graviton environment does \emph{not} induce decoherence.
\footnote{If the limit $\hbar_{G} \to 0$ is taken prematurely, the oscillatory
structure of the interaction picture is lost, preventing the cancellation of
intermediate $\hbar_{G}$ factors. The classical limit must therefore be
applied only after the full expression for the reduced dynamics has been
obtained. In particular, taking $|\alpha_k| \to 0$ correctly reproduces the
pure vacuum--induced dynamics discussed in the previous section.}

\section{The decoherence time scale}

In both the quantum and classical analyses, the decoherence rate is controlled by
the spectral weight of the gravitational environment evaluated at the detector
frequency together with the detector--bath coupling. From the perturbative
master equation, the leading contribution to the decay of coherences is
proportional to
\begin{equation}
\frac{1}{\tau_{\alpha}}\;\propto\; f(\omega_{c})\, C_{\omega_{c}}^{2},
\label{tau_general}
\end{equation}
where \(f(\omega_c)\) denotes the density of graviton modes per unit frequency and
\(C_{\omega_{c}}\) is the mode-dependent coupling entering the interaction
Hamiltonian.

From Eq.~\eqref{f}, the coupling coefficient is given by
\(
C_{\omega}=\sqrt{\omega c\,\ell_p^2/(2L^3)}
\),
and therefore scales as
\begin{equation}
C_{\omega}\propto \sqrt{\omega}.
\end{equation}
Physically, this scaling reflects the fact that higher-frequency gravitational
modes carry larger tidal gradients and therefore couple more strongly to the
quadratic detector observable appearing in the interaction Hamiltonian.
Substituting this into Eq.~\eqref{tau_general}, we obtain the general scaling
\begin{equation}
\frac{1}{\tau_{\alpha}}\;\propto\; f(\omega_{c})\,\omega_{c}.
\label{tau_scaling}
\end{equation}
This dependence on \(\omega_{c}\) is physically natural: decoherence arises from
energy- and phase-exchange processes between the detector mode and gravitational
modes of comparable frequency. The available phase space for such processes is
set by the density of graviton modes \(f(\omega_{c})\), while the interaction
strength contributes the additional factor of \(\omega_{c}\).

\begin{figure}[t]
    \centering
   \includegraphics[width=\linewidth]{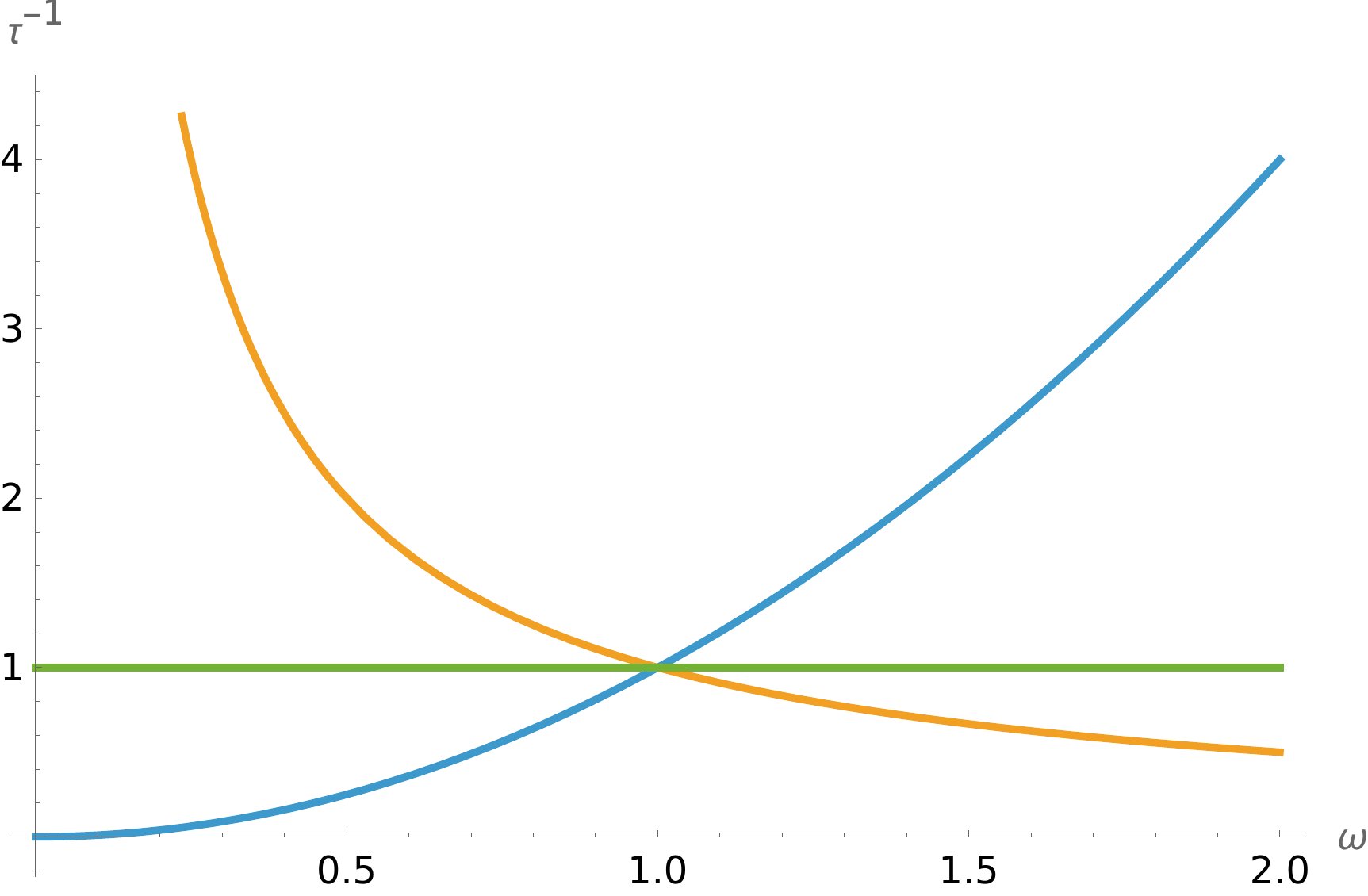}
    \caption{Inverse decoherence time \(\tau_{\alpha}^{-1}\) as a function of the detector
   frequency \(\omega_{c}\) for different spectral profiles
    \(f(\omega_{c})\propto\omega_{c}^{s}\). The green, blue, and yellow
    curves correspond respectively to \(s=-1, 1,\) and \(-2\), illustrating
    the strong sensitivity of the decoherence rate to the gravitational spectral
    density.}
    \label{fig1}
\end{figure}

To make the discussion concrete, let us consider a general power-law form for the
gravitational spectral density,
\begin{equation}
f(\omega)\propto \omega^{s},
\end{equation}
which yields
\begin{equation}
\frac{1}{\tau_{\alpha}}\;\propto\; \omega_{c}^{\,s+1}.
\end{equation}
Different values of the exponent \(\alpha\) lead to qualitatively different
behaviour:
\begin{itemize}
    \item For \(s=-1\), the decoherence rate is independent of the detector
    frequency, \(\tau_{\alpha}^{-1}=\mathrm{const}\), as shown by the green curve in
    Fig.~\ref{fig1}.
    \item For \(s>-1\), the decoherence rate increases with \(\omega_{c}\),
    illustrated by the blue curve (\(s=1\)).
    \item For \(s<-1\), the decoherence rate decreases with increasing
    \(\omega_{c}\), as shown by the yellow curve (\(s=-2\)).
\end{itemize}

For free gravitons propagating in $3+1$ dimensions, the spectral density follows
from the density of states of a massless field. Since $\omega = c|\mathbf{k}|$,
the number of modes in a shell $k \to k+dk$ scales as $4\pi k^2 dk$,
implying
\begin{equation}
f(\omega) \propto \omega^2 .
\end{equation}

This implies that, all else being equal, higher-frequency detectors experience
stronger gravitationally induced decoherence. In practice, however, this
advantage is counterbalanced by experimental constraints such as increased
thermal noise, reduced coherence times, and limitations in state preparation
and readout at high frequencies.

More importantly, Eq.~\eqref{tau_scaling} underscores a key conceptual point: the decoherence time cannot be predicted independently without specifying the spectral profile of the gravitational environment. Since $f(\omega)$ depends on the astrophysical or cosmological origin of the gravitational background, the same detector may experience vastly different decoherence rates in different gravitational settings. Moreover, the ultimate manifestation of gravitationally induced decoherence depends not only on the spectral structure of the environment but also on the quantum state of the gravitons and on the initially prepared state of the detector system.

This observation reinforces the central theme of the present work---\emph{it is the structure and selectivity of decoherence, rather than its absolute magnitude, that encodes information about the underlying gravitational background}.



\section{Experimental protocol: state-selective test of gravitational decoherence}

The results obtained in this work suggest a conceptually robust strategy for
probing the microscopic nature of gravitational waves using quantum-controlled
mechanical systems. A central outcome of our analysis is that the
\emph{structure} of gravitationally induced decoherence—rather than its absolute
magnitude—provides the relevant experimental diagnostic. In particular, the
perturbative master equation derived in
Sec.~III [see Eq.~\eqref{Lg_PRD}] predicts a qualitatively
different response of low-lying Fock manifolds depending on whether the
gravitational background is a quantum vacuum or a classical stochastic field.

This distinction follows directly from the explicit solutions of the reduced
dynamics. For vacuum gravitons, the quadratic detector–gravity coupling imposes
selection rules that suppress all leading-order dissipative contributions within
the lowest Fock manifold $\{|0\rangle,|1\rangle\}$ [see Eq.~(\ref{vc})]. As a
result, coherent superpositions supported entirely within this subspace do not
decohere at zero temperature in the absence of other environments. By contrast,
classical stochastic gravitational-wave backgrounds—including phase-randomized
coherent states and thermal graviton ensembles—produce finite decoherence rates
already in the lowest Fock sector. This qualitative difference motivates the
state-selective experimental protocol outlined below.

Crucially, however, the absence of observable decoherence in the
$\{|0\rangle,|1\rangle\}$ subspace cannot by itself establish the quantum nature
of gravity, since unavoidable environmental channels will always contribute
residual damping and dephasing. The discriminatory power of our proposal instead
lies in a \emph{comparative, state-selective strategy} that exploits the
different scaling of decoherence across Fock manifolds.

\subsection*{Step 1: Preparation of low-lying phonon superpositions}

Consider a single mechanical mode whose phonon states can be prepared and
measured with high fidelity. Modern hybrid quantum platforms—including
GHz-frequency bulk-acoustic-wave resonators, membrane-in-the-middle systems, and
optomechanical crystal cavities coupled to superconducting qubits—permit
deterministic preparation of low-lying phonon number states and their coherent
superpositions via qubit–mechanical or optical state-transfer protocols
\cite{vonLupke2022NatPhys,Marti2024NatPhys,PhysRevLett.130.133604}.

Under otherwise identical experimental conditions, prepare the two initial
states
\begin{equation}
\ket{\psi_{01}}=\frac{1}{\sqrt{2}}(\ket{0}+\ket{1}),\qquad
\ket{\psi_{02}}=\frac{1}{\sqrt{2}}(\ket{0}+\ket{2}).
\end{equation}

\subsection*{Step 2: Measurement of relaxation and coherence times}

A first experimental probe consists of preparing the mechanical mode in the
state $\ket{1}$, allowing it to evolve freely, and measuring the population
$\langle 1 | \rho_d(t) | 1 \rangle$ as a function of time. The observed decay
defines an effective relaxation time $T_1$, which at short times is well
described by a single exponential,
\begin{equation}
T_1^{-1} = \gamma + \Gamma^{\mathrm{grav}}_{11},
\end{equation}
where $\gamma$ denotes the intrinsic (non-gravitational) mechanical damping rate,
and $\Gamma^{\mathrm{grav}}_{11} = 3/\tau_G$ is the gravitationally induced
contribution, as obtained from Eq.~\eqref{eq:T1extract}.

An independent probe is obtained by preparing the coherent superposition
$(\ket{0} + \ket{1})/\sqrt{2}$ and performing a phase-sensitive Ramsey-type
measurement~\cite{vonLupke2022NatPhys,Marti2024NatPhys} to monitor the decay of the
off-diagonal density matrix element $\langle 0 | \rho_d(t) | 1 \rangle$. The decay
of the Ramsey fringe envelope defines an effective coherence time $T_2$, which at
short times obeys
\begin{equation}
T_2^{-1} = \frac{\gamma}{2} + \Gamma^{\mathrm{grav}}_{01},
\end{equation}
where $\Gamma^{\mathrm{grav}}_{01} = 2/\tau_G$ follows from
Eq.~\eqref{eq:T2extract}.

Both $T_1$ and $T_2$ are therefore experimentally accessible observables
characterizing decoherence within the lowest Fock manifold. Importantly, neither
quantity alone suffices to disentangle gravitationally induced decoherence from
ordinary environmental noise, motivating the comparative strategy described
below.

\subsection*{Step 3: Gravitational decoherence rates from theory}

From the perturbative solution of the master equation, the
gravitationally induced decoherence rates for a thermal graviton
environment with mean occupation number $\bar n_c$ are obtained as
\begin{align}
\Gamma^{\mathrm{grav,th}}_{01}
&= 4\pi\, \bar n_c\, f(\omega_c)\, C_{\omega_c}^2, \\
\Gamma^{\mathrm{grav,th}}_{11}
&= 6\pi\, \bar n_c\, f(\omega_c)\, C_{\omega_c}^2,
\end{align}
as follows from Eq.~\eqref{tauG}.  For superpositions involving the
second excited state one similarly finds
\begin{align}
\Gamma^{\mathrm{grav,th}}_{02}
&= \pi\, f(\omega_c)\, C_{\omega_c}^2 \left(1 + 8\bar n_c\right), \\
\Gamma^{\mathrm{grav,th}}_{22}
&= 2\pi\, f(\omega_c)\, C_{\omega_c}^2 \left(1 + 6\bar n_c\right),
\end{align}
as obtained from Eq.~\eqref{fd}.

Taking the vacuum limit $\bar n_c \to 0$ yields the key structural
result
\begin{equation}
\Gamma^{\mathrm{grav,vac}}_{01} = 0,
\qquad
\Gamma^{\mathrm{grav,vac}}_{02}
= \pi\, f(\omega_c)\, C_{\omega_c}^2 \neq 0.
\end{equation}
Thus, at leading order the vacuum gravitational environment does not
induce decoherence for superpositions supported entirely within the
$\{\ket{0},\ket{1}\}$ manifold, while decoherence appears once states
separated by two quanta are involved. This behavior follows directly
from the quadratic detector--gravity coupling, which enforces the
selection rule $\Delta n=\pm2$.

\begin{figure}[t]
\centering
\includegraphics[width=0.9\linewidth]{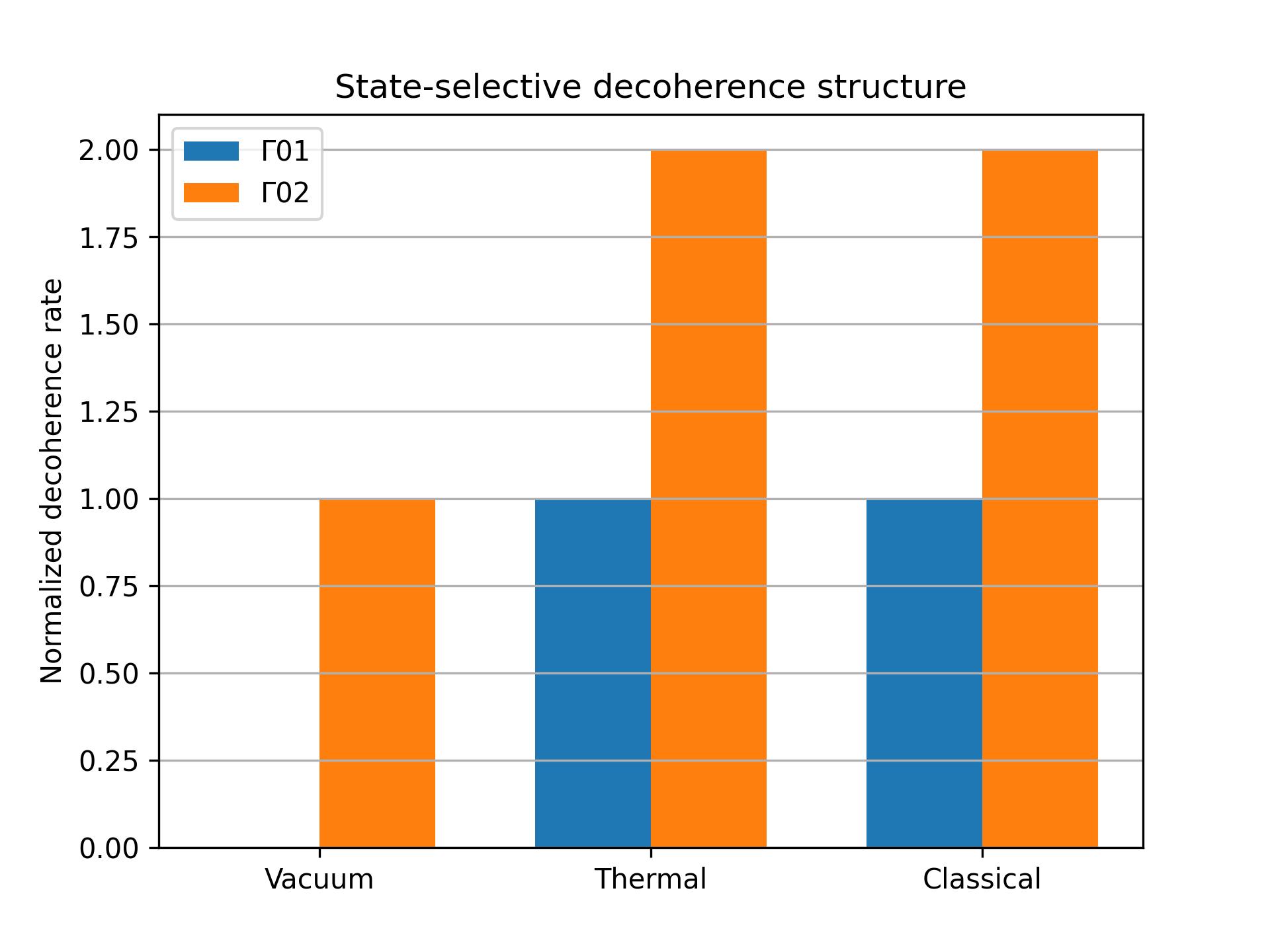}
\caption{
Comparison of gravitationally induced decoherence rates for the
lowest Fock superpositions. In a quantum-vacuum gravitational
environment the rate $\Gamma_{01}$ vanishes while $\Gamma_{02}$
remains finite, reflecting the $\Delta n=\pm2$ selection rule of the
quadratic coupling. Thermal and classical phase-randomized
gravitational backgrounds instead exhibit the universal scaling
$\Gamma_{02}=2\Gamma_{01}$, indicating the absence of a protected
subspace.}
\label{fig:bar_gamma}
\end{figure}

By contrast, thermal graviton states and classical stochastic
gravitational backgrounds generate decoherence already in the lowest
Fock manifold and obey the universal scaling
$\Gamma_{02}=2\Gamma_{01}$ at small occupation numbers.  The resulting
structural difference between the vacuum and stochastic cases is
illustrated in Fig.~\ref{fig:bar_gamma}, which compares the rates
$\Gamma_{01}$ and $\Gamma_{02}$ across the different gravitational
environments.

\subsection*{Step 4: Construction of a dimensionless selectivity ratio}

Because gravitational decoherence cannot be isolated from environmental noise
at the level of absolute rates, we introduce a dimensionless observable that
depends only on the \emph{relative scaling} of decoherence across Fock
manifolds:
\begin{equation}
R \equiv \frac{\Gamma^{\mathrm{tot}}_{02}}{2\,\Gamma^{\mathrm{tot}}_{01}},
\end{equation}
where $\Gamma^{\mathrm{tot}}_{ij}=\Gamma^{\mathrm{env}}_{ij}
+\Gamma^{\mathrm{grav}}_{ij}$ are the experimentally measured total decoherence
rates. 
For generic Markovian non-gravitational environments described by Lindblad
operators that are linear in the oscillator variables, the decoherence rates
exhibit a universal scaling,
\(
\Gamma^{\mathrm{env}}_{n_1 n_2} \propto n_1 + n_2
\)
(see Eq.~\eqref{ds}). As a result,
\(
\Gamma^{\mathrm{env}}_{02} = 2\,\Gamma^{\mathrm{env}}_{01},
\)
yielding $R = 1$ independently of the microscopic details of the environment.

Importantly, this behavior is both qualitatively and quantitatively identical to that obtained for thermal graviton backgrounds and for phase-randomized coherent gravitational states in the regime of sufficiently large occupation numbers. Although these two states represent distinct physical descriptions --- a mixed quantum Gaussian state and a classical stochastic ensemble, respectively --- they yield identical second-order correlation functions at the detector frequency. Consequently, the induced decoherence rates obey the universal scaling
\begin{equation}
\Gamma^{\mathrm{grav}}_{02} \simeq 2\, \Gamma^{\mathrm{grav}}_{01},
\end{equation}
with the overall magnitude controlled by the occupation number of the background: $\bar n_{c}$ for a thermal state and $|\alpha|^2$ for a phase--randomized
coherent state. Indeed, the latter case can be obtained from the former by the
formal replacement $\bar n_{c} \rightarrow |\alpha|^2$,
as follows from Eq.~\eqref{x}.

\subsection*{Step 5: Interpretation and no-go criterion}

\begin{figure}[t]
\centering
\includegraphics[width=0.9\linewidth]{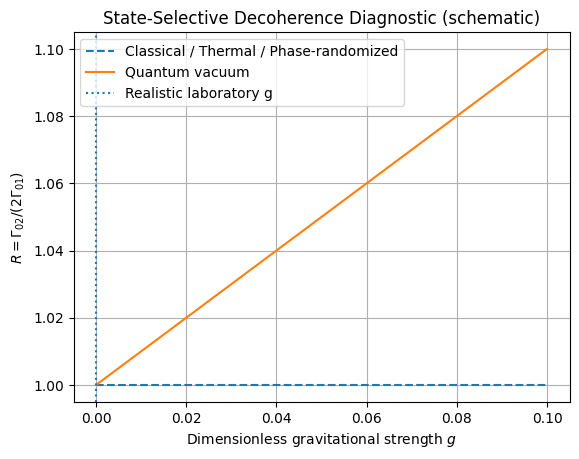}
\caption{
Schematic illustration of the state-selective diagnostic.
Classical stochastic, thermal, and phase-randomized coherent
gravitational backgrounds yield the universal scaling
$\Gamma_{02}=2\Gamma_{01}$ and therefore $R=1$.
Vacuum fluctuations of a quantized gravitational field
instead produce $R=1+g$, reflecting suppression of
decoherence within the $\{|0\rangle,|1\rangle\}$ manifold.
The horizontal axis is displayed schematically for clarity;
realistic laboratory parameters correspond to $g\ll 1$.
}
\label{fig:selectivity}
\end{figure}

The experimentally measured ratio
\begin{equation}
R \equiv \frac{\Gamma_{02}}{2\Gamma_{01}}
\end{equation}
provides a state-selective diagnostic of the gravitational environment.
Its interpretation is summarized in Fig.~\ref{fig:selectivity} and can be
stated as follows:

\begin{itemize}
\item $R\simeq 1$: consistent with ordinary environmental decoherence,
classical stochastic gravitational-wave backgrounds, thermal graviton
states, or phase-randomized coherent gravitational radiation. In all
these cases the decoherence rates obey the universal scaling
$\Gamma_{02}=2\Gamma_{01}$, so that no protected subspace exists in the
lowest Fock manifold.

\item $R>1$: indicates suppression of decoherence within the
$\{|0\rangle,|1\rangle\}$ manifold while decoherence persists for
superpositions involving higher Fock states. This behavior arises from
the quadratic detector--gravity coupling and is uniquely consistent, at
leading order, with vacuum fluctuations of a quantized gravitational
field.
\end{itemize}

To make this statement explicit, consider a quantum gravitational vacuum
in the presence of a generic Markovian environment. The total decoherence
rates can be written as
\begin{equation}
\Gamma^{\mathrm{tot}}_{01}=\Gamma^{\mathrm{env}}_{01},\qquad
\Gamma^{\mathrm{tot}}_{02}=\Gamma^{\mathrm{env}}_{02}
+\Gamma^{\mathrm{grav}}_{02},
\end{equation}
where the vacuum graviton contribution satisfies
$\Gamma^{\mathrm{grav}}_{01}=0$ and $\Gamma^{\mathrm{grav}}_{02}\neq0$.
For standard environmental channels described by operators linear in the
oscillator variables one has $\Gamma^{\mathrm{env}}_{02}=2\Gamma^{\mathrm{env}}_{01}$,
yielding
\begin{equation}
R
= \frac{\Gamma^{\mathrm{tot}}_{02}}{2\Gamma^{\mathrm{tot}}_{01}}
= 1 + \frac{\Gamma^{\mathrm{grav}}_{02}}{2\Gamma^{\mathrm{env}}_{01}}
> 1.
\end{equation}

Any statistically significant deviation of $R$ from unity therefore
signals a structural suppression of decoherence in the lowest Fock
manifold. Such a deviation cannot be reproduced by classical stochastic
gravitational-wave backgrounds, thermal graviton states, or
phase-randomized coherent radiation derived from the same microscopic
coupling. Within the perturbative regime considered here, it constitutes
an operational signature consistent only with vacuum fluctuations of a
quantized gravitational field.

\subsection*{Feasibility and outlook}

The absolute magnitude of gravitationally induced decoherence
predicted by the present framework is strongly suppressed by the
Planck scale. In a quantum-vacuum gravitational environment,
the leading nonvanishing contribution arises only for
superpositions differing by two quanta and is parametrically
proportional to $f(\omega_c) C_{\omega_c}^2 \sim G \omega_c^5$.
The corresponding decoherence times therefore exceed present
mechanical coherence times by many orders of magnitude,
making direct observation of vacuum graviton emission in
mesoscopic systems unrealistic in the near term.

The experimentally relevant question is instead whether
non-vacuum gravitational backgrounds contribute measurable
decoherence beyond well-characterized environmental channels.
The detector frequency considered here lies in the GHz regime,
corresponding to the operational band of current high-frequency
mechanical resonators such as HBAR devices.
Standard slow-roll inflation produces a squeezed vacuum spectrum
of tensor modes whose present-day frequencies typically do not
extend far beyond $\mathcal{O}(10^{7}\,\mathrm{Hz})$ after
cosmological redshift \cite{Ito:2021JCAPbreak}.
Thus conventional inflationary vacuum fluctuations do not
naturally populate the GHz band \cite{PhysRevD.109.096023}. Any nonzero occupation number
at GHz frequencies  must therefore originate from alternative
high-frequency production mechanisms \cite{Kanno:2023whr}, such as post-inflationary
preheating dynamics, high-scale phase transitions, compact
sources (e.g.\ primordial black hole binaries), or other exotic
gravitational processes \cite{Maggiore:2000review,Ema:2020inflatonGW,Domcke:2020radio}.
Our analysis does not assume any specific cosmological origin
and constrains only the occupation number at the detector frequency.

For thermal or phase-randomized coherent graviton states,
the decoherence rate scales linearly with the mean occupation
number $\bar n_c$ (or $|\alpha|^2$). Using the continuum
spectral density $f(\omega_c)=\omega_c^2$ and the
coupling $C_{\omega_c}^2=\omega_c \hbar G/(2c^2)$, the
gravitational contribution to the lowest-manifold decoherence
rate becomes
\begin{equation}
\Gamma^{\mathrm{grav}}_{01}
=
\frac{\hbar G}{\pi c^2}\,
\omega_c^3\,\bar n_c .
\end{equation}

For an HBAR device with
$\omega_c = 2\pi\times 5.96\,\mathrm{GHz}$ 
\cite{PhysRevD.111.026009,Chu:2017koi},
this evaluates numerically to
\begin{equation}
\Gamma^{\mathrm{grav}}_{01}
\approx
1.3\times10^{-30}\,\bar n_c
\;\mathrm{s}^{-1}.
\end{equation}
Demanding that no anomalous decoherence exceed the measured
mechanical dephasing rate
$\Gamma_{\mathrm{env}}\sim10^{4}\,\mathrm{s}^{-1}$
implies the bound
\begin{equation}
\bar n_c \lesssim 8\times10^{33}.
\end{equation}
Thus present GHz mechanical experiments already constrain
non-vacuum gravitational radiation backgrounds to have
effective occupation numbers below $\sim10^{34}$ near
$\omega_c$. Although numerically large, this bound directly
reflects the Planck suppression of the gravitational coupling
at GHz frequencies and is independent of cosmological model
assumptions within the perturbative regime considered here.

Complementary information is obtained from the
state-selective ratio
\(
R=\Gamma_{02}/(2\Gamma_{01}).
\)
Thermal graviton states, classical stochastic backgrounds,
and phase-randomized coherent states all obey the universal
scaling $\Gamma_{02}=2\Gamma_{01}$, implying
\begin{equation}
R=1.
\end{equation}
If an experiment verifies $|R-1|<\varepsilon$
within statistical precision $\varepsilon$,
any vacuum-induced contribution must satisfy
\begin{equation}
\Gamma_{02}^{\mathrm{grav,vac}}
<
2\,\varepsilon\,\Gamma_{01}^{\mathrm{env}}.
\end{equation}
Conversely, a statistically significant deviation of $R$
from unity would indicate suppression of decoherence
within the $\{|0\rangle,|1\rangle\}$ manifold and
cannot be reproduced by classical or stochastic
gravitational backgrounds derived from the same
microscopic coupling. Within the perturbative regime,
such a deviation would therefore be consistent only
with vacuum-induced two-quantum processes.

The protocol thus provides two independent constraints:
(i) an absolute-rate bound on non-vacuum graviton occupation,
and (ii) a structural bound on vacuum-induced decoherence
derived from deviations of $R$ from unity. Improvements in
mechanical coherence time, state-preparation fidelity, and
statistical precision translate directly into progressively
stronger bounds on both quantities.

While Planck-suppressed vacuum effects remain far below
current experimental sensitivity, continued advances in
cryogenic optomechanics \cite{PhysRevA.93.013836}, circuit quantum acoustodynamics \cite{hxvq-vvnv},
and high-overtone bulk acoustic resonator \cite{Chu:2017koi} platforms steadily
extend coherence times and quantum control capabilities.
Within this trajectory, the framework developed here provides
a concrete operational roadmap for constraining the quantum
state of gravitational radiation using mesoscopic quantum
systems.

\begin{table*}[t]
\centering
\begin{tabular}{l l c c}
\hline\hline
Environment & Physical interpretation & Decoherence? & Protected? \\
\hline

Pure coherent state (fixed phase) &
Deterministic classical gravitational wave &
No & N/A \\

Phase-randomized coherent ensemble &
Classical stochastic gravitational-wave background &
Yes & No \\

Thermal graviton state &
Gaussian state (parametrization of non-vacuum quantum backgrounds) &
Yes & No \\

Quantum vacuum &
Quantum vacuum fluctuations of the gravitational field &
Selective & Yes \\

\hline\hline
\end{tabular}
\caption{Classification of gravitational environments analyzed in this work.
Only the vacuum state leads to a protected low-phonon manifold at leading order.}
\label{tab:GWclassification}
\end{table*}

\section{Conclusion}

In this work we have developed a unified and internally consistent framework
for analysing how gravitational waves—treated either as a quantized field or
as a classical stochastic background—affect the coherence properties of a
mesoscopic quantum system. Starting from a common microscopic coupling
between a harmonic oscillator and a weak gravitational perturbation, we
derived the reduced dynamics in both the quantum and classical descriptions
and compared the resulting master equations on equal footing. This unified
treatment eliminates model-dependent assumptions and makes the dynamical
differences between distinct gravitational environments directly visible
at the level of observable decoherence.

Our central finding is that \emph{decoherence by itself is not a signature of
quantum gravity}. Classical stochastic gravitational-wave backgrounds,
quantum thermal graviton fields, and quantum vacuum fluctuations can all
induce decoherence in an open quantum system. What distinguishes these
scenarios is instead the \emph{structure} of the induced reduced dynamics.
In particular, in the quantum-vacuum case decoherence arises solely through
spontaneous emission of real, on-shell gravitons, leading to a two-quantum
selection rule that suppresses transitions within the lowest-excitation
manifold $\{|0\rangle,|1\rangle\}$ at leading order. By contrast, a classical
stochastic gravitational background generically induces phase diffusion even
within this same subspace, thereby removing the protection.

This structural distinction provides a sharp operational criterion (see Table~I).
The absence of decoherence in the $\{|0\rangle,|1\rangle\}$ manifold is
consistent only with a quantum gravitational vacuum, whereas the observation
of decoherence within this nominally protected sector would rule out a
purely vacuum description of the gravitational field, independently of the
absolute magnitude of the decoherence rate. Deterministic classical
gravitational waves—described by coherent states of the field—produce
unitary dynamics without intrinsic decoherence, highlighting that
stochasticity rather than classicality per se is the essential ingredient
behind classical gravitational decoherence.

At the level of linearized quantum field theory there exists a single
dynamical gravitational field $h_{\mu\nu}$, and the distinction between
``external'' and ``self'' gravity is not fundamental. Static gravitational
interactions arise from virtual graviton exchange at tree level and
correspond to conservative forces such as the Newtonian potential, while
gravitational radiation corresponds to the emission of real, on-shell
gravitons. The vacuum state is simply the ground state of this field.
Consequently, whether gravity acts as an external environment or as a
channel for spontaneous emission is determined by the quantum state in
which the field is prepared, rather than by a structural difference in
the interaction itself.

In the vacuum setting, coherence protection within
$\{|0\rangle,|1\rangle\}$ follows from the two-quantum transition
mechanism associated with spontaneous graviton radiation, as shown
in quantum-field-theoretic analyses of harmonic systems
\cite{PhysRevD.109.084050}. Our contribution is to demonstrate that this
protection is not a universal consequence of weak gravitational
coupling, but a property specific to the vacuum state of the
gravitational field. When the same microscopic coupling is considered
for classical stochastic or non-vacuum gravitational configurations,
the protected subspace generically disappears. In this way, coherence
protection is elevated from a radiation-theory result to a structural
diagnostic of gravitational `quantum'-ness.

We emphasize that the Born--Markov treatment adopted here does not exclude
spontaneous graviton emission by the detector \cite{PhysRevD.110.026008}. Consistency requires only
that the detector-induced modification of the gravitational bath remain
perturbatively small. As shown in Appendix~E, the emission rate is
suppressed by the Planck scale, leading to an occupation number
$\langle b_k^\dagger b_k \rangle \sim \gamma_{\mathrm{em}} t \ll 1$
over experimentally relevant timescales. The gravitational bath therefore
remains effectively in its initial state to leading order, ensuring that
the factorization assumption underlying the master-equation derivation is
self-consistent.

Quantitatively, present GHz mechanical experiments already constrain
non-vacuum gravitational radiation backgrounds in the quantum regime.
Using experimentally accessible parameters for high-overtone bulk acoustic
resonators, the absence of excess decoherence beyond measured
environmental dephasing implies an upper bound on the effective graviton
occupation number at GHz frequencies of order $\bar n_c \lesssim 10^{34}$.
Although numerically large, this bound directly reflects the extreme
Planck suppression of gravitational couplings at these frequencies and
constitutes a direct quantum-regime constraint on non-vacuum gravitational
backgrounds within the perturbative framework considered here.

Complementary information is obtained from the state-selective ratio
$R=\Gamma_{02}/(2\Gamma_{01})$.
Thermal graviton states, classical stochastic backgrounds, and
phase-randomized coherent states all yield the universal scaling
$\Gamma_{02}=2\Gamma_{01}$ and therefore $R=1$.
If an experiment were to observe a statistically significant deviation
$|R-1|>\varepsilon$, such a deviation could not be reproduced by these
non-vacuum gravitational configurations derived from the same microscopic
interaction and would instead indicate vacuum-induced two-quantum
processes at leading order.
In this sense, improvements in coherence time and statistical precision
translate directly into progressively tighter bounds on both non-vacuum
gravitational occupation and vacuum-induced decoherence contributions.

Conceptually, the present work shifts the focus of gravitational
decoherence studies from the magnitude of decay rates to the qualitative
structure of reduced dynamics. Rather than asking whether gravity induces
decoherence, we ask how the pattern of decoherence depends on the
microscopic state of the gravitational field. This reframing provides a
clear and experimentally falsifiable criterion rooted in symmetry and
selection rules rather than absolute rate estimates.

An interesting extension of the present framework would be to analyze
gravitational environments prepared in squeezed quantum states.
Time-dependent gravitational backgrounds, such as cosmological
expansion or dynamical matter sources, are known to generically
generate squeezed states of graviton modes in perturbative quantum
gravity~\cite{Das:2025SqueezedGravity}. Since squeezed states introduce
additional anomalous correlations in the environmental two-point
functions, incorporating such gravitational environments within the
open-system treatment developed here could modify the structure of the
induced master equation and potentially lead to new signatures in the
decoherence pattern of mesoscopic quantum probes. Exploring these
effects would provide a natural generalization of the present analysis
and further clarify how different quantum states of gravitational
radiation are reflected in detector dynamics.

Experimentally, quantum systems are inevitably open, and environmental
decoherence from electromagnetic, thermal, and material noise cannot be
fully eliminated. However, advances in cryogenic optomechanics,
circuit–mechanical architectures, and high-overtone bulk acoustic
resonators (HBARs) coupled to superconducting qubits already enable
preparation of lowest-excitation states, including coherent
superpositions of $|0\rangle$ and $|1\rangle$, as well as nonclassical
mechanical states. The state family required for the present proposal is
therefore technologically accessible, independent of the specific hardware
implementation. While current coherence times remain insufficient to
observe Planck-suppressed vacuum effects directly, the identification of
protected and unprotected manifolds enables differential measurements
robust against unknown environmental noise. Within this trajectory, the
framework developed here establishes a concrete operational roadmap for
probing and constraining the microscopic quantum state of gravitational
radiation using mesoscopic quantum systems.

In summary, while decoherence is a generic feature of open quantum
systems, its \emph{state-dependent structure} can serve as an operational
probe of the microscopic nature of gravitational radiation. The
persistence or breakdown of coherence within the lowest-excitation
manifold provides a precise and experimentally meaningful discriminator
between a quantum gravitational vacuum and classical or stochastic
gravitational backgrounds. This state-selective approach offers a
conceptually distinct route to probing the quantumness of gravity—one
rooted not in the magnitude of energy exchange, but in the qualitative
structure of quantum coherence.

\begin{acknowledgments}
P.N. acknowledges support from the National Institute for Theoretical and Computational Sciences (NITheCS) through the Rector’s Postdoctoral Fellowship Programme (RPFP). He thanks Prof.\ Douglas Singleton and Prof.\ Kazuhiro Yamamoto for their helpful comments and keen interest in this work.

P.N. and S.S. also thanks Dr.\ Sandro Donadi and Dr.\ Matteo Fadel for useful discussions and correspondence. P.N. is grateful to Prof.\ Ananda Das Gupta and Prof.\ Golam Mortuza Hossain for constructive comments on the Born--Markov approximation and for arranging an invited lecture at the Indian Institute of Science Education and Research (IISER) Kolkata.
\end{acknowledgments}

\appendix

\section{Derivation of the Hamiltonian for graviton modes} \label{AppA}
\label{Appendix A}
The analysis is being followed from \cite{Nandi:2024jyf,Nandi:2024zxp}
We start from linearized gravity about Minkowski space,
\begin{equation}
g_{\mu\nu}=\eta_{\mu\nu}+h_{\mu\nu}, \qquad |h_{\mu\nu}|\ll 1,
\end{equation}
and impose the transverse--traceless (TT) gauge on the spatial components:
\begin{equation}
h_{0\mu}=0,\qquad \partial^i h_{ij}=0,\qquad h^i{}_{i}=0.
\end{equation}
In this gauge the quadratic (free) gravitational (Einstein-Hilbert) action is
\begin{align}
S_{\mathrm{grav}} &= -\frac{1}{64 \pi G} \int d^4x \, (\partial_\alpha h_{ij}) (\partial^\alpha h^{ij}) \nonumber\\
&= \frac{1}{64 \pi G} \int d^4x \, \Big( \dot{h}_{ij} \dot{h}^{ij} - \partial_k h_{ij} \, \partial_k h^{ij} \Big).
\end{align}

To exhibit the canonical variables, we decompose the metric deformation into TT plane waves in a cubic box of side $L$ (volume $V=L^3$):
\begin{equation}
h_{ij}(t,\vec x)=\frac{1}{\sqrt{\hbar G}}\sum_{\vec k,\,s=\{+,\times\}} 
q_{\vec k,s}(t)\,e^{i\vec k\cdot\vec x}\,\epsilon^{(s)}_{ij}(\vec k),
\end{equation}
where $\epsilon^{(s)}_{ij}(\vec k)$ are real, transverse and traceless polarization tensors, orthonormal on the TT subspace. 
The ``canonical coordinate" for each graviton degree of freedom is the TT-projected Fourier amplitude of the metric deformation:
\begin{equation}
q_{\vec k,s}(t)=\sqrt{\hbar G}\,\frac{1}{V}\int d^3x\;e^{-i\vec k\cdot\vec x}\,
\epsilon^{(s)}_{ij}(\vec k)\,h_{ij}(t,\vec x),
\end{equation}
so that the metric perturbation is reconstructed from the set $\{q_{\vec k,s}\}$ by the inverse expansion above. 
Reality of $h_{ij}$ implies $q^{\ast}_{\vec k,s}(t)\epsilon^{(s)}_{ij}(\vec k)=q_{-\vec k,s}(t)\epsilon^{(s)}_{ij}(-\vec k)$. 

Substituting the mode expansion and using 
$\int_V d^3x\,e^{i(\vec k-\vec k')\cdot\vec x}=V\,\delta_{\vec k,\vec k'}$ 
and polarization orthonormality, the action becomes a sum of decoupled oscillators: 
\begin{equation}
S = \frac{L^3}{32\pi\,\hbar\,G^2} \int dt \sum_{\vec k, s} \Big( \dot q_{\vec k,s}^2 - \omega_k^2 \, q_{\vec k,s}^2 \Big) \equiv \int dt~ \mathcal{L}, 
\qquad \omega_k \equiv |\vec k|.
\end{equation}
Here $L$ denotes the linear size of the normalization volume $V=L^3$
introduced in the canonical quantization procedure. In the continuum
limit, the explicit dependence on $L$ cancels against the density of
states. For a gravitational mode of frequency $\omega_g$ (setting $c=1$),
the physically relevant scale governing the interaction is the
corresponding wavelength $\lambda_g \sim 1/\omega_g$.

The canonical momentum conjugate to \(q_{\vec k,s}\) is
\begin{equation}
p_{\vec k,s} \equiv \frac{\partial \mathcal{L}}{\partial \dot q_{\vec k,s}} = \frac{L^3}{16\pi\,\hbar\,G^2} \, \dot q_{\vec k,s},
\end{equation}
so that
\begin{align}
\dot{q}_{\vec{k},s} &= \frac{16 \pi \, \hbar \, G^2}{L^3} \, p_{\vec{k},s}, \\
p_{\vec{k},s} &= \frac{L^3}{16 \pi \, \hbar \, G^2} \, \dot{q}_{\vec{k},s}.
\end{align}
It is now convenient to introduce a mode-independent ``mass'' parameter
\begin{equation}
m \equiv \frac{L^3}{16 \pi \, \hbar \, G^2},
\end{equation}
so that the Hamiltonian for each mode takes the standard harmonic-oscillator form:
\begin{equation}
H_{\vec k,s} = \frac{p_{\vec k,s}^2}{2 m} + \frac{1}{2} m \, \omega_k^2 \, q_{\vec k,s}^2, 
\qquad H = \sum_{\vec k, s} H_{\vec k,s}.
\end{equation}
For concreteness, let us focus on a single mode with wave vector along the $+\hat{z}$ direction and frequency
\begin{equation}
\omega_k= \omega^{g}_{k}= |\vec k|,
\end{equation}
restricted to the $+$ polarization for simplicity (the $\times$ polarization being dynamically identical at the free-field level).
In this case, the effective action and Hamiltonian reduce to
The action and Hamiltonian for a single graviton mode are
\begin{align}
S_{\omega_g} &= \int dt \, \frac{1}{2} m 
\left( \dot{q}_{+}^{2} - (\omega^{g}_{k})^2 q_{+}^{2} \right), \\
H_{\omega_g} &= \frac{p_{+}^2}{2m} + \frac{1}{2} m (\omega^{g}_{k})^2 q_{+}^{2}.
\end{align}
where \(q_{+}\) and \(p_{+}\) are the canonical coordinate and momentum of the graviton mode with frequency \(\omega^{g}_k\) and effective mass \(m\).

where, for brevity, we have denoted $q_{\vec{k},+} \to q_{+}$ and $p_{\vec{k},+} \to p_{+}$.
Canonical quantization promotes $(q_{+},p_{+})$ to operators with $[\hat q_{+},\hat p_{+}] = i\hbar$. Introducing the ladder operators
\begin{align}
\hat b_k &= \sqrt{\frac{m \, \omega^g_{k}}{2\hbar}} \, \hat q_{+} 
         + \frac{i}{\sqrt{2 m \hbar \omega^g_{k}}} \, \hat p_{+}, \nonumber\\
\hat b_k^\dagger &= \sqrt{\frac{m \, \omega^g_{k}}{2\hbar}} \, \hat q_{+} 
                 - \frac{i}{\sqrt{2 m \hbar \omega^g_{k}}} \, \hat p_{+}.
\end{align}

which satisfy $[\hat b_k, \hat b_{k}^{\dagger}] = 1$, the Hamiltonian for a single graviton mode can be expressed in normal-ordered form as
\begin{equation}
\hat H_G \equiv \hat H_{\omega^g_{k}} = \hbar \omega^g_{k} \, :\hat b_k^\dagger \hat b_k: ,
\end{equation}
where the colons denote normal ordering, ensuring that the vacuum contribution is omitted. Each transverse-traceless (TT) graviton mode $(\vec k, s)$ thus corresponds to a quantum harmonic oscillator of frequency $\omega_g = |\vec k|$, with canonical coordinate given by the TT-projected Fourier amplitude of the metric perturbation, and canonical momentum proportional to its time derivative, as defined above.

As we have chosen only the plus polarization for the GW, for each mode of the GW i.e. for each $\vec{k}$, we can have
\begin{equation}\label{mode}
    h_{ij}(t)=2\chi(t)\epsilon_+\sigma_{ij}^3=\frac{1}{\sqrt{\hbar G}}q_{+}(t)e^{i\vec{k}.\vec{x}}\epsilon_{ij}^{+}(\vec{k})~,
\end{equation}
where  $\epsilon_{ij}^{+}(\vec{k})$ is the polarization tensor corresponding to the plus polarization. Then from \eqref{mode}, we get
\begin{equation}
    \chi(t)=\frac{1}{2\sqrt{\hbar G}}q_{+}(t)e^{i\vec{k}.\vec{x}}~,
\end{equation}
which for GWs moving along z-direction reduces to $\chi(t)=\frac{1}{2\sqrt{\hbar G}}q_{+}(t)e^{ikz}$. Again, we have already promoted $q_+\rightarrow\hat{q}_+$ i.e to its quantum status. Then, we can expand $\hat{q}_+$ in terms of its modes as: $
\hat{q}_+=\sqrt{\frac{\hbar}{2m\omega_g}}(\hat{b}_ge^{-i\omega_gt}+\hat{b}_g^\dagger e^{i\omega_gt})$. Then, elevating $\chi(t)$ to its corresponding operator and without loss of generality choosing $z=0$, we get
\begin{equation}
    \hat{\chi}(t)=\frac{1}{2\sqrt{\hbar G}}\sqrt{\frac{\hbar}{2m\omega_g}}(\hat{b}_ge^{-i\omega_gt}+\hat{b}_g^\dagger e^{i\omega_gt})~.
\end{equation}
This will yield \eqref{f}.

\section{Computation of Eq~(\ref{eq:term1})} \label{AppB}

In this appendix, we derive the terms appearing in Eq.~\eqref{eq:term1}.  


Starting from the interaction Hamiltonian, we have
\begin{align}
& \int_0^\infty ds~ \langle \hat{H}_{\rm int}(t) \hat{H}_{\rm int}(t-s) \hat{\rho}_d(t) \rangle_g \nonumber\\
&= \hbar^2 \int_0^\infty ds \sum_{l,m} \mathcal{C}_{\omega^g_l} \mathcal{C}_{\omega^g_m} 
\Big\langle 
\big[
\hat{b}_l \hat{a}_1^{\dagger 2} e^{i(2\omega_c-2\omega^g_l)t}\nonumber\\
&\quad 
- \hat{b}_l \hat{a}_1^2 e^{-i(2\omega_c+2\omega^g_l)t} \nonumber\\
&\quad
+ \text{h.c.}
\big] \nonumber\\
&\quad \times 
\big[
\hat{b}_m \hat{a}_1^{\dagger 2} e^{i(2\omega_c-2\omega^g_m)(t-s)} 
- \hat{b}_m \hat{a}_1^2 e^{-i(2\omega_c+2\omega^g_m)(t-s)} 
+ \text{h.c.}
\big] \nonumber\\
&\quad\hat{\rho}_d(t)
\Big\rangle_g.
\end{align}

Evaluating the expectation values and keeping only non-vanishing contributions, we obtain
\begin{align}
& \int_0^\infty ds~ \langle \hat{H}_{\rm int}(t) \hat{H}_{\rm int}(t-s) \hat{\rho}_d(t) \rangle_g \nonumber\\
&= \hbar^2 \int_0^\infty ds \sum_l \mathcal{C}_{\omega^g_l}^2 
\Big[ 
\hat{a}_1^{\dagger 2} \hat{a}_1^2 \, \hat{\rho}_d(t) \, (1+\bar{n}_l) \, e^{-i(2\omega^g_l-2\omega_c)s} \nonumber\\
&\quad - \hat{a}_1^2 \hat{a}_1^2 \, \hat{\rho}_d(t) \, (1+\bar{n}_l) \, e^{-i 4\omega_c t} \, e^{i(2\omega^g_l-2\omega_c)s} \nonumber\\
&\quad + \hat{a}_1^2 \hat{a}_1^{\dagger 2} \, \hat{\rho}_d(t) \, \bar{n}_l \, e^{-i(2\omega_c-2\omega^g_l)s} \nonumber\\
&\quad - \hat{a}_1^{\dagger 2} \hat{a}_1^{\dagger 2} \, \hat{\rho}_d(t) \, \bar{n}_l \, e^{i 4 \omega_c t} \, e^{-i(2\omega_c-2\omega^g_l)s} 
\Big].
\end{align}

Assuming a dense distribution of reservoir modes with density $f(\omega^g_l)$, the sum $\sum_l$ can be replaced by an integral $\int d\omega^g_l f(\omega^g_l)$. Using the Markov approximation and performing the $s$-integration, we get
\begin{align}
& \int_0^\infty ds~ \langle \hat{H}_{\rm int}(t) \hat{H}_{\rm int}(t-s) \hat{\rho}_d(t) \rangle_g \nonumber\\
&= \frac{\pi}{2} \hbar^2 f(\omega_c) C_{\omega_c}^2
\Big[
(1+\bar{n}_c) (\hat{a}_1^{\dagger 2} \hat{a}_1^2 \hat{\rho}_d(t) - \hat{a}_1^2 \hat{a}_1^2 \hat{\rho}_d(t) e^{-i4\omega_c t}) \nonumber\\
&\quad + \bar{n}_c (\hat{a}_1^2 \hat{a}_1^{\dagger 2} \hat{\rho}_d(t) - \hat{a}_1^{\dagger 2} \hat{a}_1^{\dagger 2} \hat{\rho}_d(t) e^{i4\omega_c t})
\Big].
\end{align}

Similarly, the other correlation terms appearing in the right-hand side of Eq.~\eqref{FE} are

\begin{align}
& \int_0^\infty ds~ \langle \hat{H}_{\rm int}(t) \hat{\rho}_d(t) \hat{H}_{\rm int}(t-s) \rangle^{*}_g \nonumber\\
&= \frac{\pi}{2} \hbar^2 f(\omega_c) C_{\omega_c}^2 
\Big[
\bar{n}_c \big( 
\hat{a}_1^{\dagger 2} \hat{\rho}_d(t) \hat{a}_1^2 
- \hat{a}_1^2 \hat{\rho}_d(t) \hat{a}_1^2 e^{-i 4 \omega_c t} 
\big) \nonumber\\
&\quad +(1+ \bar{n}_c) \big( 
\hat{a}_1^2 \hat{\rho}_d(t) \hat{a}_1^{\dagger 2} 
- \hat{a}_1^{\dagger 2} \hat{\rho}_d(t) \hat{a}_1^{\dagger 2} e^{i 4 \omega_c t} 
\big)
\Big],\\[1mm]
& \int_0^\infty ds~ \langle \hat{H}_{\rm int}(t-s) \hat{\rho}_d(t) \hat{H}_{\rm int}(t) \rangle^{*}_g \nonumber\\
&= \frac{\pi}{2} \hbar^2 f(\omega_c) C_{\omega_c}^2 
\Big[
\bar{n}_c\big( 
\hat{a}_1^{\dagger 2} \hat{\rho}_d(t) \hat{a}_1^2 
- \hat{a}_1^{\dagger 2} \hat{\rho}_d(t) \hat{a}_1^{\dagger 2} e^{i 4 \omega_c t} 
\big) \nonumber\\
&\quad + (1+\bar{n}_c) \big( 
\hat{a}_1^2 \hat{\rho}_d(t) \hat{a}_1^{\dagger 2} 
- \hat{a}_1^2 \hat{\rho}_d(t) \hat{a}_1^2 e^{-i 4 \omega_c t} 
\big)
\Big],\\[1mm]
& \int_0^\infty ds~ \langle \hat{\rho}_d(t) \hat{H}_{\rm int}(t-s) \hat{H}_{\rm int}(t) \rangle_g \nonumber\\
&= \frac{\pi}{2} \hbar^2 f(\omega_c) C_{\omega_c}^2 
\Big[
(1+\bar{n}_c) \big( 
\hat{\rho}_d(t) \hat{a}_1^{\dagger 2} \hat{a}_1^2 
- \hat{\rho}_d(t) \hat{a}_1^{\dagger 2} \hat{a}_1^{\dagger 2} e^{i 4 \omega_c t} 
\big) \nonumber\\
&\quad + \bar{n}_c \big( 
\hat{\rho}_d(t) \hat{a}_1^2 \hat{a}_1^{\dagger 2} 
- \hat{\rho}_d(t) \hat{a}_1^2 \hat{a}_1^2 e^{-i 4 \omega_c t} 
\big)
\Big].
\end{align}

\section{Decoherence with an Energy–Relaxation Channel} \label{AppC}

In this appendix, we extend our analysis by including an energy–relaxation channel in the master equation. 
Specifically, we modify Eq.~\eqref{eom_total_PRD} with the Lindblad contribution
$\gamma\,\hat{a}_{1}\hat{\rho}_{d}(t)\hat{a}_{1}^{\dagger}
-\tfrac{\gamma}{2}\{\hat{a}^{\dagger}\hat{a},\hat{\rho}_{d}(t)\}$,
where $\gamma$ is the energy-relaxation rate. The resulting master equation reads
\begin{align}
\label{NFE3}
\frac{d}{dt}\hat{\rho}_{d}(t)
&=
-\frac{i}{\hbar}[\hat{H}_{0},\hat{\rho}_{d}(t)]
+\gamma\,\hat{a}_{1}\hat{\rho}_{d}(t)\hat{a}_{1}^{\dagger}
-\frac{\gamma}{2}\{\hat{a}^{\dagger}\hat{a},\hat{\rho}_{d}(t)\}
\nonumber\\
&\quad
-\frac{\pi}{2}f(\omega_{c})C_{\omega_{c}}^{2}
\Big[
(1+\bar{n}_{c})\,\mathcal{D}_{+}[\hat{\rho}_{d}(t)]
+
\bar{n}_{c}\,\mathcal{D}_{-}[\hat{\rho}_{d}(t)]
\Big],
\end{align}
where the dissipators $\mathcal{D}_{\pm}$ are defined explicitly as
\begin{align}
\mathcal{D}_{+}[\hat{\rho}_{d}(t)] 
&= 
\{\hat{a}_{1}^{\dagger 2} \hat{a}_{1}^{2}, \hat{\rho}_{d}(t)\} 
- 2 \hat{a}_{1}^{2} \hat{\rho}_{d}(t) \hat{a}_{1}^{\dagger 2} \nonumber\\
&\quad
- \hat{a}_{1}^{2} \hat{a}_{1}^{2} \hat{\rho}_{d}(t) 
- \hat{\rho}_{d}(t) \hat{a}_{1}^{\dagger 2} \hat{a}_{1}^{\dagger 2} 
+ \hat{a}_{1}^{2} \hat{\rho}_{d}(t) \hat{a}_{1}^{2}  \nonumber\\
&\quad
+ \hat{a}_{1}^{\dagger 2} \hat{\rho}_{d}(t) \hat{a}_{1}^{\dagger 2}, \\[2mm]
\mathcal{D}_{-}[\hat{\rho}_{d}(t)] 
&= 
\{\hat{a}_{1}^{2} \hat{a}_{1}^{\dagger 2}, \hat{\rho}_{d}(t)\} 
- 2 \hat{a}_{1}^{\dagger 2} \hat{\rho}_{d}(t) \hat{a}_{1}^{ 2} \nonumber\\
&\quad
- \hat{a}_{1}^{\dagger 2} \hat{a}_{1}^{\dagger 2} \hat{\rho}_{d}(t) 
- \hat{\rho}_{d}(t) \hat{a}_{1}^{2} \hat{a}_{1}^{2} 
+ \hat{a}_{1}^{2} \hat{\rho}_{d}(t) \hat{a}_{1}^{2}   \nonumber\\
&\quad
+ \hat{a}_{1}^{\dagger 2} \hat{\rho}_{d}(t) \hat{a}_{1}^{\dagger 2}.
\label{de}
\end{align}

Following Sec.~III, we write
\begin{equation}
\frac{d}{dt}\hat{\rho}_{d}(t) = (\hat{\mathcal{L}}_1 + \hat{\mathcal{L}}_g) \hat{\rho}_{d}(t), \quad
\hat{\mathcal{L}}_1[\hat{\rho}_{d}] = -\frac{i}{\hbar}[\hat{H}_0, \hat{\rho}_{d}] + \mathcal{L}_{\gamma}[\hat{\rho}_{d}].
\end{equation}
The first-order formal solution is
\begin{equation}
\hat{\rho}_{d}(t) = e^{\hat{\mathcal{L}}_1 t} \left[ 1 + \int_0^t d\tau\, e^{-\hat{\mathcal{L}}_1 \tau} \hat{\mathcal{L}}_g e^{\hat{\mathcal{L}}_1 \tau} \right] \hat{\rho}_{d}(0).
\end{equation}

To leading order in $\gamma$, the free evolution operator acts as
\begin{equation}
e^{\hat{\mathcal{L}}_1 t} A = e^{-i \hat{H}_0 t/\hbar} A e^{i \hat{H}_0 t/\hbar} + O(\gamma),
\end{equation}
and the matrix elements evolve as
\begin{equation}
\langle n_1|\hat{\rho}_{d}(t)|n_2\rangle = \langle n_1|\hat{\rho}_{d}(0)|n_2\rangle \exp\left[ \int_0^t f(n_1,n_2,t') dt'\right] + O(\gamma),
\end{equation}
with
\begin{equation}
f(n_1,n_2,t) = -\frac{i}{\hbar} (E_1 - E_2) - \frac{\gamma}{2} (n_1+n_2).
\label{ds}
\end{equation}

\subsection{Examples}

For an initial superposition $\hat{\rho}_{d}(0) = \tfrac{1}{2} (\ket{0}+\ket{1})(\bra{0}+\bra{1})$,
\begin{align}
\langle 0|\hat{\rho}_{d}(t)|0\rangle &= \frac{1}{2} \left(1 - \frac{t}{\tau_G}\right) + O(\gamma), \\
\langle 1|\hat{\rho}_{d}(t)|1\rangle &= \frac{1}{2} \left(1 - 3\frac{t}{\tau_G}\right) e^{-\gamma t} + O(\gamma), \\
\langle 0|\hat{\rho}_{d}(t)|1\rangle &= \frac{1}{2} \left(1 - 2\frac{t}{\tau_G}\right) e^{i \omega_c t} e^{-\gamma t/2} + O(\gamma), \\
\langle 1|\hat{\rho}_{d}(t)|0\rangle &= \frac{1}{2} \left(1 - 2\frac{t}{\tau_G}\right) e^{-i \omega_c t} e^{-\gamma t/2} + O(\gamma).
\end{align}

For the initially excited state $\hat{\rho}_{d}(0) = \ket{2}\bra{2}$,
\begin{equation}
\langle 2|\hat{\rho}_{d}(t)|2\rangle = \left[ 1 - 2 \pi f(\omega_c) C_{\omega_c}^2 t (1 + 7 \bar n_c) \right] e^{-2 \gamma t} + O(\gamma).
\end{equation}

\subsection{Extraction of effective decay times}

In an experiment, population and coherence decay rates are typically 
obtained by fitting the measured signals to single exponentials of the 
form $e^{-t/T_1}$ and $e^{-t/T_2}$.  
To connect these experimentally defined quantities with the microscopic 
parameters $(\gamma,\tau_{G})$ appearing in our master equation, we now 
derive the corresponding relations directly from Eqs.~(C10) and~(C11).

\paragraph*{Population decay.}
For short evolution times $t\ll \tau_{G}$ and $\gamma t\ll 1$, 
Eq.~(C10) gives
\begin{equation}
\langle 1|\hat\rho_d(t)|1\rangle
= \frac12\!\left(1 - 3\frac{t}{\tau_{G}}\right)e^{-\gamma t}
+ \mathcal{O}(t^2).
\label{eq:PopExact}
\end{equation}
Using $e^{-\gamma t} = 1 - \gamma t + \mathcal{O}(t^2)$, we obtain
\begin{align}
\langle 1|\hat\rho_d(t)|1\rangle
&= \frac12 \left[(1 - 3t/\tau_{G})(1 - \gamma t)\right]
+ \mathcal{O}(t^2) \nonumber\\[4pt]
&= \frac12\left[1 - \left(\gamma + \frac{3}{\tau_{G}}\right)t 
+ \mathcal{O}(t^2)\right].
\label{eq:PopExpand}
\end{align}
A single exponential fitted to this decay obeys
\(
e^{-t/T_1} = 1 - t/T_1 + \mathcal{O}(t^2).
\)
Matching the linear terms in $t$ between Eqs.~\eqref{eq:PopExpand} 
and $e^{-t/T_1}$ yields
\begin{equation}
\boxed{
T_1^{-1} = \gamma + \frac{3}{\tau_{G}}.
}
\label{eq:T1extract}
\end{equation}

\paragraph*{Coherence decay.}
Similarly, Eq.~(C11) gives for the magnitude of the coherence
\begin{equation}
\left|\langle 0|\hat\rho_d(t)|1\rangle\right|
= \frac12 \!\left(1 - 2\frac{t}{\tau_{G}}\right)e^{-\gamma t/2}
+ \mathcal{O}(t^2).
\label{eq:CoherenceExact}
\end{equation}
Expanding $e^{-\gamma t/2} = 1 - \frac{\gamma}{2}t + 
\mathcal{O}(t^2)$ and multiplying out,
\begin{align}
\big|\langle 0|\hat\rho_d(t)|1\rangle\big|
&= \frac12\left[(1 - 2t/\tau_{G})
\left(1 - \frac{\gamma}{2}t\right)\right]
+ \mathcal{O}(t^2) \nonumber\\[4pt]
&= \frac12\left[
1 - \left(\frac{\gamma}{2} + \frac{2}{\tau_{G}}\right)t
+ \mathcal{O}(t^2)
\right].
\label{eq:CoherenceExpand}
\end{align}
The Ramsey-envelope fit has the form 
$e^{-t/T_2} = 1 - t/T_2 + \mathcal{O}(t^2)$.
Matching linear coefficients gives
\begin{equation}
\boxed{
T_2^{-1} = \frac{\gamma}{2} + \frac{2}{\tau_{G}}.
}
\label{eq:T2extract}
\end{equation}

Eqs.~\eqref{eq:T1extract} and~\eqref{eq:T2extract} show that the 
experimentally accessible decay times $(T_1,T_2)$ encode two 
independent linear combinations of the intrinsic damping rate $\gamma$ 
and the gravitationally induced decoherence rate $1/\tau$.  
Consequently, simultaneous measurement of $(T_1,T_2)$ allows one to 
solve for $(\gamma,\tau)$ uniquely.  
In particular, in the quantum--vacuum case the coefficients $(3,2)$ imply 
that the $\{|0\rangle,|1\rangle\}$ subspace is partially protected 
against gravitational decoherence, whereas in the classical stochastic 
case the same subspace decoheres at a rate proportional to $1/\tau_\alpha$.  
This provides a clear experimental discriminant between quantum and 
classical gravitational-wave backgrounds.

\section{Generalized coherent states}
\label{AppBB}

In the main text, in order to model the classical limit of
gravitationally induced decoherence, we chose to work with an ensemble
of phase--randomized coherent states, as defined in
Eq.~\eqref{Gch}. This choice was made primarily for calculational
simplicity and to allow a transparent one-to-one correspondence with
the fully quantum treatment. Importantly, however, the physical
conclusions of our analysis do not rely on the phase randomization
itself.

Indeed, one may equally well consider a generalized multimode coherent
state of the gravitational field, in which the graviton density matrix
is taken to be a pure product state,
\begin{equation}
\rho_g \;\rightarrow\; \rho^{\rm ch}_{g}
=
\bigotimes_{k} |\alpha_k\rangle\langle \alpha_k|,
\qquad
\alpha_k = |\alpha_k| e^{i\phi_k},
\label{pc}
\end{equation}
with fixed (but otherwise arbitrary) phases $\phi_k$.

Adopting this description modifies the graviton bilinear expectation
values entering the master equation. In particular, the relations in
Eqs.~\eqref{bilina} are replaced by
\begin{align}\label{bilina appen}
\langle \hat{b}_l^{\dagger}\hat{b}_m^{\dagger} \rangle_g^{\alpha}
&=\alpha^{*}_{l}\alpha^{*}_{m}, \qquad
\langle \hat{b}_l^{\dagger}\hat{b}_m \rangle_g^{\alpha}
=\alpha^{*}_{l}\alpha_{m},\nonumber\\
\langle \hat{b}_l \hat{b}_m^{\dagger} \rangle_g^{\alpha}
&=\delta_{lm}+\alpha_{l}\alpha^{*}_{m}, \qquad
\langle \hat{b}_l \hat{b}_m \rangle_g^{\alpha}
=\alpha_{l}\alpha_{m},\nonumber\\
\langle \hat{b}_l^{\dagger}\hat{b}_m^{\dagger} \rangle_g^{*\alpha}
&=\alpha^{*}_{l}\alpha^{*}_{m}, \qquad
\langle \hat{b}_l^{\dagger}\hat{b}_m \rangle_g^{*\alpha}
=\delta_{lm}+\alpha^{*}_{l}\alpha_{m},\nonumber\\
\langle \hat{b}_l \hat{b}_m^{\dagger} \rangle_g^{*\alpha}
&=\alpha_{l}\alpha^{*}_{m}, \qquad
\langle \hat{b}_l \hat{b}_m \rangle_g^{*\alpha}
=\alpha_{l}\alpha_{m}.
\end{align}

Using these relations instead of Eqs.~\eqref{bilina} leads to a master
equation that is algebraically more involved than
Eq.~\eqref{coherME}. Nevertheless, its overall structure remains
unchanged: the reduced dynamics of the detector (Eqs.~\eqref{FE}) can still be written as
the sum of a vacuum contribution and a coherent-state-dependent
contribution,
\begin{equation}
\dot{\rho}_d = \dot{\rho}_{d}^{\rm vac} + \dot{\rho}_{d}^{(\alpha)} .
\end{equation}

The vacuum contribution originates from the $\delta_{km}$ term in
$\langle \hat{b}_k \hat{b}_m^\dagger \rangle_g$ and gives rise to the
genuine Lindblad dissipator

\begin{align}
\dot{\rho}_{d}^{\rm vac}
&=
\Gamma
\left(
a^2\rho_{d} a^{\dagger 2}
-\tfrac12\{a^{\dagger 2}a^2,\rho_{d}\}
\right)\nonumber\\
& - \hat{a}_{1}^{2} \hat{a}_{1}^{2} \hat{\rho}_{d}(t) e^{-i4\omega_{c}t}
- \hat{\rho}_{d}(t) \hat{a}_{1}^{\dagger 2} \hat{a}_{1}^{\dagger 2} e^{i4\omega_{c}t}
+ \hat{a}_{1}^{2} \hat{\rho}_{d}(t) \hat{a}_{1}^{2}e^{-i4\omega_{c}t}\nonumber\\
&+ \hat{a}_{1}^{\dagger 2} \hat{\rho}_{d}(t) e^{i4\omega_{c}t}\hat{a}_{1}^{\dagger 2}
,
\end{align}
with $\Gamma = 2\pi f(\omega_c) C_{\omega_c}^2$.


The $\alpha$-dependent contribution takes the form
\begin{equation}
\begin{aligned}
\dot{\rho}_{d}^{(\alpha)}(t)
&=
|\alpha_{\omega_c}|^2
\sum_k
\Big[
e^{+i\Delta_k t}
\big(a^{\dagger 2}\rho_{d} a^2 - a^2 a^{\dagger 2}\rho_{d}\big)
\\
&\qquad\qquad
+
e^{-i\Delta_k t}
\big(a^2\rho_{d} a^{\dagger 2} - \rho_{d} a^{\dagger 2}a^2\big)
\Big].
\end{aligned}
\end{equation}

with
\begin{equation}
\Delta_k = 2(\omega_c-\omega_k).
\end{equation}

Crucially, all $\alpha$-dependent terms are purely oscillatory in time.
Upon performing the long-time (secular) average,
\begin{equation}
\frac{1}{T}\int_{t}^{t+T} dt'\, e^{\pm i \Delta t'}
\;\xrightarrow[T\to\infty]{}\; 0 ,
\end{equation}
these contributions vanish identically. Consequently, a purely coherent
gravitational-wave background with fixed phase does not induce
decoherence in the detector, and the reduced evolution remains unitary.

Decoherence therefore arises solely from vacuum graviton fluctuations or
from genuinely stochastic gravitational-wave backgrounds, such as those
described by phase–randomized coherent states.

\section{Spontaneous Graviton Emission and Consistency of the Born--Markov Approximation}

In deriving the master equation we have employed the Born--Markov approximation,
\begin{equation}
\rho_{\mathrm{tot}}(t)
\approx
\rho_d(t)\otimes\rho_g,
\end{equation}
where the gravitational bath is assumed to remain in its initial state $\rho_g$, taken to be the vacuum. Since the detector--gravity interaction permits spontaneous graviton emission, it is necessary to verify that this factorization remains self-consistent.

Consider the transition
\begin{equation}
|i\rangle = |2\rangle \otimes |0_g\rangle
\longrightarrow
|f\rangle = |0\rangle \otimes |1_k\rangle,
\end{equation}
in which the detector loses energy $2\hbar\omega_c$ while emitting a graviton of energy $\hbar\omega_k$. Energy conservation requires
\begin{equation}
2\hbar\omega_c = \hbar\omega_k,
\qquad
\omega_k = 2\omega_c.
\end{equation}

For the interaction Hamiltonian
\begin{equation}
\begin{aligned}
H_{\mathrm{int}}
&=
\hbar \sum_k \mathcal{C}_{\omega_k^{g}}
\Big[
e^{i \omega_k^{g} t}\, b_k^\dagger
\left(\hat{a}_1^{\dagger 2} - \hat{a}_1^{2}\right)
\\
&\quad
-
e^{-i \omega_k^{g} t}\, b_k
\left(\hat{a}_1^{\dagger 2} - \hat{a}_1^{2}\right)
\Big].
\end{aligned}
\end{equation}
the relevant matrix element is
\begin{equation}
\langle 0,1_k|H_{\mathrm{int}}|2,0\rangle.
\end{equation}
Using
\begin{equation}
a^2 |2\rangle = \sqrt{2}\,|0\rangle,
\qquad
b_k^\dagger |0\rangle = |1_k\rangle,
\end{equation}
one obtains
\begin{equation}
|\langle f|H_{\mathrm{int}}|i\rangle|^2
=
2 \hbar^{2}\mathcal{C}^2_{\omega^g_k}.
\end{equation}

Replacing the discrete mode sum by the continuum prescription used in the main text,
\begin{equation}
\sum_k \rightarrow \frac{V}{2\pi^2}\int_0^\infty d\omega_g\, f(\omega_g),
\end{equation}
the spontaneous graviton emission rate in the long-time limit can be derived using Fermi’s Golden Rule. 
Here the ``long-time limit'' refers to times large compared to the correlation 
time of the gravitational reservoir, so that the oscillatory time integrals 
generate an energy-conserving delta function. Importantly, this limit does 
\emph{not} require times comparable to the system lifetime, but only 
$t \gg \tau_{\mathrm{bath}}$, where $\tau_{\mathrm{bath}}$ denotes the 
decay time of the bath two-point correlation function.

Under this condition one obtains
\begin{equation}
\gamma_{\mathrm{em}}
=
\frac{2\pi}{\hbar}
\left|
\langle f|H_{\mathrm{int}}|i\rangle
\right|^2
f(\omega_k),
\end{equation}
which yields
\begin{equation}
\gamma_{\mathrm{em}}
= \frac{2\hbar}{\pi}\,
C_{\omega_k^{g}}^{2}\,
f(2\omega_c).
\end{equation}

For gravitational interactions one has
$\tau_{\mathrm{bath}} \ll \tau_G$
by many orders of magnitude, so there exists a broad 
intermediate regime 
$\tau_{\mathrm{bath}} \ll t \ll \tau_G$
in which both the Golden-Rule treatment and the perturbative 
Born--Markov approximation are simultaneously valid.

In our model, the coupling coefficient satisfies
\begin{equation}
C_\omega^2
=
\frac{\omega \ell_p^2}{2}
=
\frac{\hbar G\omega}{2c^3}.
\end{equation}
For free gravitons in $3+1$ dimensions, the density of modes obeys $f(\omega)\propto \omega^2$. Consequently,
\begin{equation}
\gamma_{\mathrm{em}}
\sim
\omega_c
\left(
\frac{\omega_c}{\omega_{\mathrm{Pl}}}
\right)^2,
\end{equation}
where $\omega_{\mathrm{Pl}}=\sqrt{c^{5}/(\hbar G)}$
denotes the Planck angular frequency.
For any laboratory mechanical frequency
$\omega_c \ll \omega_{\mathrm{Pl}}$,
the factor $(\omega_c/\omega_{\mathrm{Pl}})^2$
renders the emission rate negligibly small.

Initially the gravitational field is in the vacuum state,
\begin{equation}
\langle b_k^\dagger b_k \rangle_{t=0}=0.
\end{equation}
In the weak-coupling regime the emission probability grows linearly,
\begin{equation}
P_{\mathrm{emit}}(t)=\gamma_{\mathrm{em}} t.
\end{equation}
Since $\gamma_{\mathrm{em}} t \ll 1$ for all experimentally relevant times, the probability distribution satisfies
\begin{equation}
P(0)\simeq 1-\gamma_{\mathrm{em}} t,
\qquad
P(1)\simeq \gamma_{\mathrm{em}} t,
\qquad
P(n\ge2)\approx 0.
\end{equation}
The average graviton occupation therefore obeys
\begin{equation}
\langle b_k^\dagger b_k \rangle
=
\gamma_{\mathrm{em}} t
\ll 1.
\end{equation}

Thus, although spontaneous graviton emission is permitted by the interaction Hamiltonian, the detector-induced modification of the gravitational bath remains perturbatively small. The Born approximation does not require the absence of emission, but only that the bath state remain effectively unchanged to leading order. Since $\gamma_{\mathrm{em}} t \ll 1$, system--bath correlations are of higher order in the coupling, and the Born--Markov treatment employed in the main text is self-consistent.



\bibliographystyle{apsrev4-1}
\bibliography{gw_phases}

\begin{thebibliography}{72}%
\makeatletter
\providecommand \@ifxundefined [1]{%
 \@ifx{#1\undefined}
}%
\providecommand \@ifnum [1]{%
 \ifnum #1\expandafter \@firstoftwo
 \else \expandafter \@secondoftwo
 \fi
}%
\providecommand \@ifx [1]{%
 \ifx #1\expandafter \@firstoftwo
 \else \expandafter \@secondoftwo
 \fi
}%
\providecommand \natexlab [1]{#1}%
\providecommand \enquote  [1]{``#1''}%
\providecommand \bibnamefont  [1]{#1}%
\providecommand \bibfnamefont [1]{#1}%
\providecommand \citenamefont [1]{#1}%
\providecommand \href@noop [0]{\@secondoftwo}%
\providecommand \href [0]{\begingroup \@sanitize@url \@href}%
\providecommand \@href[1]{\@@startlink{#1}\@@href}%
\providecommand \@@href[1]{\endgroup#1\@@endlink}%
\providecommand \@sanitize@url [0]{\catcode `\\12\catcode `\$12\catcode `\&12\catcode `\#12\catcode `\^12\catcode `\_12\catcode `\%12\relax}%
\providecommand \@@startlink[1]{}%
\providecommand \@@endlink[0]{}%
\providecommand \url  [0]{\begingroup\@sanitize@url \@url }%
\providecommand \@url [1]{\endgroup\@href {#1}{\urlprefix }}%
\providecommand \urlprefix  [0]{URL }%
\providecommand \Eprint [0]{\href }%
\providecommand \doibase [0]{http://dx.doi.org/}%
\providecommand \selectlanguage [0]{\@gobble}%
\providecommand \bibinfo  [0]{\@secondoftwo}%
\providecommand \bibfield  [0]{\@secondoftwo}%
\providecommand \translation [1]{[#1]}%
\providecommand \BibitemOpen [0]{}%
\providecommand \bibitemStop [0]{}%
\providecommand \bibitemNoStop [0]{.\EOS\space}%
\providecommand \EOS [0]{\spacefactor3000\relax}%
\providecommand \BibitemShut  [1]{\csname bibitem#1\endcsname}%
\let\auto@bib@innerbib\@empty
\bibitem [{\citenamefont {Abbott}\ \emph {et~al.}(2016)\citenamefont {Abbott}, \citenamefont {\textit{et. al.} (LIGO Scientific~Collaboration},\ and\ \citenamefont {Collaboration)}}]{abbott}%
  \BibitemOpen
  \bibfield  {author} {\bibinfo {author} {\bibfnamefont {B.~P.}\ \bibnamefont {Abbott}}, \bibinfo {author} {\bibnamefont {\textit{et. al.} (LIGO Scientific~Collaboration}}, \ and\ \bibinfo {author} {\bibfnamefont {V.}~\bibnamefont {Collaboration)}},\ }\href {\doibase 10.1103/PhysRevLett.116.061102} {\bibfield  {journal} {\bibinfo  {journal} {Phys. Rev. Lett.}\ }\textbf {\bibinfo {volume} {116}},\ \bibinfo {pages} {061102} (\bibinfo {year} {2016})}\BibitemShut {NoStop}%
\bibitem [{\citenamefont {Abbott}\ \emph {et~al.}(2023)\citenamefont {Abbott} \emph {et~al.}}]{KAGRA:2021vkt}%
  \BibitemOpen
  \bibfield  {author} {\bibinfo {author} {\bibfnamefont {R.}~\bibnamefont {Abbott}} \emph {et~al.} (\bibinfo {collaboration} {KAGRA, VIRGO, LIGO Scientific}),\ }\href {\doibase 10.1103/PhysRevX.13.041039} {\bibfield  {journal} {\bibinfo  {journal} {Phys. Rev. X}\ }\textbf {\bibinfo {volume} {13}},\ \bibinfo {pages} {041039} (\bibinfo {year} {2023})},\ \Eprint {http://arxiv.org/abs/2111.03606} {arXiv:2111.03606 [gr-qc]} \BibitemShut {NoStop}%
\bibitem [{\citenamefont {Foo}\ \emph {et~al.}(2025)\citenamefont {Foo}, \citenamefont {Mann},\ and\ \citenamefont {Zych}}]{Foo:2025}%
  \BibitemOpen
  \bibfield  {author} {\bibinfo {author} {\bibfnamefont {J.}~\bibnamefont {Foo}}, \bibinfo {author} {\bibfnamefont {R.~B.}\ \bibnamefont {Mann}}, \ and\ \bibinfo {author} {\bibfnamefont {M.}~\bibnamefont {Zych}},\ }\href@noop {} {\bibfield  {journal} {\bibinfo  {journal} {arXiv preprint}\ } (\bibinfo {year} {2025})},\ \Eprint {http://arxiv.org/abs/2302.03259} {arXiv:2302.03259 [gr-qc]} \BibitemShut {NoStop}%
\bibitem [{\citenamefont {Manikandan}\ and\ \citenamefont {Wilczek}(2025{\natexlab{a}})}]{Manikandan:2025ykr}%
  \BibitemOpen
  \bibfield  {author} {\bibinfo {author} {\bibfnamefont {S.~K.}\ \bibnamefont {Manikandan}}\ and\ \bibinfo {author} {\bibfnamefont {F.}~\bibnamefont {Wilczek}},\ }\href {\doibase 10.1103/PhysRevA.111.033705} {\bibfield  {journal} {\bibinfo  {journal} {Phys. Rev. A}\ }\textbf {\bibinfo {volume} {111}},\ \bibinfo {pages} {033705} (\bibinfo {year} {2025}{\natexlab{a}})}\BibitemShut {NoStop}%
\bibitem [{\citenamefont {Manikandan}\ and\ \citenamefont {Wilczek}(2025{\natexlab{b}})}]{Manikandan:2025hlz}%
  \BibitemOpen
  \bibfield  {author} {\bibinfo {author} {\bibfnamefont {S.~K.}\ \bibnamefont {Manikandan}}\ and\ \bibinfo {author} {\bibfnamefont {F.}~\bibnamefont {Wilczek}},\ }\href {\doibase 10.1142/S0218271825430011} {\bibfield  {journal} {\bibinfo  {journal} {Int. J. Mod. Phys. D}\ }\textbf {\bibinfo {volume} {34}},\ \bibinfo {pages} {2543001} (\bibinfo {year} {2025}{\natexlab{b}})},\ \Eprint {http://arxiv.org/abs/2505.11407} {arXiv:2505.11407 [gr-qc]} \BibitemShut {NoStop}%
\bibitem [{\citenamefont {Manikandan}\ and\ \citenamefont {Wilczek}(2025{\natexlab{c}})}]{Manikandan:2025qgv}%
  \BibitemOpen
  \bibfield  {author} {\bibinfo {author} {\bibfnamefont {S.~K.}\ \bibnamefont {Manikandan}}\ and\ \bibinfo {author} {\bibfnamefont {F.}~\bibnamefont {Wilczek}},\ }\href {\doibase 10.1103/83tt-tt57} {\bibfield  {journal} {\bibinfo  {journal} {Phys. Rev. A}\ }\textbf {\bibinfo {volume} {112}},\ \bibinfo {pages} {043716} (\bibinfo {year} {2025}{\natexlab{c}})},\ \Eprint {http://arxiv.org/abs/2505.11422} {arXiv:2505.11422 [gr-qc]} \BibitemShut {NoStop}%
\bibitem [{\citenamefont {Marti}\ \emph {et~al.}(2024{\natexlab{a}})\citenamefont {Marti}, \citenamefont {von L{\"u}pke}, \citenamefont {Joshi}, \citenamefont {Yang}, \citenamefont {Bild}, \citenamefont {Omahen}, \citenamefont {Chu},\ and\ \citenamefont {Fadel}}]{Marti:2023abu}%
  \BibitemOpen
  \bibfield  {author} {\bibinfo {author} {\bibfnamefont {S.}~\bibnamefont {Marti}}, \bibinfo {author} {\bibfnamefont {U.}~\bibnamefont {von L{\"u}pke}}, \bibinfo {author} {\bibfnamefont {O.}~\bibnamefont {Joshi}}, \bibinfo {author} {\bibfnamefont {Y.}~\bibnamefont {Yang}}, \bibinfo {author} {\bibfnamefont {M.}~\bibnamefont {Bild}}, \bibinfo {author} {\bibfnamefont {A.}~\bibnamefont {Omahen}}, \bibinfo {author} {\bibfnamefont {Y.}~\bibnamefont {Chu}}, \ and\ \bibinfo {author} {\bibfnamefont {M.}~\bibnamefont {Fadel}},\ }\href {\doibase 10.1038/s41567-024-02545-6} {\bibfield  {journal} {\bibinfo  {journal} {Nature Phys.}\ }\textbf {\bibinfo {volume} {20}},\ \bibinfo {pages} {1448} (\bibinfo {year} {2024}{\natexlab{a}})},\ \Eprint {http://arxiv.org/abs/2312.16169} {arXiv:2312.16169 [quant-ph]} \BibitemShut {NoStop}%
\bibitem [{\citenamefont {Breuer}\ and\ \citenamefont {Petruccione}(2006)}]{BreuerPetruccione2006}%
  \BibitemOpen
  \bibfield  {author} {\bibinfo {author} {\bibfnamefont {H.-P.}\ \bibnamefont {Breuer}}\ and\ \bibinfo {author} {\bibfnamefont {F.}~\bibnamefont {Petruccione}},\ }\href {\doibase 10.1093/acprof:oso/9780199213900.001.0001} {\emph {\bibinfo {title} {The Theory of Open Quantum Systems}}}\ (\bibinfo  {publisher} {Oxford University Press},\ \bibinfo {address} {New York},\ \bibinfo {year} {2006})\BibitemShut {NoStop}%
\bibitem [{\citenamefont {Ghosh}\ \emph {et~al.}(1977)\citenamefont {Ghosh}, \citenamefont {Dutta-Roy},\ and\ \citenamefont {Dey}}]{Ghosh:1977SchrodingerFluid}%
  \BibitemOpen
  \bibfield  {author} {\bibinfo {author} {\bibfnamefont {G.}~\bibnamefont {Ghosh}}, \bibinfo {author} {\bibfnamefont {B.}~\bibnamefont {Dutta-Roy}}, \ and\ \bibinfo {author} {\bibfnamefont {M.}~\bibnamefont {Dey}},\ }\href {\doibase 10.1088/0305-4616/3/8/006} {\bibfield  {journal} {\bibinfo  {journal} {Journal of Physics G: Nuclear Physics}\ }\textbf {\bibinfo {volume} {3}},\ \bibinfo {pages} {1077} (\bibinfo {year} {1977})}\BibitemShut {NoStop}%
\bibitem [{\citenamefont {Nandi}\ \emph {et~al.}(2025{\natexlab{a}})\citenamefont {Nandi}, \citenamefont {Bhattacharyya}, \citenamefont {Majumdar}, \citenamefont {Pleasance},\ and\ \citenamefont {Petruccione}}]{Nandi:2025qyj}%
  \BibitemOpen
  \bibfield  {author} {\bibinfo {author} {\bibfnamefont {P.}~\bibnamefont {Nandi}}, \bibinfo {author} {\bibfnamefont {T.}~\bibnamefont {Bhattacharyya}}, \bibinfo {author} {\bibfnamefont {A.~S.}\ \bibnamefont {Majumdar}}, \bibinfo {author} {\bibfnamefont {G.}~\bibnamefont {Pleasance}}, \ and\ \bibinfo {author} {\bibfnamefont {F.}~\bibnamefont {Petruccione}},\ }\href@noop {} {\bibfield  {journal} {\bibinfo  {journal} {Phys. Rev. Research}\ } (\bibinfo {year} {2025}{\natexlab{a}})},\ \Eprint {http://arxiv.org/abs/2503.13061} {arXiv:2503.13061 [hep-th]} \BibitemShut {NoStop}%
\bibitem [{\citenamefont {Nandi}\ and\ \citenamefont {Petruccione}()}]{Nandi:OpenUniverse2026}%
  \BibitemOpen
  \bibfield  {author} {\bibinfo {author} {\bibfnamefont {P.}~\bibnamefont {Nandi}}\ and\ \bibinfo {author} {\bibfnamefont {F.}~\bibnamefont {Petruccione}},\ }\href@noop {} {\enquote {\bibinfo {title} {Quantum openness of the universe: Emergent dark matter and dark energy},}\ }\bibinfo {note} {In preparation}\BibitemShut {NoStop}%
\bibitem [{\citenamefont {Nandi}\ \emph {et~al.}(2025{\natexlab{b}})\citenamefont {Nandi}, \citenamefont {Ghose},\ and\ \citenamefont {Petruccione}}]{Nandi:2025hqe}%
  \BibitemOpen
  \bibfield  {author} {\bibinfo {author} {\bibfnamefont {P.}~\bibnamefont {Nandi}}, \bibinfo {author} {\bibfnamefont {P.}~\bibnamefont {Ghose}}, \ and\ \bibinfo {author} {\bibfnamefont {F.}~\bibnamefont {Petruccione}},\ }\href@noop {} {\  (\bibinfo {year} {2025}{\natexlab{b}})},\ \Eprint {http://arxiv.org/abs/2510.10836} {arXiv:2510.10836 [gr-qc]} \BibitemShut {NoStop}%
\bibitem [{\citenamefont {Zhang}(2025)}]{Zhang:2025ozb}%
  \BibitemOpen
  \bibfield  {author} {\bibinfo {author} {\bibfnamefont {C.}~\bibnamefont {Zhang}},\ }\href {\doibase 10.1103/1dfk-qqgj} {\bibfield  {journal} {\bibinfo  {journal} {Phys. Rev. D}\ }\textbf {\bibinfo {volume} {112}},\ \bibinfo {pages} {104022} (\bibinfo {year} {2025})},\ \Eprint {http://arxiv.org/abs/2511.21403} {arXiv:2511.21403 [gr-qc]} \BibitemShut {NoStop}%
\bibitem [{\citenamefont {Miki}\ \emph {et~al.}(2021)\citenamefont {Miki}, \citenamefont {Matsumura},\ and\ \citenamefont {Yamamoto}}]{PhysRevD.103.026017}%
  \BibitemOpen
  \bibfield  {author} {\bibinfo {author} {\bibfnamefont {D.}~\bibnamefont {Miki}}, \bibinfo {author} {\bibfnamefont {A.}~\bibnamefont {Matsumura}}, \ and\ \bibinfo {author} {\bibfnamefont {K.}~\bibnamefont {Yamamoto}},\ }\href {\doibase 10.1103/PhysRevD.103.026017} {\bibfield  {journal} {\bibinfo  {journal} {Phys. Rev. D}\ }\textbf {\bibinfo {volume} {103}},\ \bibinfo {pages} {026017} (\bibinfo {year} {2021})}\BibitemShut {NoStop}%
\bibitem [{\citenamefont {Guerreiro}(2025)}]{Guerreiro:2025sge}%
  \BibitemOpen
  \bibfield  {author} {\bibinfo {author} {\bibfnamefont {T.}~\bibnamefont {Guerreiro}},\ }\href {\doibase 10.1103/fn5d-mrsj} {\bibfield  {journal} {\bibinfo  {journal} {Phys. Rev. D}\ }\textbf {\bibinfo {volume} {112}},\ \bibinfo {pages} {L101904} (\bibinfo {year} {2025})},\ \Eprint {http://arxiv.org/abs/2501.17043} {arXiv:2501.17043 [gr-qc]} \BibitemShut {NoStop}%
\bibitem [{\citenamefont {Kaku}\ and\ \citenamefont {Nambu}(2025)}]{Kaku:2024lgs}%
  \BibitemOpen
  \bibfield  {author} {\bibinfo {author} {\bibfnamefont {Y.}~\bibnamefont {Kaku}}\ and\ \bibinfo {author} {\bibfnamefont {Y.}~\bibnamefont {Nambu}},\ }\href {\doibase 10.1103/PhysRevD.111.046026} {\bibfield  {journal} {\bibinfo  {journal} {Phys. Rev. D}\ }\textbf {\bibinfo {volume} {111}},\ \bibinfo {pages} {046026} (\bibinfo {year} {2025})},\ \Eprint {http://arxiv.org/abs/2411.12997} {arXiv:2411.12997 [gr-qc]} \BibitemShut {NoStop}%
\bibitem [{\citenamefont {Jones}\ \emph {et~al.}(2025)\citenamefont {Jones}, \citenamefont {Bailey}, \citenamefont {Gretarsson},\ and\ \citenamefont {Poon}}]{Jones:2024npd}%
  \BibitemOpen
  \bibfield  {author} {\bibinfo {author} {\bibfnamefont {P.}~\bibnamefont {Jones}}, \bibinfo {author} {\bibfnamefont {Q.~G.}\ \bibnamefont {Bailey}}, \bibinfo {author} {\bibfnamefont {A.}~\bibnamefont {Gretarsson}}, \ and\ \bibinfo {author} {\bibfnamefont {E.}~\bibnamefont {Poon}},\ }\href {\doibase 10.1016/j.physletb.2025.139628} {\bibfield  {journal} {\bibinfo  {journal} {Phys. Lett. B}\ }\textbf {\bibinfo {volume} {868}},\ \bibinfo {pages} {139628} (\bibinfo {year} {2025})},\ \Eprint {http://arxiv.org/abs/2411.15632} {arXiv:2411.15632 [gr-qc]} \BibitemShut {NoStop}%
\bibitem [{\citenamefont {Lin}\ and\ \citenamefont {Mondal}(2025)}]{Lin:2025ipk}%
  \BibitemOpen
  \bibfield  {author} {\bibinfo {author} {\bibfnamefont {F.-L.}\ \bibnamefont {Lin}}\ and\ \bibinfo {author} {\bibfnamefont {S.}~\bibnamefont {Mondal}},\ }\href@noop {} {\  (\bibinfo {year} {2025})},\ \Eprint {http://arxiv.org/abs/2510.23584} {arXiv:2510.23584 [hep-th]} \BibitemShut {NoStop}%
\bibitem [{\citenamefont {Singh}\ and\ \citenamefont {Padmanabhan}(1987)}]{SinghPadmanabhan1987}%
  \BibitemOpen
  \bibfield  {author} {\bibinfo {author} {\bibfnamefont {T.~P.}\ \bibnamefont {Singh}}\ and\ \bibinfo {author} {\bibfnamefont {T.}~\bibnamefont {Padmanabhan}},\ }\href {\doibase 10.1103/PhysRevD.35.2993} {\bibfield  {journal} {\bibinfo  {journal} {Physical Review D}\ }\textbf {\bibinfo {volume} {35}},\ \bibinfo {pages} {2993} (\bibinfo {year} {1987})}\BibitemShut {NoStop}%
\bibitem [{\citenamefont {Aston}(2012)}]{LIGOSuspension2012}%
  \BibitemOpen
  \bibfield  {author} {\bibinfo {author} {\bibfnamefont {S.~M. e.~a.}\ \bibnamefont {Aston}},\ }\href {https://dcc.ligo.org/LIGO-P1200056/public} {\emph {\bibinfo {title} {Update on Quadruple Suspension Design for Advanced LIGO}}},\ \bibinfo {type} {Tech. Rep.}\ \bibinfo {number} {LIGO-P1200056-v3}\ (\bibinfo  {institution} {LIGO Scientific Collaboration},\ \bibinfo {year} {2012})\BibitemShut {NoStop}%
\bibitem [{\citenamefont {Sigg}(1998)}]{Sigg1998GravitationalWaves}%
  \BibitemOpen
  \bibfield  {author} {\bibinfo {author} {\bibfnamefont {D.}~\bibnamefont {Sigg}},\ }\href {https://dcc.ligo.org/LIGO-P980007/public} {\emph {\bibinfo {title} {Gravitational Waves (A Review of LIGO)}}},\ \bibinfo {type} {Tech. Rep.}\ \bibinfo {number} {LIGO-P980007-00-D}\ (\bibinfo  {institution} {LIGO},\ \bibinfo {year} {1998})\BibitemShut {NoStop}%
\bibitem [{\citenamefont {Nandi}\ and\ \citenamefont {Majhi}(2024)}]{Nandi:2024jyf}%
  \BibitemOpen
  \bibfield  {author} {\bibinfo {author} {\bibfnamefont {P.}~\bibnamefont {Nandi}}\ and\ \bibinfo {author} {\bibfnamefont {B.~R.}\ \bibnamefont {Majhi}},\ }\href {\doibase 10.1016/j.physletb.2024.138988} {\bibfield  {journal} {\bibinfo  {journal} {Phys. Lett. B}\ }\textbf {\bibinfo {volume} {857}},\ \bibinfo {pages} {138988} (\bibinfo {year} {2024})},\ \Eprint {http://arxiv.org/abs/2403.11253} {arXiv:2403.11253 [gr-qc]} \BibitemShut {NoStop}%
\bibitem [{\citenamefont {Nandi}\ \emph {et~al.}(2024)\citenamefont {Nandi}, \citenamefont {Majhi}, \citenamefont {Debnath},\ and\ \citenamefont {Kala}}]{Nandi:2024zxp}%
  \BibitemOpen
  \bibfield  {author} {\bibinfo {author} {\bibfnamefont {P.}~\bibnamefont {Nandi}}, \bibinfo {author} {\bibfnamefont {B.~R.}\ \bibnamefont {Majhi}}, \bibinfo {author} {\bibfnamefont {N.}~\bibnamefont {Debnath}}, \ and\ \bibinfo {author} {\bibfnamefont {S.}~\bibnamefont {Kala}},\ }\href {\doibase 10.1016/j.physletb.2024.138706} {\bibfield  {journal} {\bibinfo  {journal} {Phys. Lett. B}\ }\textbf {\bibinfo {volume} {853}},\ \bibinfo {pages} {138706} (\bibinfo {year} {2024})},\ \Eprint {http://arxiv.org/abs/2401.02778} {arXiv:2401.02778 [gr-qc]} \BibitemShut {NoStop}%
\bibitem [{\citenamefont {Arvanitaki}\ and\ \citenamefont {Geraci}(2013)}]{Arvanitaki2013}%
  \BibitemOpen
  \bibfield  {author} {\bibinfo {author} {\bibfnamefont {A.}~\bibnamefont {Arvanitaki}}\ and\ \bibinfo {author} {\bibfnamefont {A.~A.}\ \bibnamefont {Geraci}},\ }\href {\doibase 10.1103/PhysRevLett.110.071105} {\bibfield  {journal} {\bibinfo  {journal} {Phys. Rev. Lett.}\ }\textbf {\bibinfo {volume} {110}},\ \bibinfo {pages} {071105} (\bibinfo {year} {2013})}\BibitemShut {NoStop}%
\bibitem [{\citenamefont {O’Connell}\ \emph {et~al.}(2010)\citenamefont {O’Connell}, \citenamefont {Hofheinz}, \citenamefont {Ansmann}, \citenamefont {Bialczak}, \citenamefont {Lenander}, \citenamefont {Lucero}, \citenamefont {Neeley}, \citenamefont {Sank}, \citenamefont {Wang}, \citenamefont {Weides}, \citenamefont {Wenner}, \citenamefont {Martinis},\ and\ \citenamefont {Cleland}}]{OConnell2010}%
  \BibitemOpen
  \bibfield  {author} {\bibinfo {author} {\bibfnamefont {A.~D.}\ \bibnamefont {O’Connell}}, \bibinfo {author} {\bibfnamefont {M.}~\bibnamefont {Hofheinz}}, \bibinfo {author} {\bibfnamefont {M.}~\bibnamefont {Ansmann}}, \bibinfo {author} {\bibfnamefont {R.~C.}\ \bibnamefont {Bialczak}}, \bibinfo {author} {\bibfnamefont {M.}~\bibnamefont {Lenander}}, \bibinfo {author} {\bibfnamefont {E.}~\bibnamefont {Lucero}}, \bibinfo {author} {\bibfnamefont {M.}~\bibnamefont {Neeley}}, \bibinfo {author} {\bibfnamefont {D.}~\bibnamefont {Sank}}, \bibinfo {author} {\bibfnamefont {H.}~\bibnamefont {Wang}}, \bibinfo {author} {\bibfnamefont {M.}~\bibnamefont {Weides}}, \bibinfo {author} {\bibfnamefont {J.}~\bibnamefont {Wenner}}, \bibinfo {author} {\bibfnamefont {J.~M.}\ \bibnamefont {Martinis}}, \ and\ \bibinfo {author} {\bibfnamefont {A.~N.}\ \bibnamefont {Cleland}},\ }\href {\doibase 10.1038/nature08967} {\bibfield  {journal} {\bibinfo  {journal} {Nature}\ }\textbf {\bibinfo {volume} {464}},\ \bibinfo {pages} {697} (\bibinfo
  {year} {2010})}\BibitemShut {NoStop}%
\bibitem [{\citenamefont {Teufel}\ \emph {et~al.}(2011)\citenamefont {Teufel}, \citenamefont {Donner}, \citenamefont {Li}, \citenamefont {Harlow}, \citenamefont {Allman}, \citenamefont {Cicak}, \citenamefont {Sirois}, \citenamefont {Whittaker}, \citenamefont {Simmonds},\ and\ \citenamefont {Lehnert}}]{Teufel2011}%
  \BibitemOpen
  \bibfield  {author} {\bibinfo {author} {\bibfnamefont {J.~D.}\ \bibnamefont {Teufel}}, \bibinfo {author} {\bibfnamefont {T.}~\bibnamefont {Donner}}, \bibinfo {author} {\bibfnamefont {D.}~\bibnamefont {Li}}, \bibinfo {author} {\bibfnamefont {J.~W.}\ \bibnamefont {Harlow}}, \bibinfo {author} {\bibfnamefont {M.~S.}\ \bibnamefont {Allman}}, \bibinfo {author} {\bibfnamefont {K.}~\bibnamefont {Cicak}}, \bibinfo {author} {\bibfnamefont {A.~J.}\ \bibnamefont {Sirois}}, \bibinfo {author} {\bibfnamefont {J.~D.}\ \bibnamefont {Whittaker}}, \bibinfo {author} {\bibfnamefont {R.~W.}\ \bibnamefont {Simmonds}}, \ and\ \bibinfo {author} {\bibfnamefont {K.~W.}\ \bibnamefont {Lehnert}},\ }\href {\doibase 10.1038/nature10261} {\bibfield  {journal} {\bibinfo  {journal} {Nature}\ }\textbf {\bibinfo {volume} {475}},\ \bibinfo {pages} {359} (\bibinfo {year} {2011})}\BibitemShut {NoStop}%
\bibitem [{\citenamefont {Chan}\ \emph {et~al.}(2011)\citenamefont {Chan}, \citenamefont {Alegre}, \citenamefont {Safavi-Naeini}, \citenamefont {Hill}, \citenamefont {Krause}, \citenamefont {Gr\"oblacher}, \citenamefont {Aspelmeyer},\ and\ \citenamefont {Painter}}]{Chan2011}%
  \BibitemOpen
  \bibfield  {author} {\bibinfo {author} {\bibfnamefont {J.}~\bibnamefont {Chan}}, \bibinfo {author} {\bibfnamefont {T.~P.~M.}\ \bibnamefont {Alegre}}, \bibinfo {author} {\bibfnamefont {A.~H.}\ \bibnamefont {Safavi-Naeini}}, \bibinfo {author} {\bibfnamefont {J.~T.}\ \bibnamefont {Hill}}, \bibinfo {author} {\bibfnamefont {A.}~\bibnamefont {Krause}}, \bibinfo {author} {\bibfnamefont {S.}~\bibnamefont {Gr\"oblacher}}, \bibinfo {author} {\bibfnamefont {M.}~\bibnamefont {Aspelmeyer}}, \ and\ \bibinfo {author} {\bibfnamefont {O.}~\bibnamefont {Painter}},\ }\href {\doibase 10.1038/nature10461} {\bibfield  {journal} {\bibinfo  {journal} {Nature}\ }\textbf {\bibinfo {volume} {478}},\ \bibinfo {pages} {89} (\bibinfo {year} {2011})}\BibitemShut {NoStop}%
\bibitem [{\citenamefont {Miki}\ \emph {et~al.}(2024)\citenamefont {Miki}, \citenamefont {Matsumura},\ and\ \citenamefont {Yamamoto}}]{Miki:2023nce}%
  \BibitemOpen
  \bibfield  {author} {\bibinfo {author} {\bibfnamefont {D.}~\bibnamefont {Miki}}, \bibinfo {author} {\bibfnamefont {A.}~\bibnamefont {Matsumura}}, \ and\ \bibinfo {author} {\bibfnamefont {K.}~\bibnamefont {Yamamoto}},\ }\href {\doibase 10.1103/PhysRevD.109.064090} {\bibfield  {journal} {\bibinfo  {journal} {Phys. Rev. D}\ }\textbf {\bibinfo {volume} {109}},\ \bibinfo {pages} {064090} (\bibinfo {year} {2024})},\ \Eprint {http://arxiv.org/abs/2311.00563} {arXiv:2311.00563 [gr-qc]} \BibitemShut {NoStop}%
\bibitem [{\citenamefont {Wan}(2017)}]{Wan:2017thesis}%
  \BibitemOpen
  \bibfield  {author} {\bibinfo {author} {\bibfnamefont {C.}~\bibnamefont {Wan}},\ }\emph {\bibinfo {title} {Quantum Superposition on Nano-Mechanical Oscillator}},\ \href@noop {} {\bibinfo {type} {Ph.d. thesis}},\ \bibinfo  {school} {Imperial College London} (\bibinfo {year} {2017})\BibitemShut {NoStop}%
\bibitem [{\citenamefont {Chu}\ \emph {et~al.}(2017)\citenamefont {Chu}, \citenamefont {Kharel}, \citenamefont {Renninger}, \citenamefont {Burkhart}, \citenamefont {Frunzio}, \citenamefont {Rakich},\ and\ \citenamefont {Schoelkopf}}]{Chu:2017koi}%
  \BibitemOpen
  \bibfield  {author} {\bibinfo {author} {\bibfnamefont {Y.}~\bibnamefont {Chu}}, \bibinfo {author} {\bibfnamefont {P.}~\bibnamefont {Kharel}}, \bibinfo {author} {\bibfnamefont {W.~H.}\ \bibnamefont {Renninger}}, \bibinfo {author} {\bibfnamefont {L.~D.}\ \bibnamefont {Burkhart}}, \bibinfo {author} {\bibfnamefont {L.}~\bibnamefont {Frunzio}}, \bibinfo {author} {\bibfnamefont {P.~T.}\ \bibnamefont {Rakich}}, \ and\ \bibinfo {author} {\bibfnamefont {R.~J.}\ \bibnamefont {Schoelkopf}},\ }\href {\doibase 10.1126/science.aao1511} {\bibfield  {journal} {\bibinfo  {journal} {Science}\ }\textbf {\bibinfo {volume} {358}},\ \bibinfo {pages} {199} (\bibinfo {year} {2017})},\ \Eprint {http://arxiv.org/abs/1703.00342} {arXiv:1703.00342 [quant-ph]} \BibitemShut {NoStop}%
\bibitem [{\citenamefont {Donadi}\ and\ \citenamefont {Fadel}(2025)}]{PhysRevD.111.026009}%
  \BibitemOpen
  \bibfield  {author} {\bibinfo {author} {\bibfnamefont {S.}~\bibnamefont {Donadi}}\ and\ \bibinfo {author} {\bibfnamefont {M.}~\bibnamefont {Fadel}},\ }\href {\doibase 10.1103/PhysRevD.111.026009} {\bibfield  {journal} {\bibinfo  {journal} {Phys. Rev. D}\ }\textbf {\bibinfo {volume} {111}},\ \bibinfo {pages} {026009} (\bibinfo {year} {2025})}\BibitemShut {NoStop}%
\bibitem [{\citenamefont {Nandi}\ \emph {et~al.}(2023)\citenamefont {Nandi}, \citenamefont {Pal}, \citenamefont {Pal},\ and\ \citenamefont {Majhi}}]{Nandi:2022sjy}%
  \BibitemOpen
  \bibfield  {author} {\bibinfo {author} {\bibfnamefont {P.}~\bibnamefont {Nandi}}, \bibinfo {author} {\bibfnamefont {S.}~\bibnamefont {Pal}}, \bibinfo {author} {\bibfnamefont {S.~K.}\ \bibnamefont {Pal}}, \ and\ \bibinfo {author} {\bibfnamefont {B.~R.}\ \bibnamefont {Majhi}},\ }\href {\doibase 10.1103/PhysRevD.108.124069} {\bibfield  {journal} {\bibinfo  {journal} {Phys. Rev. D}\ }\textbf {\bibinfo {volume} {108}},\ \bibinfo {pages} {124069} (\bibinfo {year} {2023})},\ \Eprint {http://arxiv.org/abs/2207.08687} {arXiv:2207.08687 [gr-qc]} \BibitemShut {NoStop}%
\bibitem [{\citenamefont {Dutta}\ \emph {et~al.}(2025{\natexlab{a}})\citenamefont {Dutta}, \citenamefont {Nandi},\ and\ \citenamefont {Majhi}}]{Dutta:2025bge}%
  \BibitemOpen
  \bibfield  {author} {\bibinfo {author} {\bibfnamefont {M.}~\bibnamefont {Dutta}}, \bibinfo {author} {\bibfnamefont {P.}~\bibnamefont {Nandi}}, \ and\ \bibinfo {author} {\bibfnamefont {B.~R.}\ \bibnamefont {Majhi}},\ }\href {\doibase 10.1007/JHEP08(2025)104} {\bibfield  {journal} {\bibinfo  {journal} {JHEP}\ }\textbf {\bibinfo {volume} {08}},\ \bibinfo {pages} {104} (\bibinfo {year} {2025}{\natexlab{a}})},\ \Eprint {http://arxiv.org/abs/2503.19688} {arXiv:2503.19688 [gr-qc]} \BibitemShut {NoStop}%
\bibitem [{\citenamefont {Dutta}\ \emph {et~al.}(2025{\natexlab{b}})\citenamefont {Dutta}, \citenamefont {Nandi},\ and\ \citenamefont {Majhi}}]{Dutta:2025ouy}%
  \BibitemOpen
  \bibfield  {author} {\bibinfo {author} {\bibfnamefont {M.}~\bibnamefont {Dutta}}, \bibinfo {author} {\bibfnamefont {P.}~\bibnamefont {Nandi}}, \ and\ \bibinfo {author} {\bibfnamefont {B.~R.}\ \bibnamefont {Majhi}},\ }\href@noop {} {\  (\bibinfo {year} {2025}{\natexlab{b}})},\ \Eprint {http://arxiv.org/abs/2510.11075} {arXiv:2510.11075 [gr-qc]} \BibitemShut {NoStop}%
\bibitem [{\citenamefont {Cho}\ and\ \citenamefont {Hu}(2022)}]{Cho:2021gvg}%
  \BibitemOpen
  \bibfield  {author} {\bibinfo {author} {\bibfnamefont {H.-T.}\ \bibnamefont {Cho}}\ and\ \bibinfo {author} {\bibfnamefont {B.-L.}\ \bibnamefont {Hu}},\ }\href {\doibase 10.1103/PhysRevD.105.086004} {\bibfield  {journal} {\bibinfo  {journal} {Phys. Rev. D}\ }\textbf {\bibinfo {volume} {105}},\ \bibinfo {pages} {086004} (\bibinfo {year} {2022})},\ \Eprint {http://arxiv.org/abs/2112.08174} {arXiv:2112.08174 [gr-qc]} \BibitemShut {NoStop}%
\bibitem [{\citenamefont {Bhattacharya}\ \emph {et~al.}(2006)\citenamefont {Bhattacharya}, \citenamefont {Mohanty},\ and\ \citenamefont {Nautiyal}}]{Bhattacharya:2006prl}%
  \BibitemOpen
  \bibfield  {author} {\bibinfo {author} {\bibfnamefont {K.}~\bibnamefont {Bhattacharya}}, \bibinfo {author} {\bibfnamefont {S.}~\bibnamefont {Mohanty}}, \ and\ \bibinfo {author} {\bibfnamefont {A.}~\bibnamefont {Nautiyal}},\ }\href@noop {} {\bibfield  {journal} {\bibinfo  {journal} {Phys. Rev. Lett.}\ }\textbf {\bibinfo {volume} {97}},\ \bibinfo {pages} {251301} (\bibinfo {year} {2006})}\BibitemShut {NoStop}%
\bibitem [{\citenamefont {Blencowe}(2013)}]{PhysRevLett.111.021302}%
  \BibitemOpen
  \bibfield  {author} {\bibinfo {author} {\bibfnamefont {M.~P.}\ \bibnamefont {Blencowe}},\ }\href {\doibase 10.1103/PhysRevLett.111.021302} {\bibfield  {journal} {\bibinfo  {journal} {Phys. Rev. Lett.}\ }\textbf {\bibinfo {volume} {111}},\ \bibinfo {pages} {021302} (\bibinfo {year} {2013})}\BibitemShut {NoStop}%
\bibitem [{\citenamefont {Antoniou}\ \emph {et~al.}(2016)\citenamefont {Antoniou}, \citenamefont {Papadopoulos},\ and\ \citenamefont {Perivolaropoulos}}]{Antoniou2016PropagationGW}%
  \BibitemOpen
  \bibfield  {author} {\bibinfo {author} {\bibfnamefont {I.}~\bibnamefont {Antoniou}}, \bibinfo {author} {\bibfnamefont {D.}~\bibnamefont {Papadopoulos}}, \ and\ \bibinfo {author} {\bibfnamefont {L.}~\bibnamefont {Perivolaropoulos}},\ }\href {\doibase 10.1103/PhysRevD.94.084018} {\bibfield  {journal} {\bibinfo  {journal} {Physical Review D}\ }\textbf {\bibinfo {volume} {94}},\ \bibinfo {pages} {084018} (\bibinfo {year} {2016})}\BibitemShut {NoStop}%
\bibitem [{\citenamefont {Maggiore}(2008)}]{maggiore2008}%
  \BibitemOpen
  \bibfield  {author} {\bibinfo {author} {\bibfnamefont {M.}~\bibnamefont {Maggiore}},\ }\href@noop {} {\emph {\bibinfo {title} {Gravitational Waves: Theory and Experiment}}},\ Vol.~\bibinfo {volume} {1}\ (\bibinfo  {publisher} {Oxford University Press},\ \bibinfo {address} {New York},\ \bibinfo {year} {2008})\BibitemShut {NoStop}%
\bibitem [{\citenamefont {Misner}\ \emph {et~al.}(1973)\citenamefont {Misner}, \citenamefont {Thorne},\ and\ \citenamefont {Wheeler}}]{MTW:1973}%
  \BibitemOpen
  \bibfield  {author} {\bibinfo {author} {\bibfnamefont {C.~W.}\ \bibnamefont {Misner}}, \bibinfo {author} {\bibfnamefont {K.~S.}\ \bibnamefont {Thorne}}, \ and\ \bibinfo {author} {\bibfnamefont {J.~A.}\ \bibnamefont {Wheeler}},\ }\href@noop {} {\emph {\bibinfo {title} {Gravitation}}}\ (\bibinfo  {publisher} {W. H. Freeman and Company},\ \bibinfo {address} {San Francisco},\ \bibinfo {year} {1973})\BibitemShut {NoStop}%
\bibitem [{\citenamefont {Galley}\ \emph {et~al.}(2013)\citenamefont {Galley}, \citenamefont {Behunin},\ and\ \citenamefont {Hu}}]{PhysRevA.87.043832}%
  \BibitemOpen
  \bibfield  {author} {\bibinfo {author} {\bibfnamefont {C.~R.}\ \bibnamefont {Galley}}, \bibinfo {author} {\bibfnamefont {R.~O.}\ \bibnamefont {Behunin}}, \ and\ \bibinfo {author} {\bibfnamefont {B.~L.}\ \bibnamefont {Hu}},\ }\href {\doibase 10.1103/PhysRevA.87.043832} {\bibfield  {journal} {\bibinfo  {journal} {Phys. Rev. A}\ }\textbf {\bibinfo {volume} {87}},\ \bibinfo {pages} {043832} (\bibinfo {year} {2013})}\BibitemShut {NoStop}%
\bibitem [{\citenamefont {Parikh}\ \emph {et~al.}(2021{\natexlab{a}})\citenamefont {Parikh}, \citenamefont {Wilczek},\ and\ \citenamefont {Zahariade}}]{Parikh:2020fhy}%
  \BibitemOpen
  \bibfield  {author} {\bibinfo {author} {\bibfnamefont {M.}~\bibnamefont {Parikh}}, \bibinfo {author} {\bibfnamefont {F.}~\bibnamefont {Wilczek}}, \ and\ \bibinfo {author} {\bibfnamefont {G.}~\bibnamefont {Zahariade}},\ }\href {\doibase 10.1103/PhysRevD.104.046021} {\bibfield  {journal} {\bibinfo  {journal} {Phys. Rev. D}\ }\textbf {\bibinfo {volume} {104}},\ \bibinfo {pages} {046021} (\bibinfo {year} {2021}{\natexlab{a}})},\ \Eprint {http://arxiv.org/abs/2010.08208} {arXiv:2010.08208 [hep-th]} \BibitemShut {NoStop}%
\bibitem [{\citenamefont {Speliotopoulos}(1995)}]{Speliotopoulos:1995}%
  \BibitemOpen
  \bibfield  {author} {\bibinfo {author} {\bibfnamefont {A.~D.}\ \bibnamefont {Speliotopoulos}},\ }\href {\doibase 10.1103/PhysRevD.51.1701} {\bibfield  {journal} {\bibinfo  {journal} {Phys. Rev. D}\ }\textbf {\bibinfo {volume} {51}},\ \bibinfo {pages} {1701} (\bibinfo {year} {1995})}\BibitemShut {NoStop}%
\bibitem [{\citenamefont {Parikh}\ \emph {et~al.}(2021{\natexlab{b}})\citenamefont {Parikh}, \citenamefont {Wilczek},\ and\ \citenamefont {Zahariade}}]{PhysRevLett.127.081602}%
  \BibitemOpen
  \bibfield  {author} {\bibinfo {author} {\bibfnamefont {M.}~\bibnamefont {Parikh}}, \bibinfo {author} {\bibfnamefont {F.}~\bibnamefont {Wilczek}}, \ and\ \bibinfo {author} {\bibfnamefont {G.}~\bibnamefont {Zahariade}},\ }\href {\doibase 10.1103/PhysRevLett.127.081602} {\bibfield  {journal} {\bibinfo  {journal} {Phys. Rev. Lett.}\ }\textbf {\bibinfo {volume} {127}},\ \bibinfo {pages} {081602} (\bibinfo {year} {2021}{\natexlab{b}})}\BibitemShut {NoStop}%
\bibitem [{\citenamefont {Saha}\ \emph {et~al.}(2018)\citenamefont {Saha}, \citenamefont {Gangopadhyay},\ and\ \citenamefont {Saha}}]{Saha:2018}%
  \BibitemOpen
  \bibfield  {author} {\bibinfo {author} {\bibfnamefont {A.}~\bibnamefont {Saha}}, \bibinfo {author} {\bibfnamefont {S.}~\bibnamefont {Gangopadhyay}}, \ and\ \bibinfo {author} {\bibfnamefont {S.}~\bibnamefont {Saha}},\ }\href {\doibase 10.1103/PhysRevD.97.044015} {\bibfield  {journal} {\bibinfo  {journal} {Phys. Rev. D}\ }\textbf {\bibinfo {volume} {97}},\ \bibinfo {pages} {044015} (\bibinfo {year} {2018})}\BibitemShut {NoStop}%
\bibitem [{\citenamefont {von Lüpke}\ \emph {et~al.}(2022)\citenamefont {von Lüpke}, \citenamefont {Yang}, \citenamefont {Bild}, \citenamefont {Michaud}, \citenamefont {Fadel},\ and\ \citenamefont {Chu}}]{vonLupke2022}%
  \BibitemOpen
  \bibfield  {author} {\bibinfo {author} {\bibfnamefont {U.}~\bibnamefont {von Lüpke}}, \bibinfo {author} {\bibfnamefont {Y.}~\bibnamefont {Yang}}, \bibinfo {author} {\bibfnamefont {M.}~\bibnamefont {Bild}}, \bibinfo {author} {\bibfnamefont {L.}~\bibnamefont {Michaud}}, \bibinfo {author} {\bibfnamefont {M.}~\bibnamefont {Fadel}}, \ and\ \bibinfo {author} {\bibfnamefont {Y.}~\bibnamefont {Chu}},\ }\href {\doibase 10.1038/s41567-022-01681-2} {\bibfield  {journal} {\bibinfo  {journal} {Nature Physics}\ }\textbf {\bibinfo {volume} {18}},\ \bibinfo {pages} {794} (\bibinfo {year} {2022})}\BibitemShut {NoStop}%
\bibitem [{\citenamefont {Aspelmeyer}\ \emph {et~al.}(2014)\citenamefont {Aspelmeyer}, \citenamefont {Kippenberg},\ and\ \citenamefont {Marquardt}}]{Aspelmeyer2014}%
  \BibitemOpen
  \bibfield  {author} {\bibinfo {author} {\bibfnamefont {M.}~\bibnamefont {Aspelmeyer}}, \bibinfo {author} {\bibfnamefont {T.~J.}\ \bibnamefont {Kippenberg}}, \ and\ \bibinfo {author} {\bibfnamefont {F.}~\bibnamefont {Marquardt}},\ }\href {\doibase 10.1103/RevModPhys.86.1391} {\bibfield  {journal} {\bibinfo  {journal} {Rev. Mod. Phys.}\ }\textbf {\bibinfo {volume} {86}},\ \bibinfo {pages} {1391} (\bibinfo {year} {2014})}\BibitemShut {NoStop}%
\bibitem [{\citenamefont {Vitali}\ \emph {et~al.}(2007)\citenamefont {Vitali}, \citenamefont {Gigan}, \citenamefont {Ferreira}, \citenamefont {B\"ohm}, \citenamefont {Tombesi}, \citenamefont {Guerreiro}, \citenamefont {Vedral}, \citenamefont {Zeilinger},\ and\ \citenamefont {Aspelmeyer}}]{Vitali2007}%
  \BibitemOpen
  \bibfield  {author} {\bibinfo {author} {\bibfnamefont {D.}~\bibnamefont {Vitali}}, \bibinfo {author} {\bibfnamefont {S.}~\bibnamefont {Gigan}}, \bibinfo {author} {\bibfnamefont {A.}~\bibnamefont {Ferreira}}, \bibinfo {author} {\bibfnamefont {H.~R.}\ \bibnamefont {B\"ohm}}, \bibinfo {author} {\bibfnamefont {P.}~\bibnamefont {Tombesi}}, \bibinfo {author} {\bibfnamefont {A.}~\bibnamefont {Guerreiro}}, \bibinfo {author} {\bibfnamefont {V.}~\bibnamefont {Vedral}}, \bibinfo {author} {\bibfnamefont {A.}~\bibnamefont {Zeilinger}}, \ and\ \bibinfo {author} {\bibfnamefont {M.}~\bibnamefont {Aspelmeyer}},\ }\href {\doibase 10.1103/PhysRevLett.98.030405} {\bibfield  {journal} {\bibinfo  {journal} {Phys. Rev. Lett.}\ }\textbf {\bibinfo {volume} {98}},\ \bibinfo {pages} {030405} (\bibinfo {year} {2007})}\BibitemShut {NoStop}%
\bibitem [{\citenamefont {Deli{\'c}}\ \emph {et~al.}(2020)\citenamefont {Deli{\'c}}, \citenamefont {Reisenbauer}, \citenamefont {Dare}, \citenamefont {Grass}, \citenamefont {Vuleti{\'c}}, \citenamefont {Kiesel},\ and\ \citenamefont {Aspelmeyer}}]{Delic2020}%
  \BibitemOpen
  \bibfield  {author} {\bibinfo {author} {\bibfnamefont {U.}~\bibnamefont {Deli{\'c}}}, \bibinfo {author} {\bibfnamefont {M.}~\bibnamefont {Reisenbauer}}, \bibinfo {author} {\bibfnamefont {K.}~\bibnamefont {Dare}}, \bibinfo {author} {\bibfnamefont {D.}~\bibnamefont {Grass}}, \bibinfo {author} {\bibfnamefont {V.}~\bibnamefont {Vuleti{\'c}}}, \bibinfo {author} {\bibfnamefont {N.}~\bibnamefont {Kiesel}}, \ and\ \bibinfo {author} {\bibfnamefont {M.}~\bibnamefont {Aspelmeyer}},\ }\href {\doibase 10.1126/science.aba3993} {\bibfield  {journal} {\bibinfo  {journal} {Science}\ }\textbf {\bibinfo {volume} {367}},\ \bibinfo {pages} {892} (\bibinfo {year} {2020})},\ \Eprint {http://arxiv.org/abs/1911.04406} {arXiv:1911.04406 [quant-ph]} \BibitemShut {NoStop}%
\bibitem [{\citenamefont {Barman}\ \emph {et~al.}(2021)\citenamefont {Barman}, \citenamefont {Barman},\ and\ \citenamefont {Majhi}}]{Barman:2021bbw}%
  \BibitemOpen
  \bibfield  {author} {\bibinfo {author} {\bibfnamefont {D.}~\bibnamefont {Barman}}, \bibinfo {author} {\bibfnamefont {S.}~\bibnamefont {Barman}}, \ and\ \bibinfo {author} {\bibfnamefont {B.~R.}\ \bibnamefont {Majhi}},\ }\href {\doibase 10.1007/JHEP07(2021)124} {\bibfield  {journal} {\bibinfo  {journal} {JHEP}\ }\textbf {\bibinfo {volume} {07}},\ \bibinfo {pages} {124} (\bibinfo {year} {2021})},\ \Eprint {http://arxiv.org/abs/2104.11269} {arXiv:2104.11269 [gr-qc]} \BibitemShut {NoStop}%
\bibitem [{\citenamefont {Agarwal}\ \emph {et~al.}(1997)\citenamefont {Agarwal}, \citenamefont {Puri},\ and\ \citenamefont {Singh}}]{PhysRevA.56.2249}%
  \BibitemOpen
  \bibfield  {author} {\bibinfo {author} {\bibfnamefont {G.~S.}\ \bibnamefont {Agarwal}}, \bibinfo {author} {\bibfnamefont {R.~R.}\ \bibnamefont {Puri}}, \ and\ \bibinfo {author} {\bibfnamefont {R.~P.}\ \bibnamefont {Singh}},\ }\href {\doibase 10.1103/PhysRevA.56.2249} {\bibfield  {journal} {\bibinfo  {journal} {Phys. Rev. A}\ }\textbf {\bibinfo {volume} {56}},\ \bibinfo {pages} {2249} (\bibinfo {year} {1997})}\BibitemShut {NoStop}%
\bibitem [{\citenamefont {Caldeira}\ and\ \citenamefont {Leggett}(1983)}]{Caldeira:1982iu}%
  \BibitemOpen
  \bibfield  {author} {\bibinfo {author} {\bibfnamefont {A.~O.}\ \bibnamefont {Caldeira}}\ and\ \bibinfo {author} {\bibfnamefont {A.~J.}\ \bibnamefont {Leggett}},\ }\href {\doibase 10.1016/0378-4371(83)90013-4} {\bibfield  {journal} {\bibinfo  {journal} {Physica A}\ }\textbf {\bibinfo {volume} {121}},\ \bibinfo {pages} {587} (\bibinfo {year} {1983})}\BibitemShut {NoStop}%
\bibitem [{\citenamefont {Ghayour}(2012)}]{Ghayour:2012Thesis}%
  \BibitemOpen
  \bibfield  {author} {\bibinfo {author} {\bibfnamefont {B.}~\bibnamefont {Ghayour}},\ }\emph {\bibinfo {title} {Gravitational Waves in Thermal States}},\ \href@noop {} {\bibinfo {type} {Ph.d. thesis}},\ \bibinfo  {school} {University of Hyderabad}, \bibinfo {address} {Hyderabad, India} (\bibinfo {year} {2012})\BibitemShut {NoStop}%
\bibitem [{\citenamefont {Birrell}\ and\ \citenamefont {Davies}(1982)}]{BirrellDavies1982}%
  \BibitemOpen
  \bibfield  {author} {\bibinfo {author} {\bibfnamefont {N.~D.}\ \bibnamefont {Birrell}}\ and\ \bibinfo {author} {\bibfnamefont {P.~C.~W.}\ \bibnamefont {Davies}},\ }\href@noop {} {\emph {\bibinfo {title} {Quantum Fields in Curved Space}}}\ (\bibinfo  {publisher} {Cambridge University Press},\ \bibinfo {year} {1982})\BibitemShut {NoStop}%
\bibitem [{\citenamefont {Botke}\ \emph {et~al.}(1974)\citenamefont {Botke}, \citenamefont {Scalapino},\ and\ \citenamefont {Sugar}}]{Botke1974Coherent}%
  \BibitemOpen
  \bibfield  {author} {\bibinfo {author} {\bibfnamefont {J.~C.}\ \bibnamefont {Botke}}, \bibinfo {author} {\bibfnamefont {D.~J.}\ \bibnamefont {Scalapino}}, \ and\ \bibinfo {author} {\bibfnamefont {R.~L.}\ \bibnamefont {Sugar}},\ }\href {\doibase 10.1103/PhysRevD.9.813} {\bibfield  {journal} {\bibinfo  {journal} {Physical Review D}\ }\textbf {\bibinfo {volume} {9}},\ \bibinfo {pages} {813} (\bibinfo {year} {1974})}\BibitemShut {NoStop}%
\bibitem [{\citenamefont {Kanno}\ \emph {et~al.}(2026)\citenamefont {Kanno}, \citenamefont {Soda},\ and\ \citenamefont {Taniguchi}}]{kv1t-j27m}%
  \BibitemOpen
  \bibfield  {author} {\bibinfo {author} {\bibfnamefont {S.}~\bibnamefont {Kanno}}, \bibinfo {author} {\bibfnamefont {J.}~\bibnamefont {Soda}}, \ and\ \bibinfo {author} {\bibfnamefont {A.}~\bibnamefont {Taniguchi}},\ }\href {\doibase 10.1103/kv1t-j27m} {\bibfield  {journal} {\bibinfo  {journal} {Phys. Rev. Lett.}\ }\textbf {\bibinfo {volume} {136}},\ \bibinfo {pages} {061404} (\bibinfo {year} {2026})}\BibitemShut {NoStop}%
\bibitem [{\citenamefont {Bhaumik}\ \emph {et~al.}(1976)\citenamefont {Bhaumik}, \citenamefont {Bhaumik},\ and\ \citenamefont {Dutta-Roy}}]{Bhaumik1976ChargedCoherent}%
  \BibitemOpen
  \bibfield  {author} {\bibinfo {author} {\bibfnamefont {D.}~\bibnamefont {Bhaumik}}, \bibinfo {author} {\bibfnamefont {K.}~\bibnamefont {Bhaumik}}, \ and\ \bibinfo {author} {\bibfnamefont {B.}~\bibnamefont {Dutta-Roy}},\ }\href {\doibase 10.1088/0305-4470/9/9/011} {\bibfield  {journal} {\bibinfo  {journal} {Journal of Physics A: Mathematical and General}\ }\textbf {\bibinfo {volume} {9}},\ \bibinfo {pages} {1507} (\bibinfo {year} {1976})}\BibitemShut {NoStop}%
\bibitem [{\citenamefont {Gomatam}(1971)}]{Gomatam1971Coherent}%
  \BibitemOpen
  \bibfield  {author} {\bibinfo {author} {\bibfnamefont {R.}~\bibnamefont {Gomatam}},\ }\href@noop {} {\bibfield  {journal} {\bibinfo  {journal} {Physical Review D}\ }\textbf {\bibinfo {volume} {3}},\ \bibinfo {pages} {2169} (\bibinfo {year} {1971})}\BibitemShut {NoStop}%
\bibitem [{\citenamefont {von L{\"u}pke}\ \emph {et~al.}(2022)\citenamefont {von L{\"u}pke}, \citenamefont {Yang}, \citenamefont {Bild}, \citenamefont {Michaud}, \citenamefont {Fadel},\ and\ \citenamefont {Chu}}]{vonLupke2022NatPhys}%
  \BibitemOpen
  \bibfield  {author} {\bibinfo {author} {\bibfnamefont {U.}~\bibnamefont {von L{\"u}pke}}, \bibinfo {author} {\bibfnamefont {Y.}~\bibnamefont {Yang}}, \bibinfo {author} {\bibfnamefont {M.}~\bibnamefont {Bild}}, \bibinfo {author} {\bibfnamefont {L.}~\bibnamefont {Michaud}}, \bibinfo {author} {\bibfnamefont {M.}~\bibnamefont {Fadel}}, \ and\ \bibinfo {author} {\bibfnamefont {Y.}~\bibnamefont {Chu}},\ }\href {\doibase 10.1038/s41567-022-01578-0} {\bibfield  {journal} {\bibinfo  {journal} {Nature Physics}\ }\textbf {\bibinfo {volume} {18}},\ \bibinfo {pages} {794} (\bibinfo {year} {2022})}\BibitemShut {NoStop}%
\bibitem [{\citenamefont {Marti}\ \emph {et~al.}(2024{\natexlab{b}})\citenamefont {Marti}, \citenamefont {von L{\"u}pke}, \citenamefont {Joshi}, \citenamefont {Yang}, \citenamefont {Bild}, \citenamefont {Omahen}, \citenamefont {Chu},\ and\ \citenamefont {Fadel}}]{Marti2024NatPhys}%
  \BibitemOpen
  \bibfield  {author} {\bibinfo {author} {\bibfnamefont {S.}~\bibnamefont {Marti}}, \bibinfo {author} {\bibfnamefont {U.}~\bibnamefont {von L{\"u}pke}}, \bibinfo {author} {\bibfnamefont {O.}~\bibnamefont {Joshi}}, \bibinfo {author} {\bibfnamefont {Y.}~\bibnamefont {Yang}}, \bibinfo {author} {\bibfnamefont {M.}~\bibnamefont {Bild}}, \bibinfo {author} {\bibfnamefont {A.}~\bibnamefont {Omahen}}, \bibinfo {author} {\bibfnamefont {Y.}~\bibnamefont {Chu}}, \ and\ \bibinfo {author} {\bibfnamefont {M.}~\bibnamefont {Fadel}},\ }\href@noop {} {\bibfield  {journal} {\bibinfo  {journal} {Nature Physics}\ }\textbf {\bibinfo {volume} {20}},\ \bibinfo {pages} {1448} (\bibinfo {year} {2024}{\natexlab{b}})}\BibitemShut {NoStop}%
\bibitem [{\citenamefont {Schrinski}\ \emph {et~al.}(2023)\citenamefont {Schrinski}, \citenamefont {Yang}, \citenamefont {von L\"upke}, \citenamefont {Bild}, \citenamefont {Chu}, \citenamefont {Hornberger}, \citenamefont {Nimmrichter},\ and\ \citenamefont {Fadel}}]{PhysRevLett.130.133604}%
  \BibitemOpen
  \bibfield  {author} {\bibinfo {author} {\bibfnamefont {B.}~\bibnamefont {Schrinski}}, \bibinfo {author} {\bibfnamefont {Y.}~\bibnamefont {Yang}}, \bibinfo {author} {\bibfnamefont {U.}~\bibnamefont {von L\"upke}}, \bibinfo {author} {\bibfnamefont {M.}~\bibnamefont {Bild}}, \bibinfo {author} {\bibfnamefont {Y.}~\bibnamefont {Chu}}, \bibinfo {author} {\bibfnamefont {K.}~\bibnamefont {Hornberger}}, \bibinfo {author} {\bibfnamefont {S.}~\bibnamefont {Nimmrichter}}, \ and\ \bibinfo {author} {\bibfnamefont {M.}~\bibnamefont {Fadel}},\ }\href {\doibase 10.1103/PhysRevLett.130.133604} {\bibfield  {journal} {\bibinfo  {journal} {Phys. Rev. Lett.}\ }\textbf {\bibinfo {volume} {130}},\ \bibinfo {pages} {133604} (\bibinfo {year} {2023})}\BibitemShut {NoStop}%
\bibitem [{\citenamefont {Ito}\ \emph {et~al.}(2021)\citenamefont {Ito}, \citenamefont {Soda},\ and\ \citenamefont {Yamaguchi}}]{Ito:2021JCAPbreak}%
  \BibitemOpen
  \bibfield  {author} {\bibinfo {author} {\bibfnamefont {A.}~\bibnamefont {Ito}}, \bibinfo {author} {\bibfnamefont {J.}~\bibnamefont {Soda}}, \ and\ \bibinfo {author} {\bibfnamefont {M.}~\bibnamefont {Yamaguchi}},\ }\href {\doibase 10.1088/1475-7516/2021/03/033} {\bibfield  {journal} {\bibinfo  {journal} {JCAP}\ }\textbf {\bibinfo {volume} {03}},\ \bibinfo {pages} {033} (\bibinfo {year} {2021})},\ \Eprint {http://arxiv.org/abs/2009.03611} {arXiv:2009.03611 [astro-ph.CO]} \BibitemShut {NoStop}%
\bibitem [{\citenamefont {Kahn}\ \emph {et~al.}(2024)\citenamefont {Kahn}, \citenamefont {Sch\"utte-Engel},\ and\ \citenamefont {Trickle}}]{PhysRevD.109.096023}%
  \BibitemOpen
  \bibfield  {author} {\bibinfo {author} {\bibfnamefont {Y.}~\bibnamefont {Kahn}}, \bibinfo {author} {\bibfnamefont {J.}~\bibnamefont {Sch\"utte-Engel}}, \ and\ \bibinfo {author} {\bibfnamefont {T.}~\bibnamefont {Trickle}},\ }\href {\doibase 10.1103/PhysRevD.109.096023} {\bibfield  {journal} {\bibinfo  {journal} {Phys. Rev. D}\ }\textbf {\bibinfo {volume} {109}},\ \bibinfo {pages} {096023} (\bibinfo {year} {2024})}\BibitemShut {NoStop}%
\bibitem [{\citenamefont {Kanno}\ \emph {et~al.}(2025)\citenamefont {Kanno}, \citenamefont {Soda},\ and\ \citenamefont {Taniguchi}}]{Kanno:2023whr}%
  \BibitemOpen
  \bibfield  {author} {\bibinfo {author} {\bibfnamefont {S.}~\bibnamefont {Kanno}}, \bibinfo {author} {\bibfnamefont {J.}~\bibnamefont {Soda}}, \ and\ \bibinfo {author} {\bibfnamefont {A.}~\bibnamefont {Taniguchi}},\ }\href {\doibase 10.1140/epjc/s10052-024-13736-z} {\bibfield  {journal} {\bibinfo  {journal} {Eur. Phys. J. C}\ }\textbf {\bibinfo {volume} {85}},\ \bibinfo {pages} {31} (\bibinfo {year} {2025})},\ \Eprint {http://arxiv.org/abs/2311.03890} {arXiv:2311.03890 [gr-qc]} \BibitemShut {NoStop}%
\bibitem [{\citenamefont {Maggiore}(2000)}]{Maggiore:2000review}%
  \BibitemOpen
  \bibfield  {author} {\bibinfo {author} {\bibfnamefont {M.}~\bibnamefont {Maggiore}},\ }\href {\doibase 10.1016/S0370-1573(99)00102-7} {\bibfield  {journal} {\bibinfo  {journal} {Phys. Rept.}\ }\textbf {\bibinfo {volume} {331}},\ \bibinfo {pages} {283} (\bibinfo {year} {2000})},\ \Eprint {http://arxiv.org/abs/gr-qc/9909001} {arXiv:gr-qc/9909001} \BibitemShut {NoStop}%
\bibitem [{\citenamefont {Ema}\ \emph {et~al.}(2020)\citenamefont {Ema}, \citenamefont {Jinno},\ and\ \citenamefont {Nakayama}}]{Ema:2020inflatonGW}%
  \BibitemOpen
  \bibfield  {author} {\bibinfo {author} {\bibfnamefont {Y.}~\bibnamefont {Ema}}, \bibinfo {author} {\bibfnamefont {R.}~\bibnamefont {Jinno}}, \ and\ \bibinfo {author} {\bibfnamefont {K.}~\bibnamefont {Nakayama}},\ }\href {\doibase 10.1088/1475-7516/2020/09/015} {\bibfield  {journal} {\bibinfo  {journal} {JCAP}\ }\textbf {\bibinfo {volume} {09}},\ \bibinfo {pages} {015} (\bibinfo {year} {2020})},\ \Eprint {http://arxiv.org/abs/2006.09972} {arXiv:2006.09972 [astro-ph.CO]} \BibitemShut {NoStop}%
\bibitem [{\citenamefont {Domcke}\ and\ \citenamefont {Garcia-Cely}(2021)}]{Domcke:2020radio}%
  \BibitemOpen
  \bibfield  {author} {\bibinfo {author} {\bibfnamefont {V.}~\bibnamefont {Domcke}}\ and\ \bibinfo {author} {\bibfnamefont {C.}~\bibnamefont {Garcia-Cely}},\ }\href {\doibase 10.1103/PhysRevLett.126.021104} {\bibfield  {journal} {\bibinfo  {journal} {Phys. Rev. Lett.}\ }\textbf {\bibinfo {volume} {126}},\ \bibinfo {pages} {021104} (\bibinfo {year} {2021})},\ \Eprint {http://arxiv.org/abs/2006.01161} {arXiv:2006.01161 [astro-ph.CO]} \BibitemShut {NoStop}%
\bibitem [{\citenamefont {MacDonald}\ \emph {et~al.}(2016)\citenamefont {MacDonald}, \citenamefont {Hauer}, \citenamefont {Rojas}, \citenamefont {Kim}, \citenamefont {Popowich},\ and\ \citenamefont {Davis}}]{PhysRevA.93.013836}%
  \BibitemOpen
  \bibfield  {author} {\bibinfo {author} {\bibfnamefont {A.~J.~R.}\ \bibnamefont {MacDonald}}, \bibinfo {author} {\bibfnamefont {B.~D.}\ \bibnamefont {Hauer}}, \bibinfo {author} {\bibfnamefont {X.}~\bibnamefont {Rojas}}, \bibinfo {author} {\bibfnamefont {P.~H.}\ \bibnamefont {Kim}}, \bibinfo {author} {\bibfnamefont {G.~G.}\ \bibnamefont {Popowich}}, \ and\ \bibinfo {author} {\bibfnamefont {J.~P.}\ \bibnamefont {Davis}},\ }\href {\doibase 10.1103/PhysRevA.93.013836} {\bibfield  {journal} {\bibinfo  {journal} {Phys. Rev. A}\ }\textbf {\bibinfo {volume} {93}},\ \bibinfo {pages} {013836} (\bibinfo {year} {2016})}\BibitemShut {NoStop}%
\bibitem [{\citenamefont {Li}\ \emph {et~al.}(2025)\citenamefont {Li}, \citenamefont {Ruan}, \citenamefont {Zhao} \emph {et~al.}}]{hxvq-vvnv}%
  \BibitemOpen
  \bibfield  {author} {\bibinfo {author} {\bibfnamefont {L.}~\bibnamefont {Li}}, \bibinfo {author} {\bibfnamefont {X.}~\bibnamefont {Ruan}}, \bibinfo {author} {\bibfnamefont {S.-L.}\ \bibnamefont {Zhao}},  \emph {et~al.},\ }\href {\doibase 10.1103/hxvq-vvnv} {\bibfield  {journal} {\bibinfo  {journal} {Phys. Rev. Appl.}\ }\textbf {\bibinfo {volume} {24}},\ \bibinfo {pages} {064064} (\bibinfo {year} {2025})}\BibitemShut {NoStop}%
\bibitem [{\citenamefont {Toro\ifmmode~\check{s}\else \v{s}\fi{}}\ \emph {et~al.}(2024)\citenamefont {Toro\ifmmode~\check{s}\else \v{s}\fi{}}, \citenamefont {Mazumdar},\ and\ \citenamefont {Bose}}]{PhysRevD.109.084050}%
  \BibitemOpen
  \bibfield  {author} {\bibinfo {author} {\bibfnamefont {M.}~\bibnamefont {Toro\ifmmode~\check{s}\else \v{s}\fi{}}}, \bibinfo {author} {\bibfnamefont {A.}~\bibnamefont {Mazumdar}}, \ and\ \bibinfo {author} {\bibfnamefont {S.}~\bibnamefont {Bose}},\ }\href {\doibase 10.1103/PhysRevD.109.084050} {\bibfield  {journal} {\bibinfo  {journal} {Phys. Rev. D}\ }\textbf {\bibinfo {volume} {109}},\ \bibinfo {pages} {084050} (\bibinfo {year} {2024})}\BibitemShut {NoStop}%
\bibitem [{\citenamefont {Sen}\ \emph {et~al.}(2024)\citenamefont {Sen}, \citenamefont {Gangopadhyay},\ and\ \citenamefont {Bhattacharyya}}]{PhysRevD.110.026008}%
  \BibitemOpen
  \bibfield  {author} {\bibinfo {author} {\bibfnamefont {S.}~\bibnamefont {Sen}}, \bibinfo {author} {\bibfnamefont {S.}~\bibnamefont {Gangopadhyay}}, \ and\ \bibinfo {author} {\bibfnamefont {S.}~\bibnamefont {Bhattacharyya}},\ }\href {\doibase 10.1103/PhysRevD.110.026008} {\bibfield  {journal} {\bibinfo  {journal} {Phys. Rev. D}\ }\textbf {\bibinfo {volume} {110}},\ \bibinfo {pages} {026008} (\bibinfo {year} {2024})}\BibitemShut {NoStop}%
\bibitem [{\citenamefont {Das}\ \emph {et~al.}(2025)\citenamefont {Das}, \citenamefont {Parikh}, \citenamefont {Wilczek},\ and\ \citenamefont {Wutte}}]{Das:2025SqueezedGravity}%
  \BibitemOpen
  \bibfield  {author} {\bibinfo {author} {\bibfnamefont {A.}~\bibnamefont {Das}}, \bibinfo {author} {\bibfnamefont {M.}~\bibnamefont {Parikh}}, \bibinfo {author} {\bibfnamefont {F.}~\bibnamefont {Wilczek}}, \ and\ \bibinfo {author} {\bibfnamefont {R.}~\bibnamefont {Wutte}},\ }\href@noop {} {\  (\bibinfo {year} {2025})},\ \Eprint {http://arxiv.org/abs/2512.20601} {arXiv:2512.20601 [gr-qc]} \BibitemShut {NoStop}%
\end{thebibliography}%



\end{document}